%% file: ms2.tex
\documentclass[usenatbib]{emulateapj}

\usepackage{tabularx,hhline,amsmath,amssymb}

\newcommand{\be}{\begin{equation}}
\newcommand{\ee}{\end{equation}}
\newcommand{\bfig}{\begin{figure}}
\newcommand{\efig}{\end{figure}}
\newcommand{\bfige}{\begin{figure*}}
\newcommand{\efige}{\end{figure*}}
\newcommand{\bea}{\begin{eqnarray}}
\newcommand{\eea}{\end{eqnarray}}

\newcommand{\chandra}{{\it Chandra}~}

\shorttitle{X-ray photometry for \chandra}
\shortauthors{Grimm et al.}

\begin{document}

\title{An X-ray photometry system I: \chandra ACIS}
\author{H.-J. Grimm\altaffilmark{1}, J.~McDowell\altaffilmark{1},
G. Fabbiano\altaffilmark{1}, M. Elvis\altaffilmark{1}}
\affil{Harvard-Smithsonian Center for Astrophysics, 60 Garden Street,
Cambridge, MA 02138}

\begin{abstract}
We present a system of X-ray photometry for the \chandra
satellite. X-ray photometry can be a powerful tool to obtain flux
estimates, hardness ratios, and colors unbiased by assumptions about
spectral shape and independent of temporal and spatial changes in
instrument characteristics. The system we have developed relies on our
knowledge of effective area and the energy-to-channel conversion to
construct filters similar to photometric filters in the optical
bandpass. We show that the filters are well behaved functions of
energy and that this X-ray photometric system is able to reconstruct
fluxes to within about 20\%, without color corrections, for
non-pathological spectra. Even in the worst cases it is better than
50\%. Our method also treats errors in a consistent manner, both
statistical as well as systematic.
\end{abstract}


\section{Introduction}
\label{sec:intro}
Photometry is one of the most widely used, relatively simple, tools
used in describing and categorizing astronomical objects.
Standardization by \citet{johnson:53} and subsequent additions in the
optical and infrared (see e.g. \citet{bessel:05}) have allowed
comparisons between measurements by different telescopes and
instruments without bias. Important applications of optical photometry
include stellar classifications (see e.g. \citet{johnson:53}), galaxy
redshifts \citep{puschell:81}, and the discovery of the most distant
quasars in the SDSS \citep{fan:99}. In particular photometry is
important for sources too faint to extract detailed spectra, i.e. most
sources, given the almost universal increase of source numbers to
faint fluxes.

Although optical astronomy is a much older discipline than X-ray
astronomy, optical photometry was established only about 20 years
before the beginning of X-ray astronomy in the 1960s. Given the
enormous success of optical photometry it seems obvious to use it as a
precedent and try to duplicate its success in other wavebands beyond
UV and infrared. The X-ray band can be defined as reaching from about
0.1 keV to a few hundred keV, spanning almost 4 decades of frequency
-- although most work concentrates on the 0.2--20 keV band -- compared
with 2 octaves in the optical. Moreover, an important difference
between optical and X-ray is the typically low number of photons in
X-ray astronomy. The X-ray range is photon starved such that sources
with a few hundred counts are considered bright in X-rays. This
limitation increases the importance of broad-band photometry in the
X-ray band.

While X-ray astronomy has used relative, mission-specific photometry
for most of its existence, there is, as yet, no standard X-ray
photometric system. The usefulness of an X-ray photometric system is
evident already from the use of these somewhat idiosyncratic energy
bands. The bands used in the past have been chosen for specific
purposes, e.g. to use color--color diagrams to diagnose X-ray binary
spectral states \citep{white:84,hasinger:89} where, in the latter, the
energy bands are different for each source. Even so, the resulting
color--color diagrams have immensely increased our knowledge of X-ray
binary spectral/accretion states
\citep[e.g.][]{prestwich:03,gierlinski:06}. Thus a standard X-ray
photometric system is highly desirable in X-ray astronomy in order to
cross-compare observations of the hundreds of thousands of sources
being cataloged by XMM \citep{watson:07}, \chandra
\citep{fabbiano:07}, and other missions. Even within a given mission
different types of CCDs (XMM) or changes in operating temperature,
gain or contamination (\chandra) mean that simple count rates cannot 
be used.

However, there have been complicating factors in establishing
photometric energy bands beyond individual observations.
One cause of this lack of standardization is that the energy
ranges covered by different X-ray satellites and instruments differ
widely. For example, RXTE/PCA, GINGA and EXOSAT bands have practically
no overlap with ROSAT bands; and ASCA, XMM and \chandra bands are
somewhere in the middle. Fig. \ref{fig:ebands} shows a selection of
energy bands used by different authors for different X-ray
satellites\footnote{For more information on X-ray satellites, see the
HEASARC web page http://heasarc.gsfc.nasa.gov/docs/observatories.html.
A list of energy bands and references is given in the on-line version
of this paper.}. Most X-ray missions with focusing optics cover the
energy range from $\sim$0.1 keV to 10 keV.

Another reason for the lack of a standard photometric system is that in
X-rays there are no bright constant point sources in the way that
stars can be used for calibration like in the optical.

But the most fundamental cause for the lack of an X-ray photometric
system has been the limited spectral resolving power (R$=$E/$\Delta
E\sim$1) of proportional counters, which were used in X-ray astronomy
from the earliest days through to ROSAT and RXTE. A resolution of
R$\sim$1 allows no clean separation of energy bands, and different
spectra with similar flux will give widely different flux estimates in
any chosen band. In optical terms, the ``color correction'' is very
large. However, with the introduction of X-ray CCDs in ASCA
\citep{burke:93} this limitation has largely gone away. X-ray CCDs
have R$>$10, so comparable to the R$\sim$6 of broad band optical
photometry. It seems thus timely to consider the introduction of an
X-ray photometric system. Therefore we have investigated how good a
photometric system can be created for \chandra ACIS observations and,
by extension, for all other X-ray CCDs. We report the encouraging
results in this paper.

\bfig[h]
  \resizebox{\hsize}{!}{\includegraphics{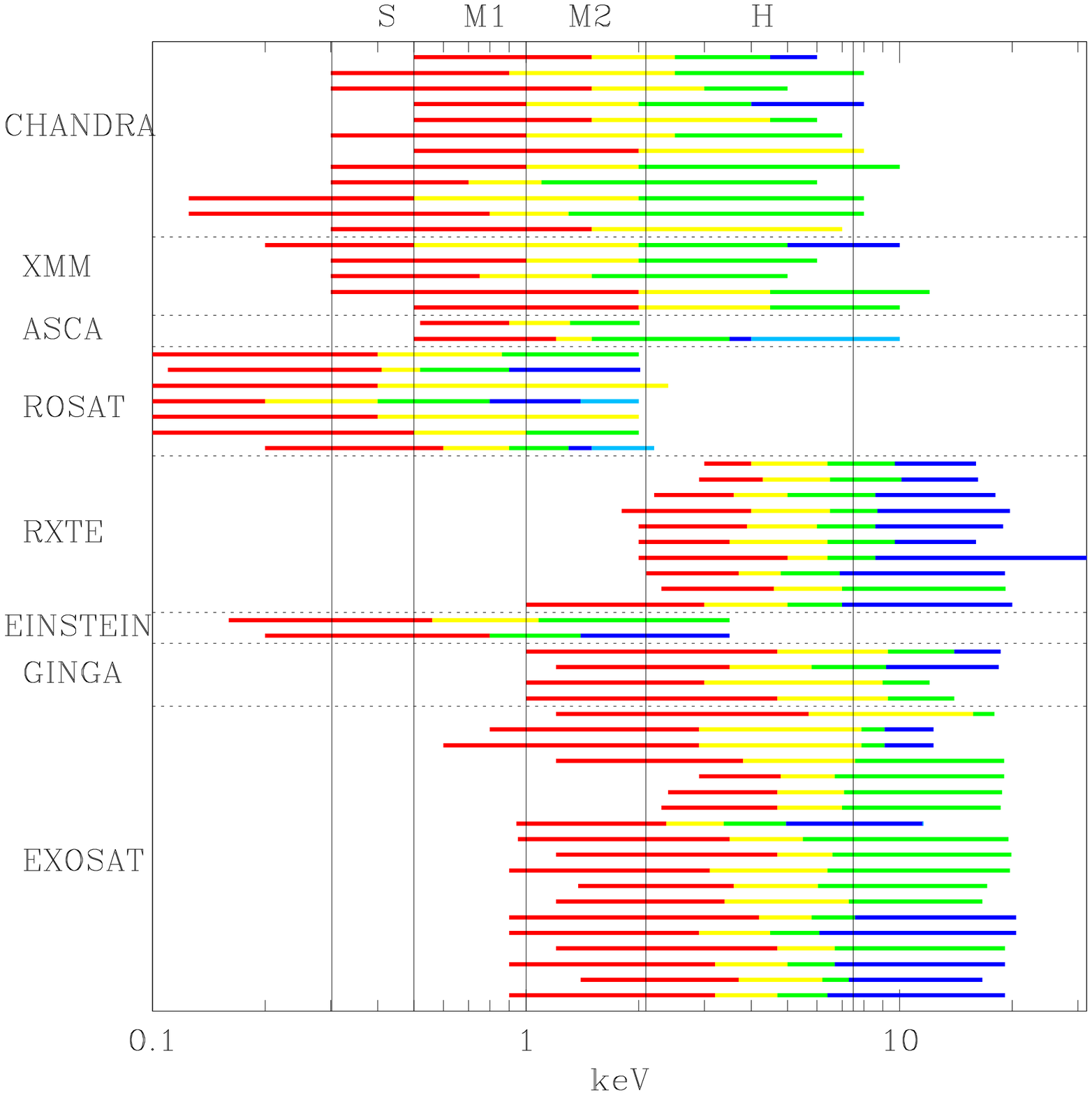}}
  \caption{Energy bands used in various publications for X-ray
satellites. The different energy bands for different satellites are
due to energy coverage of the instruments.  The colors represent the
width of the soft/medium/hard bands used in the corresponding
papers. A table of energy bands and references are given in the
on-line version of the paper.\label{fig:ebands}}
\efig

\section{Measurements with \chandra ACIS}

The ACIS (Advanced CCD Imaging Spectrometer) instrument is one of the
two detectors on \chandra. ACIS is arranged as an array of 10 1024 by
1024 pixel CCDs, one as a two by two array (ACIS-I), and one as a 1 by
six array (ACIS-S). The instrument allows simultaneous high resolution
imaging and moderate resolution spectroscopy. Two CCDs on ACIS-S are
back-illuminated (ACIS-S1 and ACIS-S3), the rest is
front-illuminated. 

\subsection{Energy bands}
\label{sec:bands}
The black vertical lines in the Fig. \ref{fig:ebands} are the energy
bands we have chosen. They are roughly an octave wide, though the hard
band is somewhat wider.

\begin{table}[h]
\caption{Standard photometric bands for \chandra}
\begin{tabular}{|c|c|c|c|}
\hline
   S     &    M1    &     M2   &   H \\
 soft    & medium-soft  & medium-hard  & hard \\
\hline
0.3--0.5 keV & 0.5--1.0 keV & 1.0--2.1 keV & 2.1--7.5 keV\\
\hline
\end{tabular}
\label{tab:bands}
\end{table}

Because of the similarity of the XMM and \chandra energy ranges wes.
will use these bands as a starting point. Most publications use
similar bands as can be seen in Fig. \ref{fig:ebands}.

The low energy bound is chosen to be 0.3 keV because the steepness of
the ACIS effective area curve for $E<0.3$ keV leads to large
dependencies on spectral shape and, secondly, the effective area and
photon energy-to-channel conversion for the S-3 chip are not well
calibrated below 0.23 keV (P. Plucinksy, priv. comm.). The iridium
edge at $\sim$2.1 keV defines the separation between the second medium
and hard band. Note that other X-ray mirrors are coated with other
high Z elements (e.g. Au, Pt) but these have edges at similar energies
(2.3 keV and 2.13 keV, respectively). The high energy boundary is
defined by the rapidly rising ACIS background and falling effective
area at energies above $\sim$7.5 keV (Chapter 6.16 in \citet{pog:08}).

The exact location of the boundaries will not strongly affect the
analysis of source colors, except for pathological spectral shapes or
strong lines at or close to boundaries. For \chandra a much stronger
effect, at least in the soft(er) band(s), is the build up of
contaminating material on the front of the ACIS, which reduces the
effective area significantly compared to the case of no contamination
(see \ref{sec:acis_arf} and \citet{plucinsky:03, marshall:04}). 

The number and boundaries of energy bands are of course not immutable,
optical photometry also has a variety of more or less different
bands for different scientific purposes. For example narrow band
filters at neutral or hydrogen-like iron lines or silicon lines can
be of interest. Our program to compute correction factors for the
count to flux conversion is not limited to the above mentioned energy
bands. However, the use of a common standard system across missions is
highly desirable as explained in Sec.\ref{sec:intro}.

\subsection{ACIS ARF}
\label{sec:acis_arf}
The ARF (Ancillary Response File) describes the effective
area of a telescope at a given energy. All effects related to the
probability of detecting a photon with the telescope and detector
(quantum efficiency, blockage, mirror area, vignetting) are
combined in the ARF. In an ideal system the effective area would be
independent of energy. However, the shape of the effective area curve
of X-ray telescopes varies far more than for optical telescopes, with
variations of factors of a few, even across the octave-wide energy
bands used here. Optical telescopes also cover a much smaller
logarithmic range of photon energies.

For the back-illuminated (BI) ACIS chips of \chandra, there is
significant time variation in the effective area at low energies. Due
to resublimation of material on the chips, sensitivity at low
energies, up to 0.6 keV, decreased by a factor of $\sim$30 since the
beginning of the mission. The sensitivity of the front-illuminated
(FI) chips also suffers from accumulation of material but the
reduction of low energy effective area is not as pronounced as the
area is already quite small for these chips \citep{plucinsky:03}. This
is shown in Fig. \ref{fig:arf}.

\bfig[h]
  \resizebox{\hsize}{!}{\includegraphics{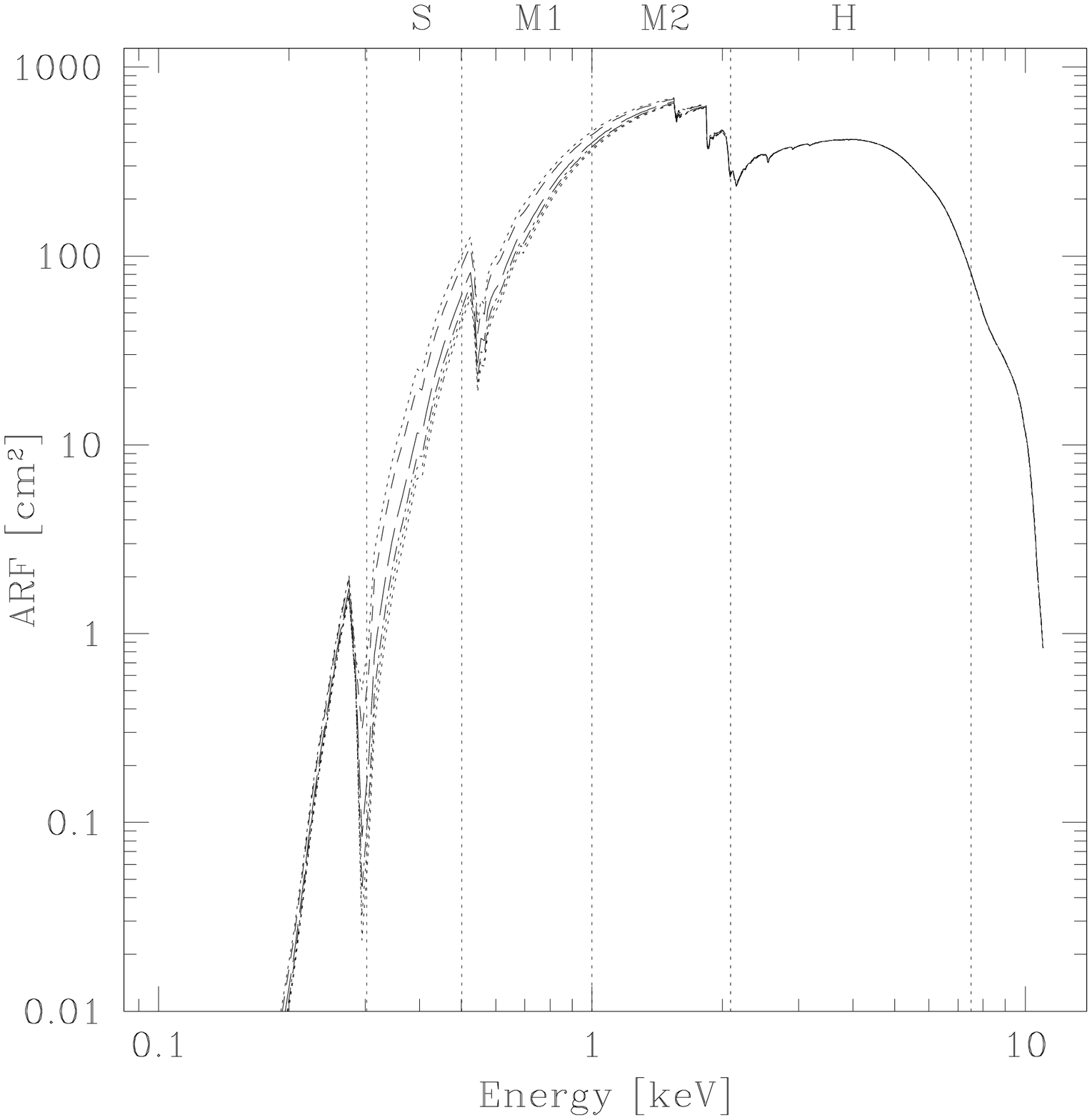}}
  \resizebox{\hsize}{!}{\includegraphics{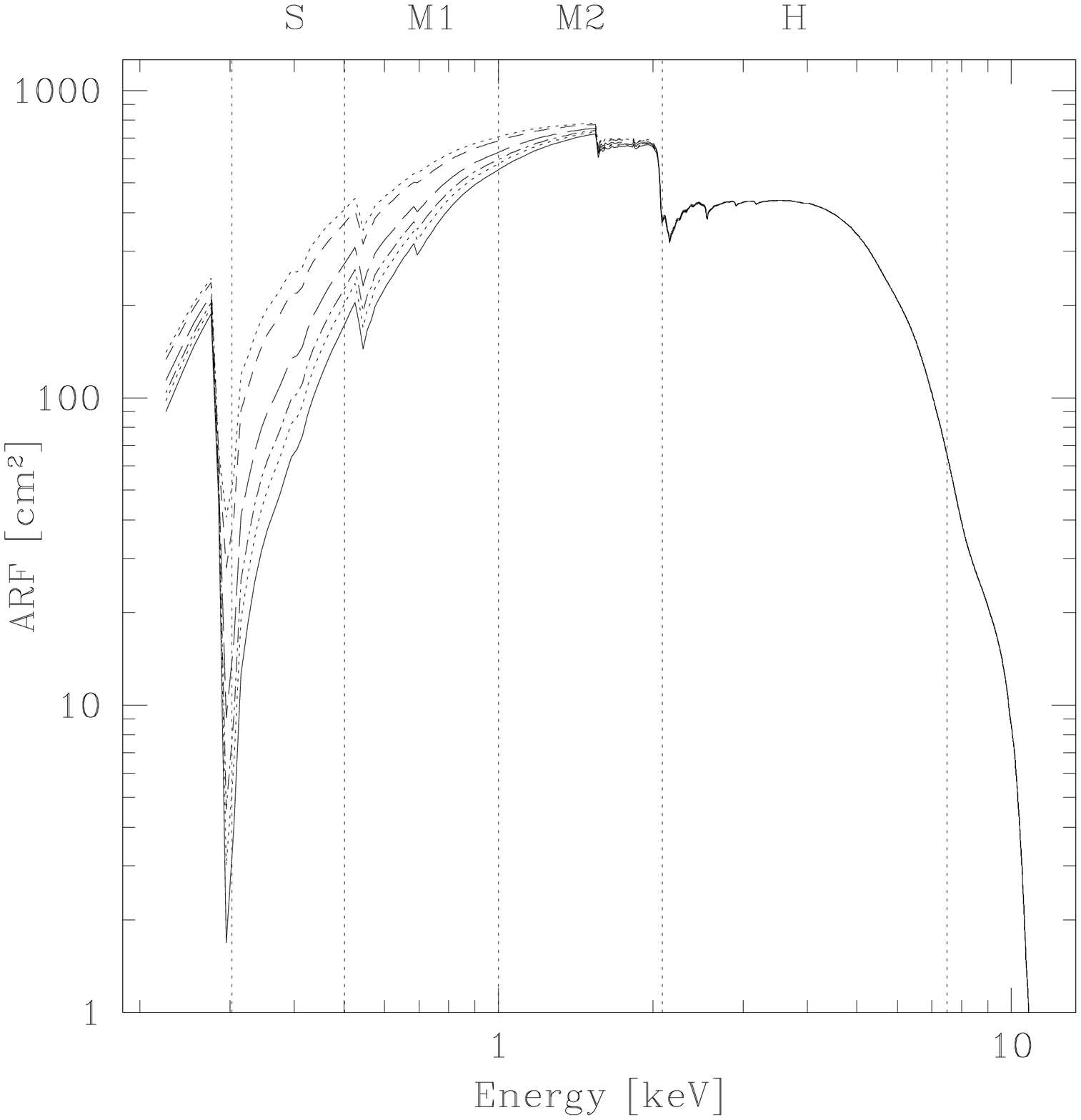}}
  \caption{Effective are curves for the front-illuminated ACIS-I3 chip
(top) and the back-illuminated ACIS-S3 chip (bottom) at the aimpoints
for the years 1999 to 2008. Years are shown in different lines
styles: dotted--1999, dashed--2000, to solid 2008. Years 2004--2007
are not shown because they are almost identical to the 2008
curve. The vertical dotted lines delimit the energy bands.\label{fig:arf}}
\efig

For more details about changes in the \chandra ARF see
\citet{schwartz:00,zhao:04}.

\subsection{ACIS RMF}
\label{sec:rmf}
The RMF (Redistribution Matrix File) is a matrix that redistributes
incident photon energies $E$ (raw energy channels) to PHA or PI
channels or observed energy $E^{\prime}$. Neglecting the effective
area the observed spectrum in a channel $Sp(E^{\prime})$ is related to
the true spectrum at a given energy $E$ by the convolution of incident
spectrum with RMF:
\bea
Sp(E^{\prime}) &=& Sp(E)\otimes RMF(E^{\prime},E) \nonumber \\ &=& \int_{0}^{\infty} Sp(E)\,
RMF(E^{\prime},E)\,dE
\eea
For \chandra the RMF matrix is not symmetric. RMFs for the aimpoints
of the main FI and BI chips are shown in Fig. \ref{fig:rmfs}. The
incident photon energies are binned from 0.1 keV to 11 keV in 10 eV
steps. Thus for FI chips there are 1090 raw energy channels; for BI
chips there are only 1078 raw channels because these chips are not
calibrated below 0.23 keV. The PHA/PI channel number is chosen to be
1024 for both kinds of chips. The values in a row, corresponding to a
single raw channel or real photon energy, correspond to the
probability of a photon with a given energy being detected in a given
PHA/PI channel.
\bfig[h]
  \resizebox{\hsize}{!}{\includegraphics{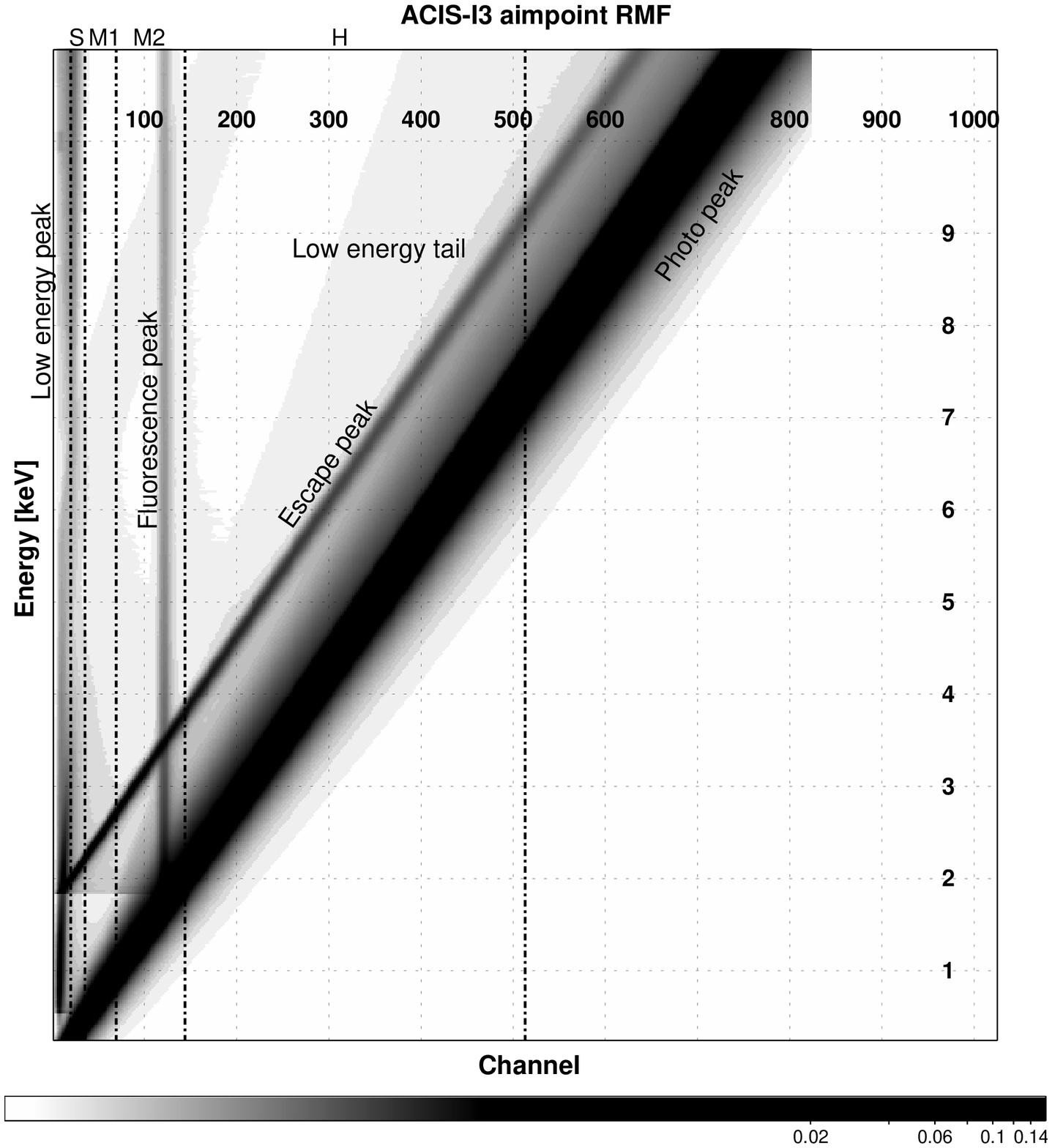}}
  \resizebox{\hsize}{!}{\includegraphics{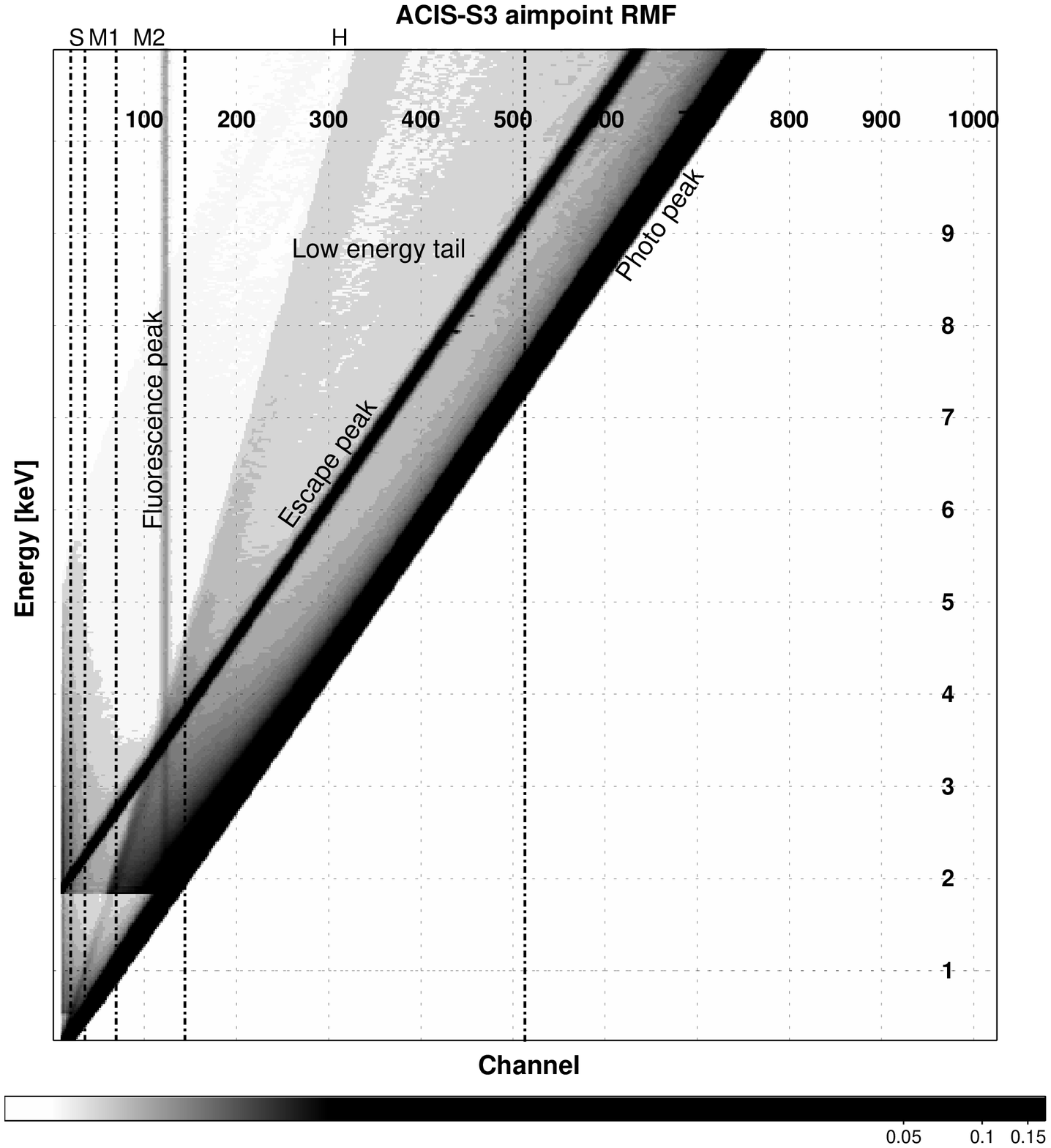}}
  \caption{Response matrices for ACIS-I3 (top) and ACIS-S3 (bottom) at
the aimpoint. Grey-scale is logarithmic. Numbers on the right are
energy in units of keV, numbers at the top are PHA/PI channel
numbers. The vertical dash-dotted lines delimit the energy
bands.\label{fig:rmfs}}
\efig

The energy to channel conversion is ideally a one to one correlation.
But the CCD has a finite energy resolution ($\Delta E\approx 40-170$
eV, depending on energy, location and CCD type). This is larger than
the PHA/PI channel width ($\sim$15 eV). Moreover, due to instrumental
effects (escape peak, fluorescence peak low energy peak and tail, etc.)
a photon of a given energy has a finite probability of being detected
in channels corresponding to a lower energy. This is comparable to
``red leak'' in optical filters.

Taking a slice of the RMF for a given energy (7.5 keV in
Fig. \ref{fig:rmf_slice}) results in an approximately Gaussian shaped
curve for the main photo peak, containing $\sim$95\% of the total,
with the FWHM of the main peak contributing about 76\% of the total.
At around channel 20 there is the so called low energy peak, with an
amplitude of about 3 orders of magnitude less than the main peak, that
contains about 0.1\% of the total. If the photon energy is high
enough, i.e. above the Si K edge, there is a secondary peak, the Si
fluorescence peak, centered on channel 119 (1.73 keV). The maximum of
this secondary peak is about 2-3 orders of magnitude smaller than the
main peak. This peak also contains about 0.1\% of the total. For
photon energies above $\sim$2 keV there is a third peak, the Si escape
peak, with an amplitude two orders of magnitude smaller than the main
peak. It contributes about 0.6\%. It follows about 100 channels behind
the main photo peak. Starting at lower energies from the main peak and
between the other peaks there is a low energy tail at a level about
3-4 orders of magnitude below the main peak. This tail contains
$\sim$4.3\% with the biggest contribution, $\sim$3\%, in between the
Si escape peak and the photopeak. The numbers are almost identical for
the two kinds of chips. All these features, peaks and tail,
are due to detector effects. In particular, the low energy tail and
peak depend on the location at which an X-ray photon interacts with
the material of the CCD and produces an electron cloud. An electron
cloud from an X-ray photon interaction completely within the gate
insulator of the CCD chip produces the low energy peak. Electron
clouds that extend partly into the actual detector material, the
depleted silicon, contribute to the low energy tail. Finally, the main
photo peak is made up of X-ray photon interactions completely within
the depleted silicon. The Si fluorescence peak is produced by X-rays
exciting Si K shell electrons in the detector material. The escape
peak is produced by fluorescence photons that leave the Si substrate
or interact away at another location with the detector. Fig.
\ref{fig:rmf_slice} shows a slice of the RMF for a photon of 7.5 keV
for the FI ACIS-I3 chip (solid line) and  for the BI ACIS-S3 chip
(dotted line).
\bfig[h]
  \resizebox{\hsize}{!}{\includegraphics{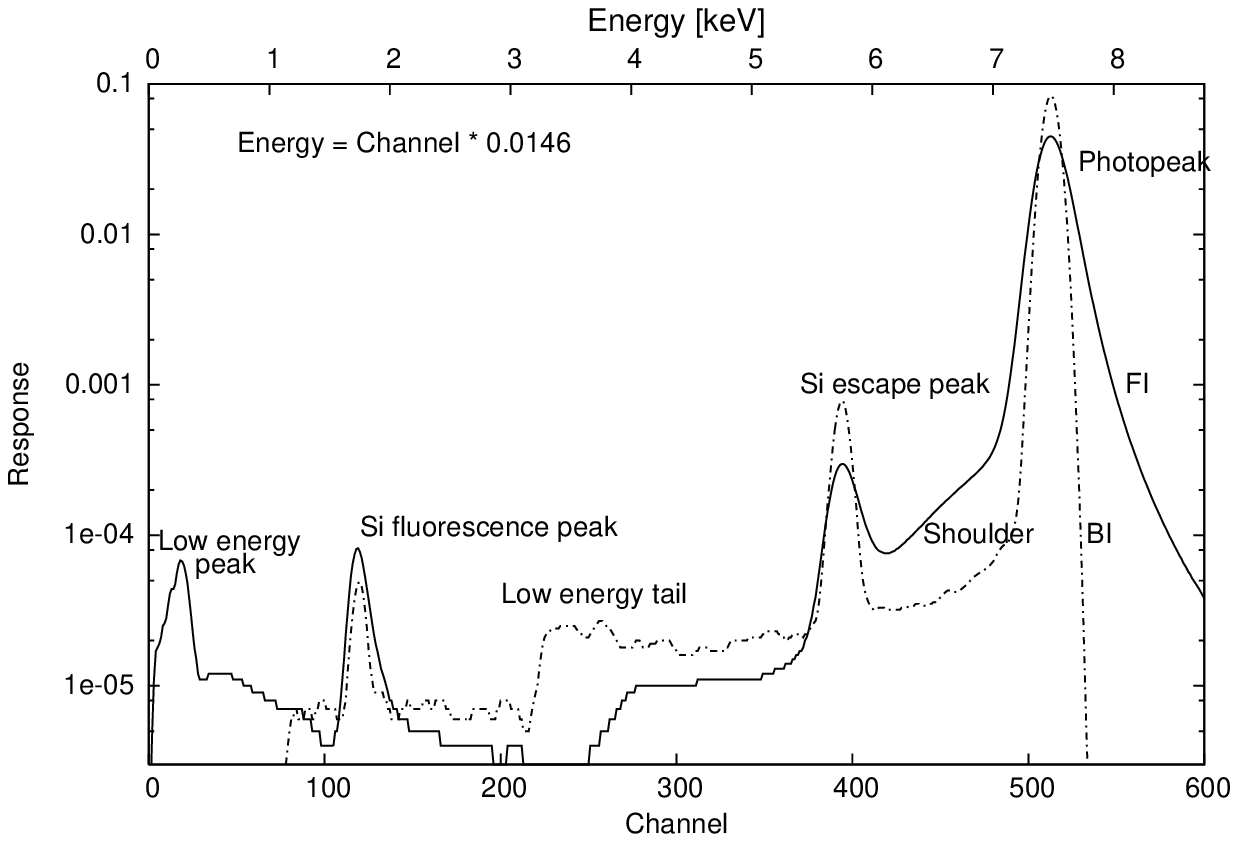}}
  \caption{Slice of the response matrices in PHA space for a photon of
energy 7.5 keV. ACIS-I3 (FI) is the solid line and ACIS-S3 (BI) is the
dotted line. Both curves are for the aimpoint. The photopeak is the
primary signal; the other features are reduced by some two orders of
magnitude or more (see text). The broader photopeak for ACIS-I3
results from the larger distance of the aimpoint to the readout
compared to ACIS-S3.\label{fig:rmf_slice}}
\efig

For calibration purposes all \chandra ACIS chips are divided into
tiles of rectangular form and three different sizes: ACIS-S3 (BI) has
a tile size of of 32$\times$32 pixels, ACIS-S1 (BI) has 64$\times$64
pixels, and all the FI chips have 32$\times$256 pixels. This
amounts to 2304 tiles over all chips, and thus, in principle 2304
possibly different RMFs. A schematic of the tiling is shown in
Fig. \ref{fig:rmftiles} in the Appendix.

The main variation among the ACIS RMFs is the widening of the main
Gaussian peaks with increasing distance from the readout for the FI
chips, i.e. the energy resolution decreases the further a source is
from the readout due to charge transfer inefficiency. The energy
resolution changes from a FWHM of $\sim$4 channels ($\sim$60 eV) at 1
keV at the readout to $\sim$9 channels ($\sim$130 eV) at the opposite
side of the chips. For BI chips spatial variation of the energy
resolution has a more complicated shape, but the variation is not as
strong as for FI chips. At 1 keV the FWHM changes from 6.3 channels at
the readout to 7.4 channels at the opposite side of the chip.

For more details about the \chandra RMFs see \citet{prigozhin:98,bautz:99}.

\subsection{Photometric bands}

The equivalent of filter shapes for the energy bands in Table
\ref{tab:bands} can be obtained by the convolution of the effective
area (ARF) with the RMF PI channels over each energy band. The
filter bands constructed for the aimpoints of ACIS-I3 and ACIS-S3 for
1999 and 2007 are shown in Fig. \ref{fig:filters}. The units are in
cm$^2$. The four bands have quite different normalizations. The soft
band, like the U band in the optical, has the smallest area. The
medium-hard and hard bands are quite symmetric and have sharp edges to
their main response, as is desirable. The soft and first medium
filters have peaks biased towards the high energy ends of their
bands. This is the result of the strong energy dependence of the
effective area. The low energy leak is quite small. The leakage
between energy bands is only in the few per cent range, and moreover
can be corrected quite easily, see below.

\bfig[h]
  \resizebox{\hsize}{!}{\includegraphics{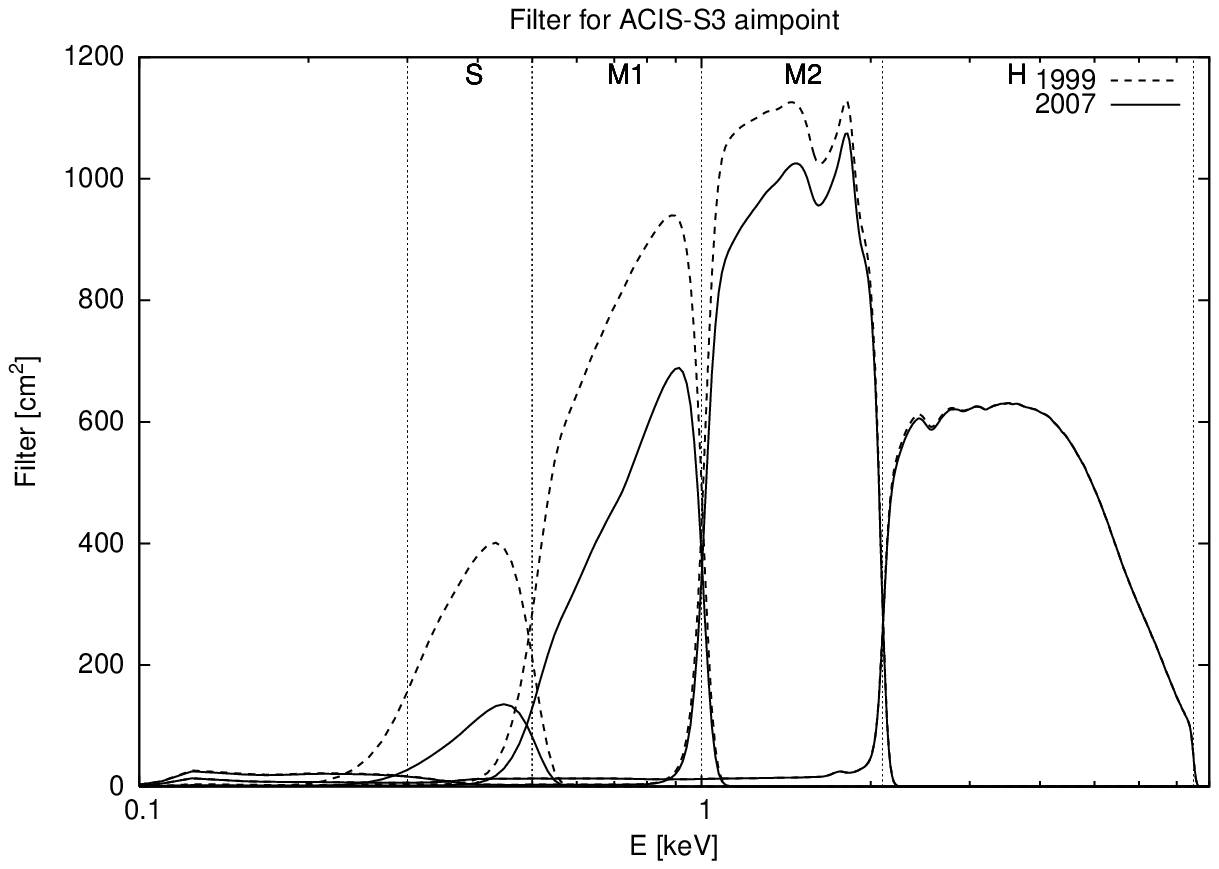}}
  \resizebox{\hsize}{!}{\includegraphics{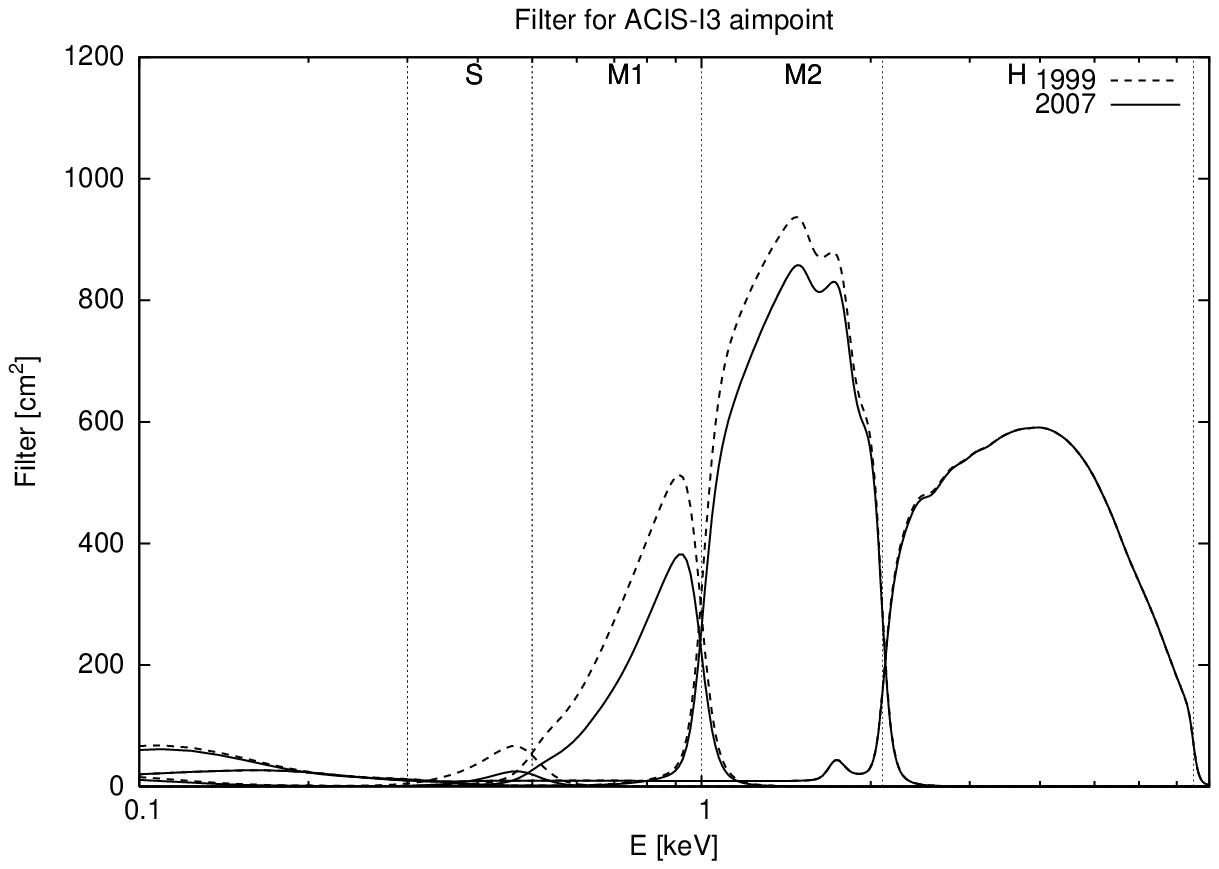}}
  \caption{X-ray filters for the passbands defined in
Sec. \ref{sec:bands} (dotted vertical lines). Filters are constructed
for the aimpoints of ACIS-I3 for 1999 and 2007 (top) and ACIS-S3 for
the same years (bottom). Dashed curves refer to 1999, solid curves to
2007.\label{fig:filters}}
\efig

The X-ray filters have a well defined spatial (from RMF) and temporal
(from ARF) dependence. The spatial dependence is related to energy
resolution, whereas the main temporal effect is decreasing sensitivity
due to accumulation of material on the CCDs.

Except for the hard band, there is a strong effect due to the increased
absorption on the CCD between year 1999 and 2007. The increasing energy
resolution towards the readout leads to a decreasing overlap between
the filter bands. However, the energy resolution of \chandra is
generally sufficiently good that this is a small effect ($\sim$3\%)
even at the opposite chip end from the readout.

These \chandra X-ray filters differ from optical filters in that they
have an area normalization given by the ARF, but they are comparable
in basic shape. Fig. \ref{fig:sdss_filter} shows a comparison between
the X-ray \chandra filters at the aimpoint of ACIS-S3 in 2008 (in keV)
and SDSS ugri filters for zero airmass (in eV). Based on the
fractional width of the filters, the X-ray filters are only a factor
of 2-4 wider than the SDSS filters. Their variation of the peak
sensitivity is about the same.

\bfig[h]
  \resizebox{\hsize}{!}{\includegraphics{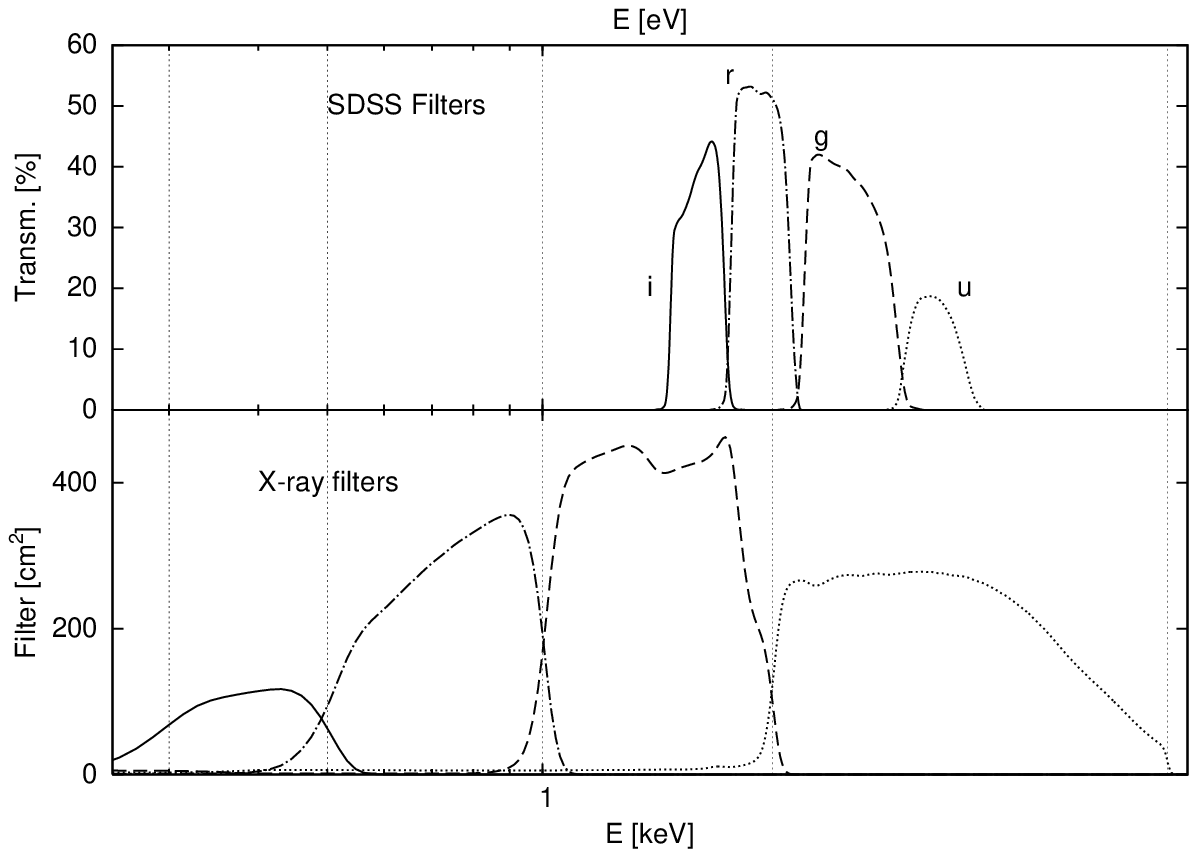}}
  \caption{Bottom: Filters for the \chandra X-ray photometry system,
i.e. combination of ARF and RMF. Abscissa in keV (log scale). Top:
SDSS ugri filters for 0 airmass. Abscissa in eV (log scale). The
variation of peak transmission among the X-ray filters is comparable
to the variation among the optical filters, with the exception of the
softest X-ray band. Measured as $\Delta E/E$ the X-ray bands are only
a factor of few broader than the optical bands. Note that the width of
the four optical filters covers about 3 eV, whereas the four X-ray
filters cover about 7000 eV. \label{fig:sdss_filter}}
\efig

Due to the effects described in Sec. \ref{sec:rmf} and illustrated in
Fig. \ref{fig:rmf_slice} each filter contributes somewhat to its
adjacent filters and to all lower energy filters. Thus the filter
areas are described with a $n\times n$ matrix where $n$ is the number
of filters. Since no photon can be detected at higher energies than
its own (plus energy resolution effects) the upper right elements of
this matrix are zero. This ignores pileup effects in which two photons
are detected as one at the sum of the individual photon energies. As
an illustration we give the area matrix at the ACIS-S3 aimpoint for
2008:
\bea
A = \left(\begin{array}{cccc}
18.7 &   2.1 &   0.0 &    0.0 \\
 4.8 & 241.1 &  11.9 &    0.0 \\
 1.2 &  13.4 & 998.0 &    9.9 \\
 2.1 &   6.3 &  30.2 & 2376.2
\end{array} \right)
\label{eq:area}
\eea
The units for the elements are in keV cm$^2$.
\section{Spectral Models}

X-ray sources can be separated in various categories, starting
with the distinction between point sources and extended sources. The
definition of point source is obviously a question of angular
resolution, e.g. we know that in X-ray binaries, which are point-like
even with \chandra, there are sometimes various regions contributing
to X-ray emission. However, given the excellent angular resolution of
\chandra (0.3'' FWHM at the aimpoint), we consider that a point source
for \chandra will be a point source in 15--20 years as well. The
approach of photometry is of limited usefulness for extended sources,
i.e. plasmas, since a definition of regions is somewhat arbitrary and
interpreting photometric differences among regions of one source would
require plasma diagnostics that are beyond the scope of this
work. Since plasmas are also present in "point sources",
e.g. extragalactic SNRs, a model is nevertheless included in the
following.

To make the photometry system useful for as large a part of the
community as possible, it is necessary to study and characterize the
accuracy of the photometry for a wide variety of commonly encountered
spectral models. Moreover, one of the motivations for a photometric
system in X-rays is to not make assumptions about spectral
models. However, the conversion of counts, effective area, and
exposure time to a flux estimate requires some knowledge of the
behavior of spectra. We thus start by taking common spectral models
and define a wide range of interest for their parameters. In effect
this wide ranging combination of spectral shapes and parameters
represents our ignorance of the true spectral shape/parameters of a
given source.

\begin{table}[h]
\caption{Spectral models and parameter ranges used to estimate
photometric accuracy} 
\begin{tabular}{|c|c|c|c|}
\hline
Component    &  Parameter & Range  & Step size \\
\hline
Absorption   & N$_{H}$$^{a}$ & $10^{20}$--$10^{24}$ & 0.2 in log\\
Power law    & $\Gamma$   & -1.0--4.0 & 0.2\\
Black Body   & kT$^{b}$    & 0.1--2.0 & 0.2 \\
Bremsstrahlung & kT$^{b}$  & 0.5--6.0 & 0.3 \\
Opt. thin plasma & kT$^{b}$ & 0.1--5.0 & 0.3 \\
             & abundance$^{c}$ & 0.1--1.0 & 0.1\\
Gaussian Line & Energies$^{b}$      & 0.8, 1.3, 1.85, 6.4 & n/a\\
             & Line widths$^{b}$    & 0.1, 0.5, 1.0 & n/a\\
\hline
\end{tabular}
\label{tab:components}
\noindent$^{a}$ -- units in cm$^{-2}$\\
\noindent$^{b}$ -- units in keV\\
\noindent$^{b}$ -- units in solar abundance\\
\end{table}

The spectral parameters and models are quite simple. For example a
flat or highly absorbed power law will produce most detected photons
in the hard band, while a steep or weakly absorbed power law will give
a spectrum dominated by the soft band. Spectral shapes outside the
ranges used are extremely hard to detect with X-ray telescopes
operating in the 0.1--10 keV range. But as we will show below the
resulting correction to the count to flux conversion is quite robust
against differences in the spectral model in a given energy band.

\section{Flux estimation}
In an ideal case the flux $F$ of a source could be simply computed from
the number of observed counts $C_s$, effective area $A_{eff}$, and
exposure time $T$ as
\begin{equation}
F = \frac{C_s}{A_{eff}T}.
\end{equation}
However this is true only for flat effective area and no cross-talk
between energy bands. To obtain a more accurate estimate of the flux
we have to correct for these effects. The effect of cross-talk between
bands can be estimated to first order by computing the contributions
of individual filters to other energy bands; in reality contributions
to adjacent bands are the dominating factor, see Eq. \ref{eq:area}.
The effect of varying effective area in a band is coupled with the
effect of the unknown spectral shape and is estimated as follows.

\subsection{Correction factors}
\label{sec:corrfac}
Having a spectrum, ARF, and RMF is sufficient to compute a theoretical
correction factor. This is not the general conversion factor from counts
to flux that we ultimately want, but is rather a theoretical value
describing how accurately finite energy bands can describe a true flux
for a given spectrum. The flux density at an observed energy
$E^{\prime}$ for a known spectral shape, ARF and RMF is 
\begin{eqnarray}
\mathcal{F}(E^{\prime}) &=& Sp(E) \, ARF(E)\otimes RMF(E^{\prime},E) \nonumber \\
&=& \int_{0}^{\infty} Sp(E) \, ARF(E)\, RMF(E^{\prime},E) dE,
\end{eqnarray}
where $Sp$ is the original incident spectrum in units of
photons/cm$^2$/s/keV, $ARF$ the effective area in units of cm$^2$,
$RMF$ is the energy-to-channel conversion, which is dimensionless.

Using a broad photometric band we can estimate the number of counts
in that band $C_s$ as the product of the integrals over the spectrum
and the effective area over the energy band.
\begin{equation}
C_s = \int_{E_1}^{E_2} Sp(E)\,dE\cdot \int_{E_1}^{E_2}ARF(E)\,dE/\Delta E,
\end{equation}
with $E_1$ and $E_2$ the lower and upper bound of the energy band, and
$\Delta E$ the width of the energy band. Note that without loss of
generality exposure time has been set to unity.

A minor complication is that the observed number of counts in a band
is not exactly the number of incident photons in that energy range due
to the redistribution effect of the RMF. This can be most easily seen
in Fig. \ref{fig:filters} as the overlap at the filter
boundaries. Therefore the observed number of counts in any band has
contributions from other bands as well, particularly for the soft
bands. In practice the strongest effect ($\sim$few to ten per cent) is
due to adjacent bands. To more distant bands the contribution is much
smaller ($\sim$1\%). Ideally the matrix of contributions should be
diagonalized. But some of the matrix elements are zero since softer
bands contribute only to the next higher band due to energy resolution
effects. Because of this limitation and since only adjacent bands are
important contributors of counts, we correct for this effect in an
iterative way starting at the highest energy band, assuming that it
does not significantly lose counts to even higher energies. Starting
at the highest band we compute the corrected number of counts as:
\begin{eqnarray}
{C_s}^{\dagger}_j &=& {C_s}_j - {C_s}_i\cdot\frac{A_{ij}}{A_{ii}}\\
{C_s}^{\dagger}_i &=& {C_s}_i + {C_s}_i\cdot\frac{A_{ij}}{A_{ii}}\\
{C_s}^{\prime}_i &=& {C_s}^{\dagger}_i - {C_s}^{\dagger}_j\cdot\frac{A_{ji}}{A_{jj}}\\
{C_s}^{\prime}_j &=& {C_s}^{\dagger}_j + {C_s}^{\dagger}_j\cdot\frac{A_{ji}}{A_{jj}}
\end{eqnarray}
where ${C_s}_{i,j}$ are the observed counts in bands $i$ and $j$ with
$j=i-1$, and $A_{ij}$ the $ij$-element of the filter area matrix (see
Eq. \ref{eq:area}). If any of the corrected counts becomes negative the
corrected counts are set to zero. This iteration can trivially be
extended to more than adjacent bands.

The correction factor for a source with known spectral shape in a
given energy band $i$ is thus defined as
\begin{equation}
K_i = \frac{{C_s}^{\prime}_i}{\int_{E_1}^{E_2} \mathcal{F}(E^{\prime})dE^{\prime}},
\label{eq:corr}
\end{equation}
The resulting correction factor is dimensionless. Note that this formula
does not use the information given by adjacent bands, which can be
used to make a 'color correction', as in optical photometry. We will
address this possibility in a subsequent paper.

Finite CCD spectral resolution, represented by the RMF, and the
separate averaging over spectrum and ARF generate deviations from the
ideal case. This separate averaging over spectrum and ARF gives values
close to unity only if both are not strongly variable within the
band. Thus the correction factor is also a measure of how flat
spectrum and ARF are within the band. In the hard and medium-hard
bands this is mostly a reasonable approximation, but in the soft, and
to some degree in the medium-soft band, the deviation from flatness is
large, see Fig. \ref{fig:arf}. Especially on the FI chips the ARF
drops precipitously below about $\sim$0.6 keV, by a factor of $\sim$30
between 0.5 and 0.3 keV. The corresponding drop for the BI chips is
only a factor $\sim$2.5. Photoelectric absorption in the incident
spectrum is also a strong factor in introducing steep slopes in the
observed spectrum at low energies, e.g. absorption of $\sim 1.5\cdot
10^{22}$ cm$^{-2}$ results in a drop of 17 orders of magnitude in flux
from 0.5 keV to 0.3 keV.

The correction factor varies as a function of spectral parameters in a
well behaved way. As an example Fig. \ref{fig:corrs} shows the
correction factor surface for a power law incident spectrum in the
four photometric bands at the aimpoint of the BI ACIS-S3 chip in 2007
as a function of absorption and photon index. 
\bfige[h]
  \resizebox{0.5\hsize}{!}{\includegraphics{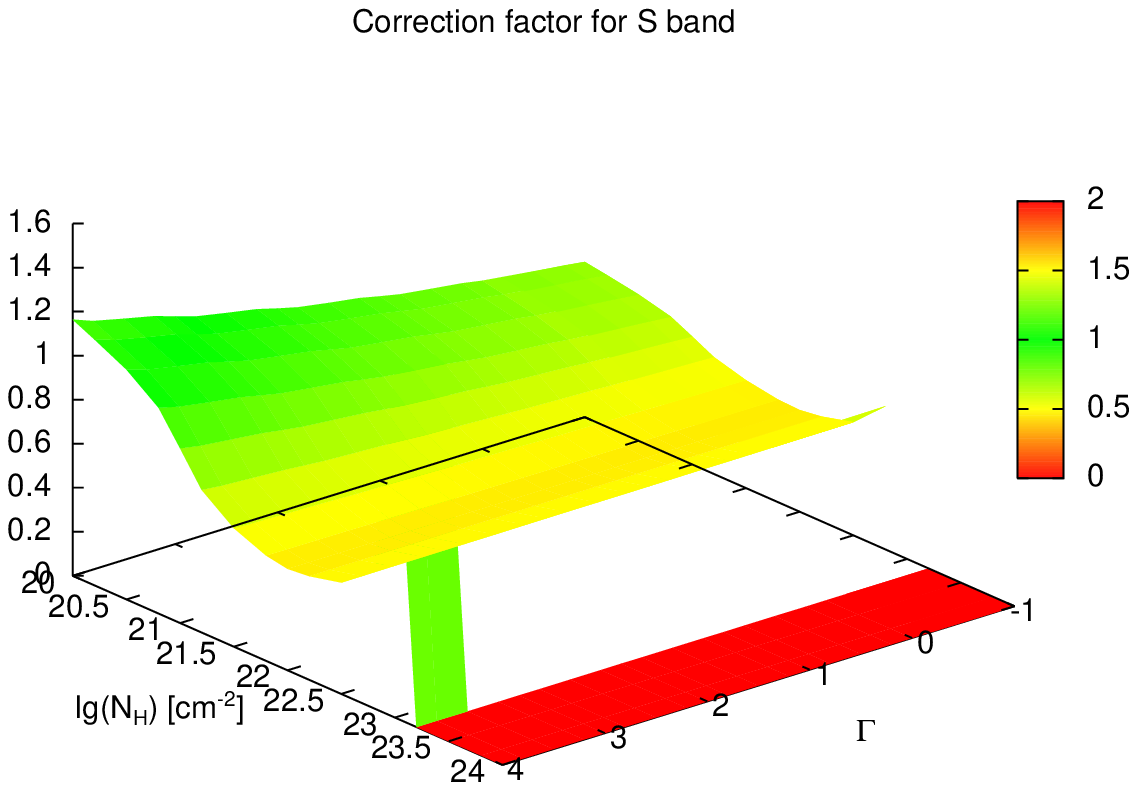}}
  \resizebox{0.5\hsize}{!}{\includegraphics{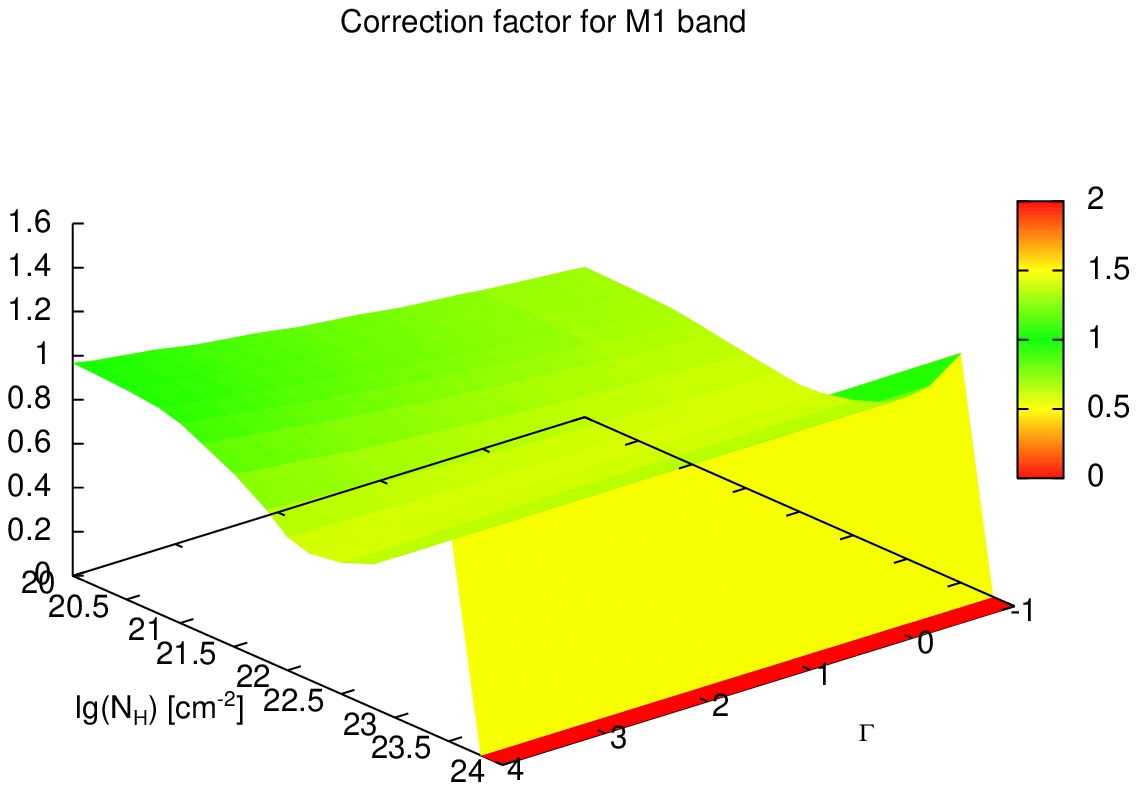}}
  \resizebox{0.5\hsize}{!}{\includegraphics{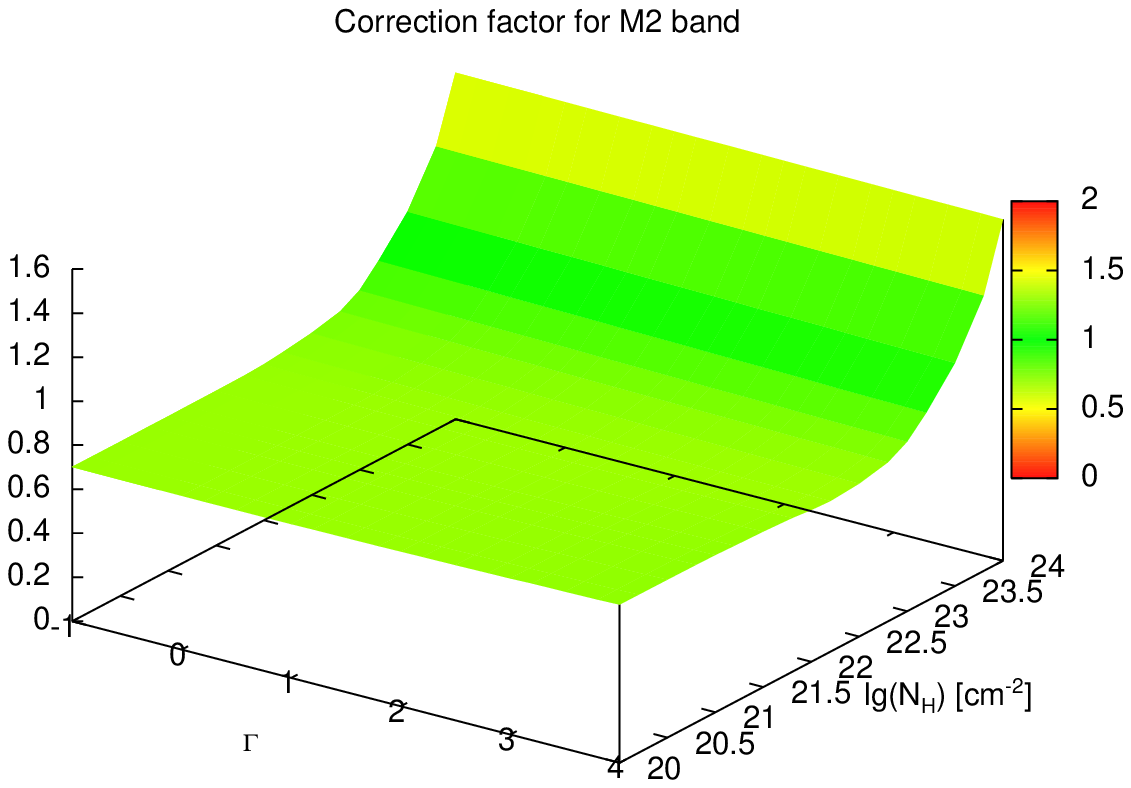}}
  \resizebox{0.5\hsize}{!}{\includegraphics{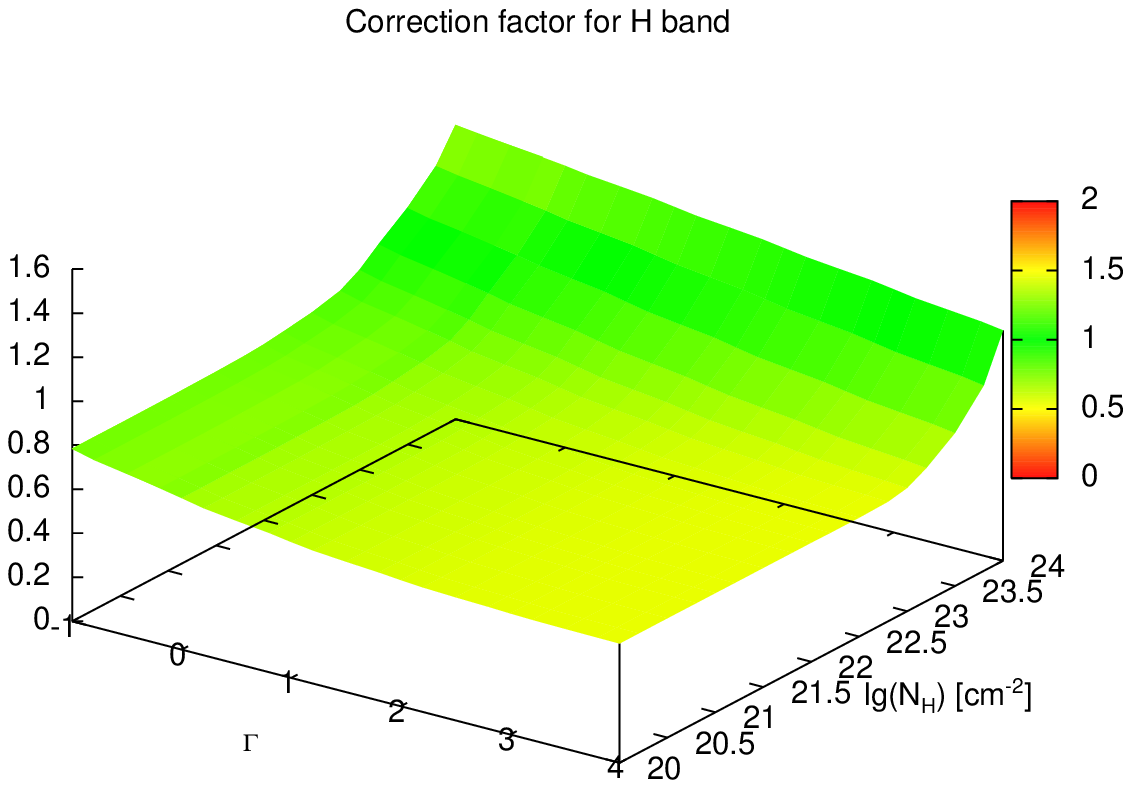}}
  \caption{Surface of correction factors versus spectral parameters
for an absorbed power law in the band passes at the aimpoint of the FI
ACIS-S3 chip in 2008.\label{fig:corrs}}
\efige

In general the correction is relatively small ($\sim$20--30\%). The
main variation occurs for large absorption values ($N_H > 10^{23}$
cm$^{-2}$) where the correction factor drops to zero in the soft and
medium-soft bands, or strongly increases in the medium-hard band. For
practical application, however, a correction factor of zero is not a
problem because a zero correction means that there is no flux in that
energy band down to computational accuracy ($\int Sp dE = 0$ in
Eq. \ref{eq:corr}), thus a correction is not useful. Practically, in
the code we set incident fluxes to zero if the integral over the
spectrum in an energy band is smaller than $10^{-30}$
photons/cm$^2$/s.

Importantly, the shapes of the correction factor surfaces in these
plots are similar for different spectral shapes, see
Figs. \ref{fig:ap}, \ref{fig:bb}, \ref{fig:br} in the Appendix. The
difference between two correction surfaces in one band is generally
less than $\sim$15\%, and never worse than $\sim$40\% for the most
extreme spectral shapes. Since the spectral shape of a source is a
priori unknown, and the point of this work is to not assume any shape,
we combine all the points on each grid for all the different spectral
shapes. The distribution of correction factors taken from all spectra
is shown in Fig.\ref{fig:corr_hist} for the aimpoint of the BI ACIS-S3
chip in 1999 and 2008. The histograms are strongly peaked, although
they have significant tails towards larger correction factors. These
tails are especially pronounced in the medium-hard band and are due to
spectra with strong absorption values.

\bfig[t]
  \resizebox{\hsize}{!}{\includegraphics{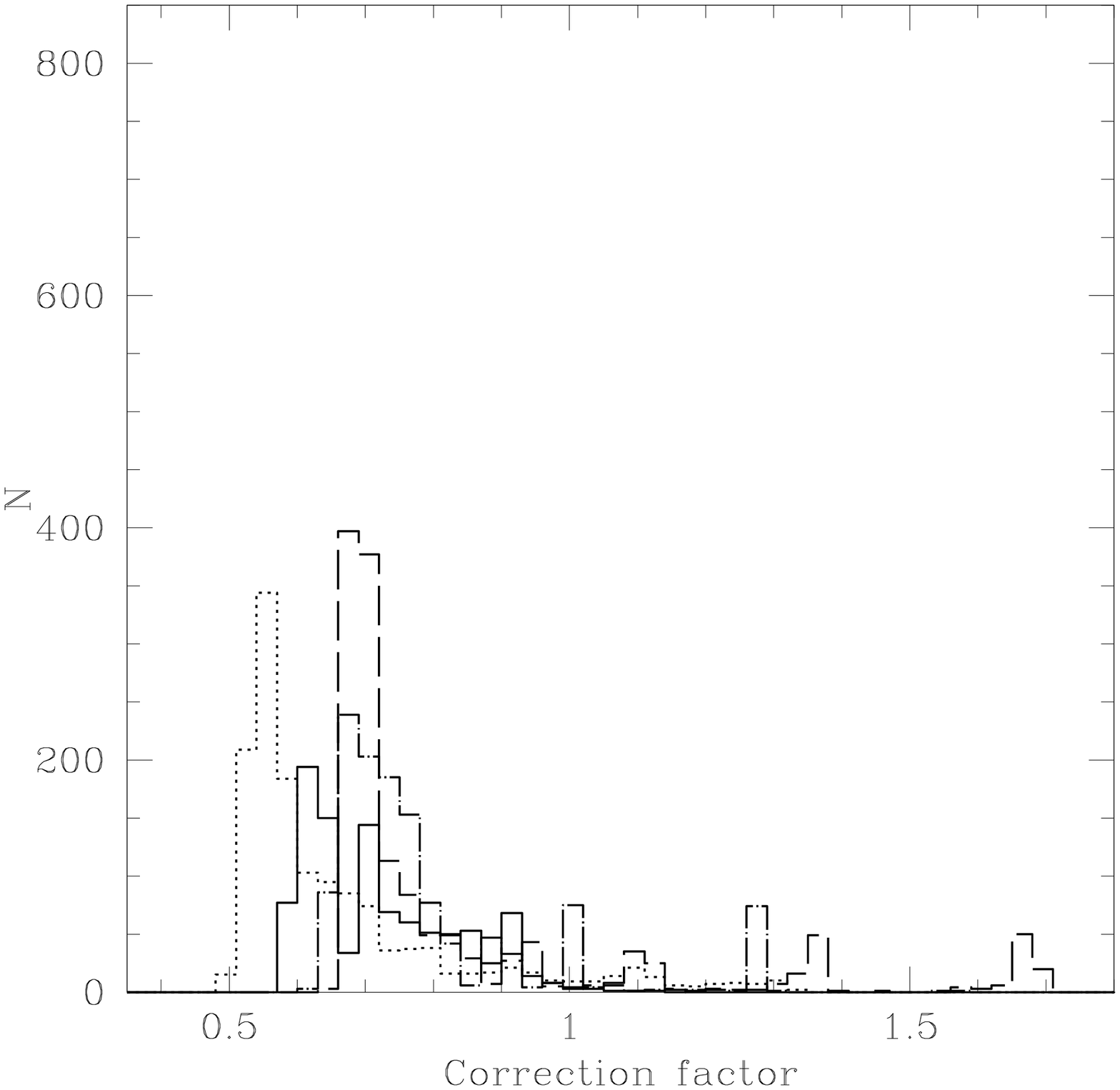}}
  \resizebox{\hsize}{!}{\includegraphics{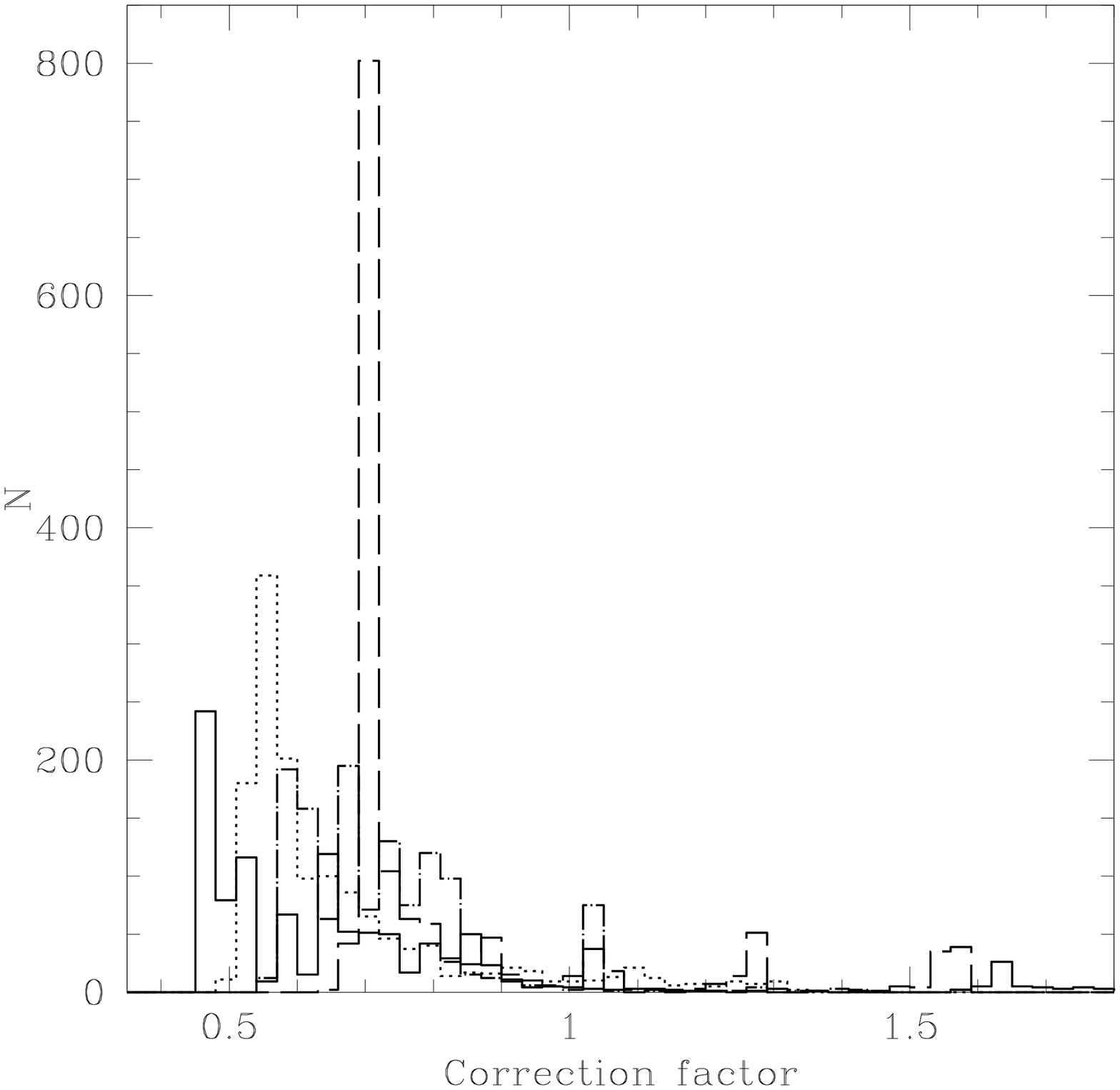}}
  \caption{Distribution of correction factors taken from all spectra
for aimpoint of ACIS-S3 in 1999 and 2008. Soft band--solid histogram,
medium-soft band--dash-dotted histogram, medium-hard band-dashed
histogram, hard band--dotted histogram.\label{fig:corr_hist}}
\efig

Ignoring spectra with large absorption values ($N_H > 10^{23}$
cm$^{-2}$) in the computation of the correction factor has no
significant impact on the peak of these distributions but considerably
reduces the asymmetry of the distribution for the medium-hard and hard
band; the tails almost disappear. This property of highly absorbed
spectra may be used in a ``color correction'' similar to optical
photometry. E.g. if there are no counts in the soft and/or medium-soft
band, this is already an indication of high absorption, which may be
used to change the correction factor distribution in the medium-soft
and/or medium-hard and hard band. We address this band ratio
correction in a subsequent paper. 

Given our ignorance about the intrinsic spectral shape of a source the
correction factor distribution can be considered the probability
density function $PDF$ of picking the right correction factor for
conversion from counts to flux. Thus the probability density function
for the correction factors is simply the conjunction of all correction
factors $K_i$ for a given spectral shape $i$:
\begin{equation}
PDF(K) = \bigwedge_{i}^{N_{sp}}K_{i}.
\label{eq:corr_pdf}
\end{equation}
with $N_{sp}$ the number of different spectra, in this case the sum of
parameter combinations from Table \ref{tab:components}, and $K_{i}$
the correction factor for spectral parameter combination $i$ according
to Eq. \ref{eq:corr}. Note that $N_{sp}$ has to be sufficiently large
to obtain a well behaved distribution. With our choice of spectral
shapes and parameters $N_{sp}$ is 1454.

It is important to note here that our selection of spectral shapes
serves effectively as a prior on this distribution. This prior is flat
in the spectral model/parameter space, i.e. all model/parameter
combinations are given equal weight in computing the correction
factor distribution. If the correct spectral shape is not covered by
this range the method is no longer valid, which is the reason for the
large parameter space we use.

Considering the correction factor distribution as a PDF we use the
mode of the distribution $\widehat{PDF}$ as the correction factor. As
an error estimate we use the range that covers 68\% of the
distribution as measured from the mode. The mode of the correction
factor distribution with the 68\% confidence level (CL) errors at the
aimpoint of ACIS-S3 for each band for the past 10 years is shown in
Fig. \ref{fig:corr_time}.

If not noted otherwise all error bars and quoted
errors in the following correspond to the 68\% confidence level.

\subsection{Flux}
Using the knowledge of the effects of separate averaging over spectrum
and ARF, our ignorance of the true spectral shape of a source, and our
estimates for correcting these effects, we can compute a flux estimate
$F$ for a source in energy band $i$ at a given location on ACIS as
\begin{equation}
F_{i} = \frac{{C'_{s}}_{i}}{\int (ARF \otimes RMF)_{i} dE \cdot T \cdot \widehat{PDF}_{i}}.
\label{eq:flux}
\end{equation}
with $C'_{s}$ as the background subtracted counts corrected for filter
overlap in energy band $i$, $ARF$ and $RMF$ the effective area and
redistribution matrix in band $i$, $T$ the exposure time, and
$\widehat{PDF}_{i}$ the mode of the correction factor distribution in
band $i$. $F_{i}$ is in units of photons/s/cm$^{2}$.

\subsection{Time variation}
Theoretically for \chandra all points on the chips have their own
value for the correction factor which in addition is time
dependent. Practically, however, the spatial variation across a chip
is much smaller than other effects. E.g. the lack of knowledge of the
correct spectral shape which is encoded in the variance of the
correction factor distribution is of order 15\%--20\%. Whereas the
variation due to changing energy resolution across a chip is about
1\%.

Instead temporal changes are important. The change in sensitivity with
time significantly affects the effective area and
thus the correction factor. Fig. \ref{fig:corr_time} shows the change
in the correction factor at the aimpoint of ACIS-S3 for the different
energy bands between 1999 and 2007. There is significant change in the
correction factor for the soft band up to about 2002 due to
accumulation of material. The total change is about 28\% between 1999
and 2008, however the medium-hard and hard band are virtually
unaffected. The variations for the ACIS-I3 aimpoint are almost
identical, except the variation in the soft band is only about 20\%.
\bfig[t]
  \resizebox{\hsize}{!}{\includegraphics{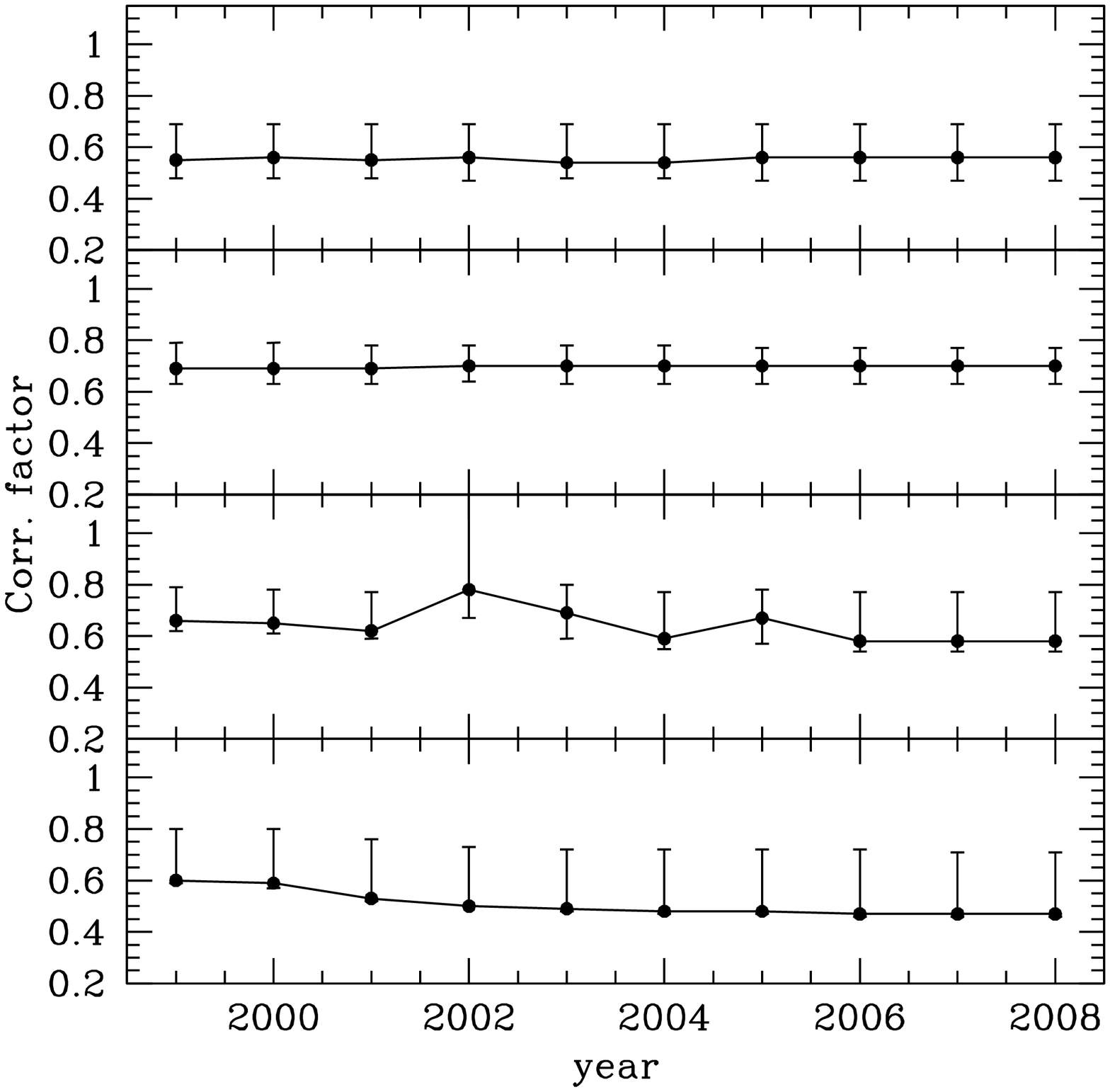}}
  \caption{Change of correction factor with 68\% error bars at the
aimpoint of ACIS-S3 between 1999 and 2008. The top panel is the
hard band correction factor, second from top medium-hard band
correction, third from top medium-soft band correction, and bottom
panel the soft band correction. For the soft band the mode of the
correction factor distribution is at the lowest value, thus there are
no lower error bars (see Fig. \ref{fig:corr_hist}). \label{fig:corr_time}}
\efig

\section{Uncertainties}
\label{sec:unc}
The uncertainties in the X-ray photometry can be separated into
statistical and systematic errors. The statistical errors are
due only to counting statistics, and so are observation dependent. Due
to the fact that our photometry method also deals with small number of
counts, Gaussian statistics and  error propagation are not
appropriate tools. Independent of counting  statistics, we consider the
distribution of correction factors for a specific chip location and
time as a systematic error. This distribution is dependent on the
number and kind of spectral shapes and the chip location in time and
space.

\subsection{Systematic uncertainties}
\label{sec:unc_sys}
Neither for ARF nor RMF values uncertainties are provided although
they are estimated to be below 10\%, and around 5\% for most of the
energy range for the ARF \citep{drake:06}. We therefore use the
correction factor distribution to estimate systematic uncertainties in
our program. Individual correction factors are the result of combining
spectral shape, ARF, and RMF. Since we use a wide range of spectral
shapes and parameters the influence of the spectral shapes/parameters
on the uncertainty in flux is certainly much larger than uncertainties
in the ARF or RMF.

To obtain the systematic error we choose the mode of the correction
factor distribution at a given location of the instrument as the
correction factor for that location and integrate from that position
in positive and negative direction until 68\% of the distribution are
covered. In general this will result in asymmetric errors since the
distribution is skewed. The flux is inversely proportional to the
correction factor, thus the systematic error does not have to be
propagated beyond inverting and multiplying by a numerical factor.

\subsection{Statistical uncertainties}
\label{sec:unc_stat}
To obtain the uncertainty on the flux in a band we propagate the
errors on the source counts through the corrections applied to the
observed counts in a band. In the presence of background we use the
method proposed by \citet{kraft:91} to compute the number of source
counts and the uncertainties in that number. Note that the
uncertainties are asymmetric and not exactly Poissonian.

Unfortunately there is no obvious or generally accepted way to
propagate asymmetric errors. Here we follow the approach of
\citet{barlow:04} for combining asymmetric errors. The approach is
based on the idea to parametrize the Log-likelihood curve of the
original probability distribution with 3 parameters: Location, scale,
and skew for each measurement. Then the Log-likelihood functions for
individual measurements are combined, in this case added. Obviously
there are numerous ways to parametrize a function through three
points. In our program we use the parametrization described as
Gaussian with linear variance. This choice is purely empirical and
made because this parametrization approximates very well a Poissonian
distribution. And although the background subtracted counts and errors
are not Poissonian, the difference is quite small. The likelihood
function for each part is then given by
\begin{equation}
f(a) = \ln(L(\vec{x};a)) = -\frac{1}{2}
\frac{a^2}{\sigma_{-}\sigma_{+} + (\sigma_{+}-\sigma_{-})a}
\end{equation}
where $a$ is the value, and $\sigma_{+-}$ are the asymmetric
uncertainties in negative and positive direction.

Combining the likelihood functions for two distributions one can then
obtain the combined uncertainties at the locations where $f =
f(\hat{a}) - \frac{1}{2}$, with $\hat{a}$ being the sum of the
counts. Using this method we obtain uncertainties that are slightly
larger than assuming Gaussian error propagation for the Gehrels
errors \citep{gehrels:86}. For large number of counts the error
approximates the Gaussian expectation value. Thus we consider our
statistical errors to be conservative.

\section{Validation}
We compare the results of our program with simulated and real data to
validate the results and check for variations.

\subsection{Simulated data}
We simulate 100 spectra each for parameter combinations of two
spectral shapes, a black body with a temperature of 0.3 keV and and
power law with a photon index of 2. Both spectral shapes have
absorption column densities of $10^{20}$ and $10^{22}$ cm$^{-2}$. All
spectra contain 1000 counts and no background. This is a good
approximation for \chandra point sources. The spectra are fitted
with Xspec and fluxes are computed using the {\it flux} command. These
are compared with results from our method.

Results from this comparison are shown in Fig. \ref{fig:fit_comp}. The
figures show  Xspec fluxes ($F_{fit}$) over photometric fluxes
($F_{phot}$) computed with our method versus photometric fluxes for
the two spectral shapes and four spectral parameter
combinations. Different bands are labeled and shown with different
symbols. The error bars used correspond to 68\% CL. Ideally all y-axis
values would be one.
\bfige[h]
  \resizebox{0.5\hsize}{!}{\includegraphics{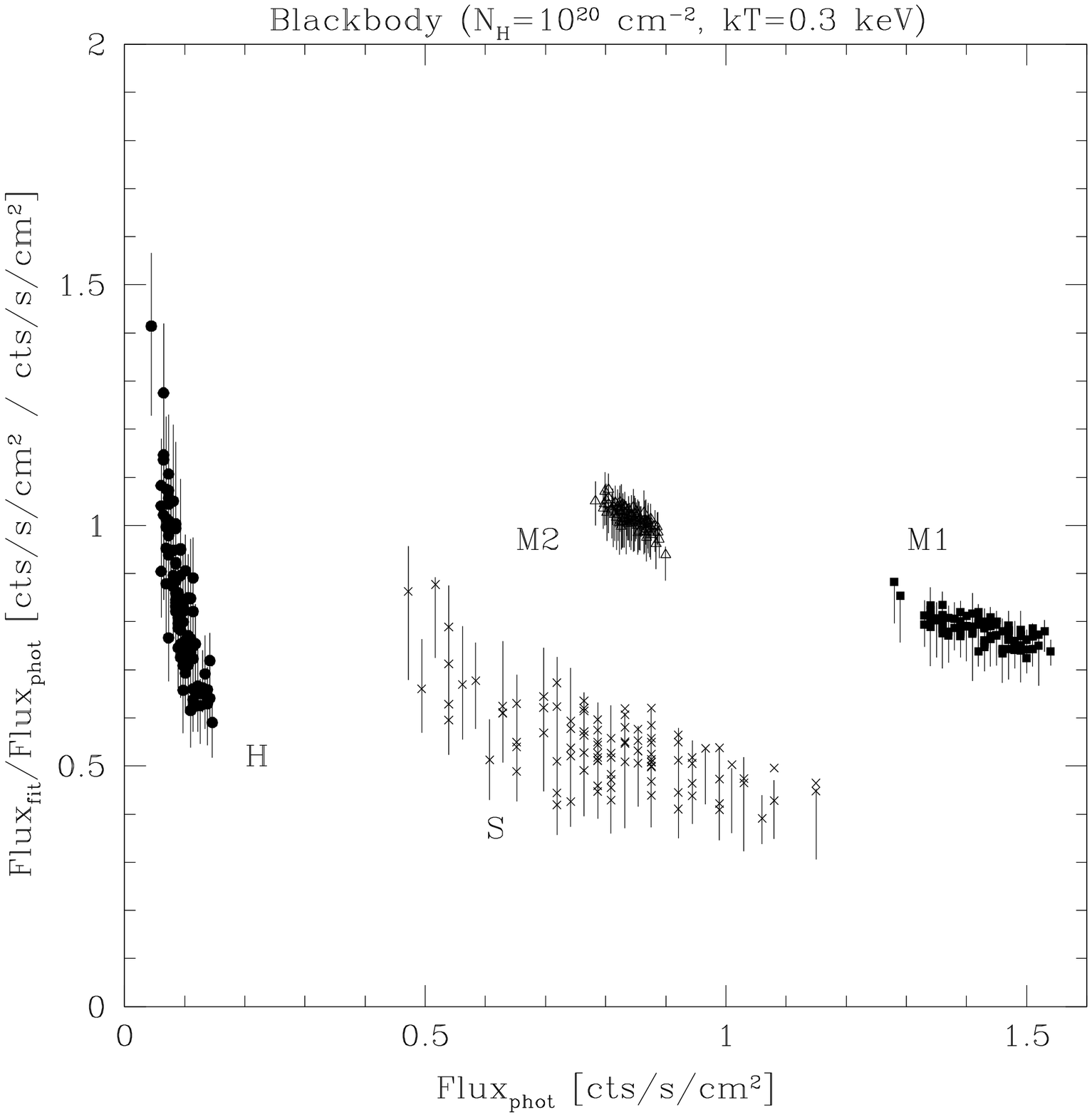}}
  \resizebox{0.5\hsize}{!}{\includegraphics{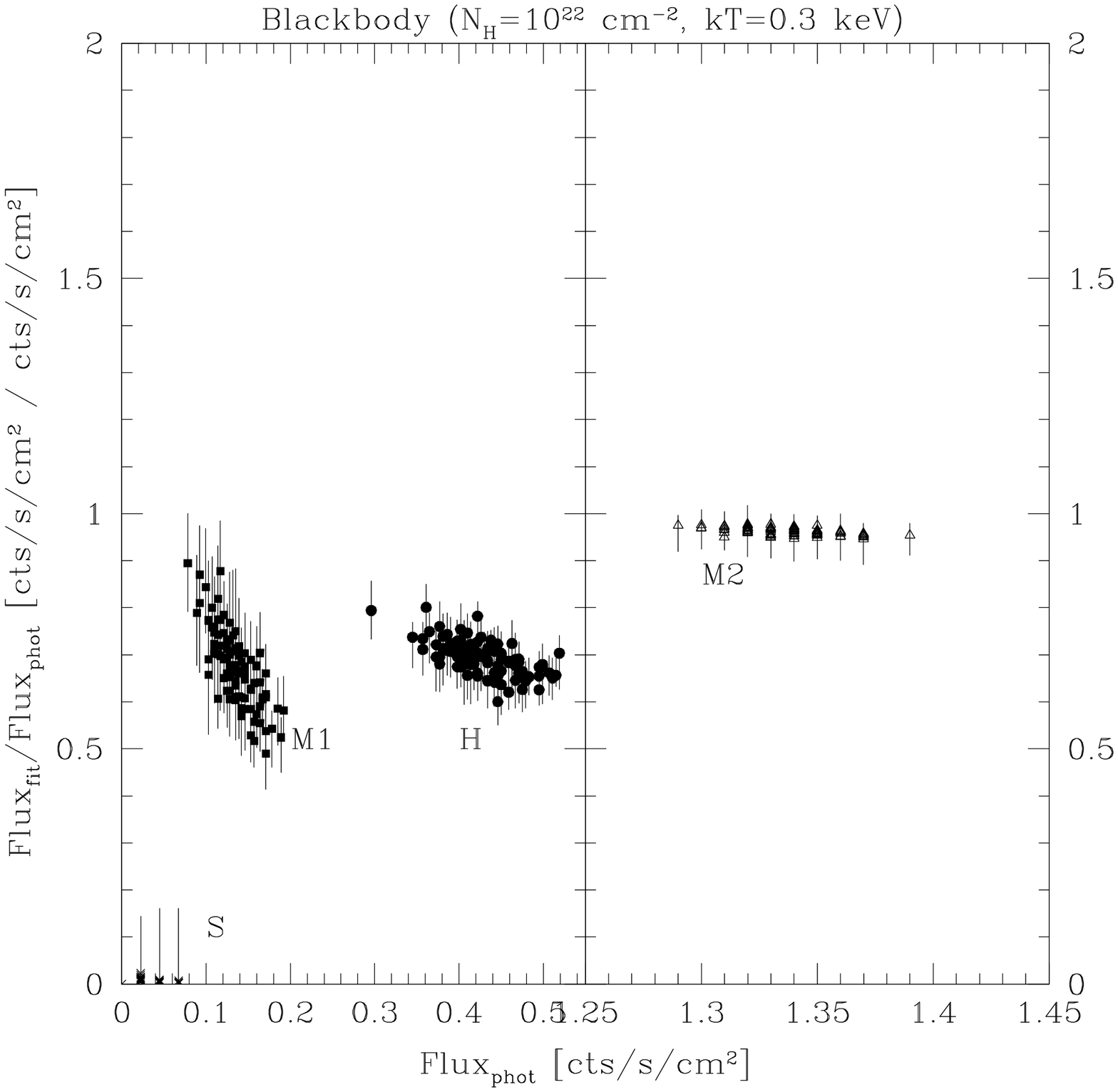}}
  \resizebox{0.5\hsize}{!}{\includegraphics{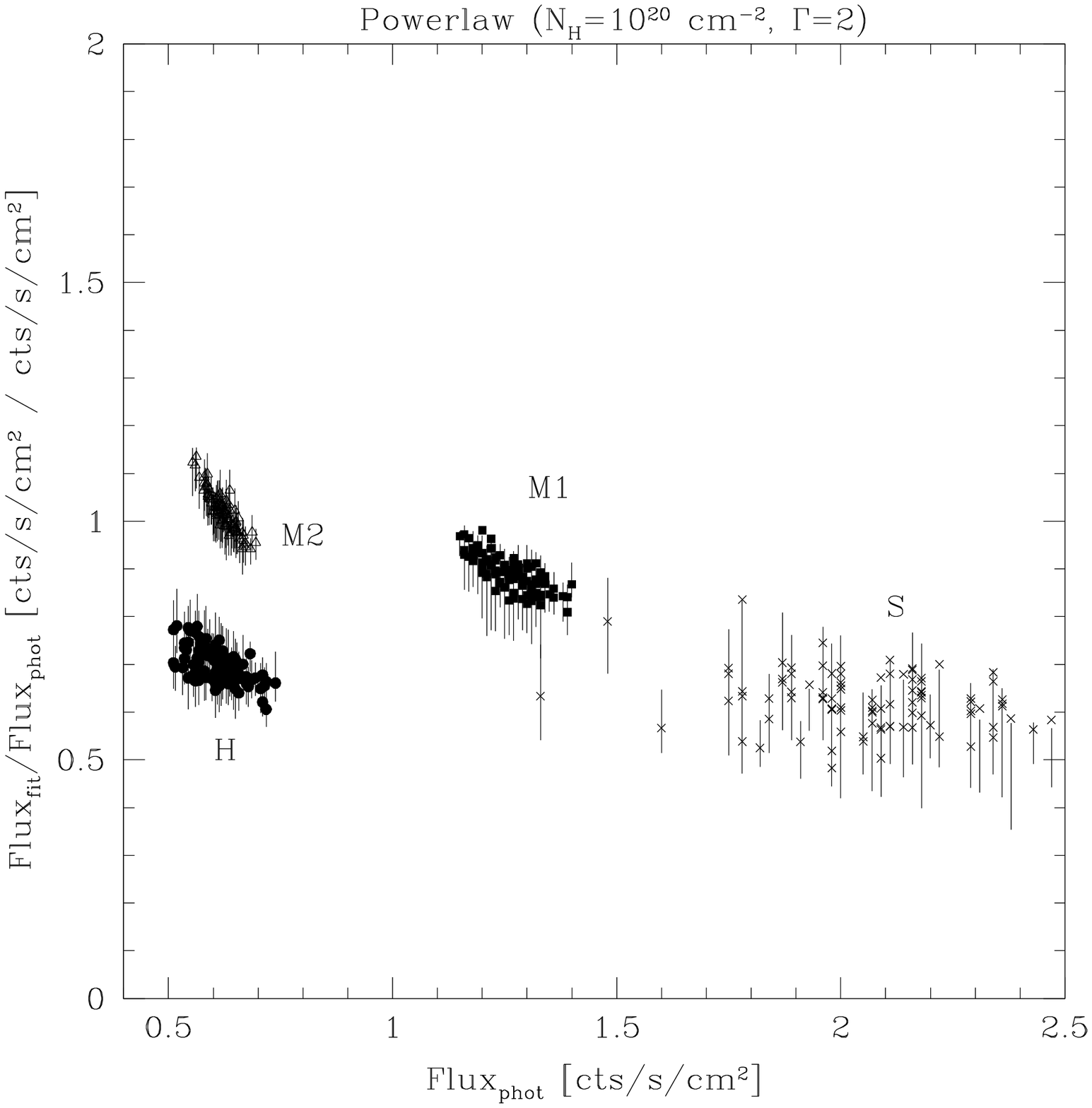}}
  \resizebox{0.5\hsize}{!}{\includegraphics{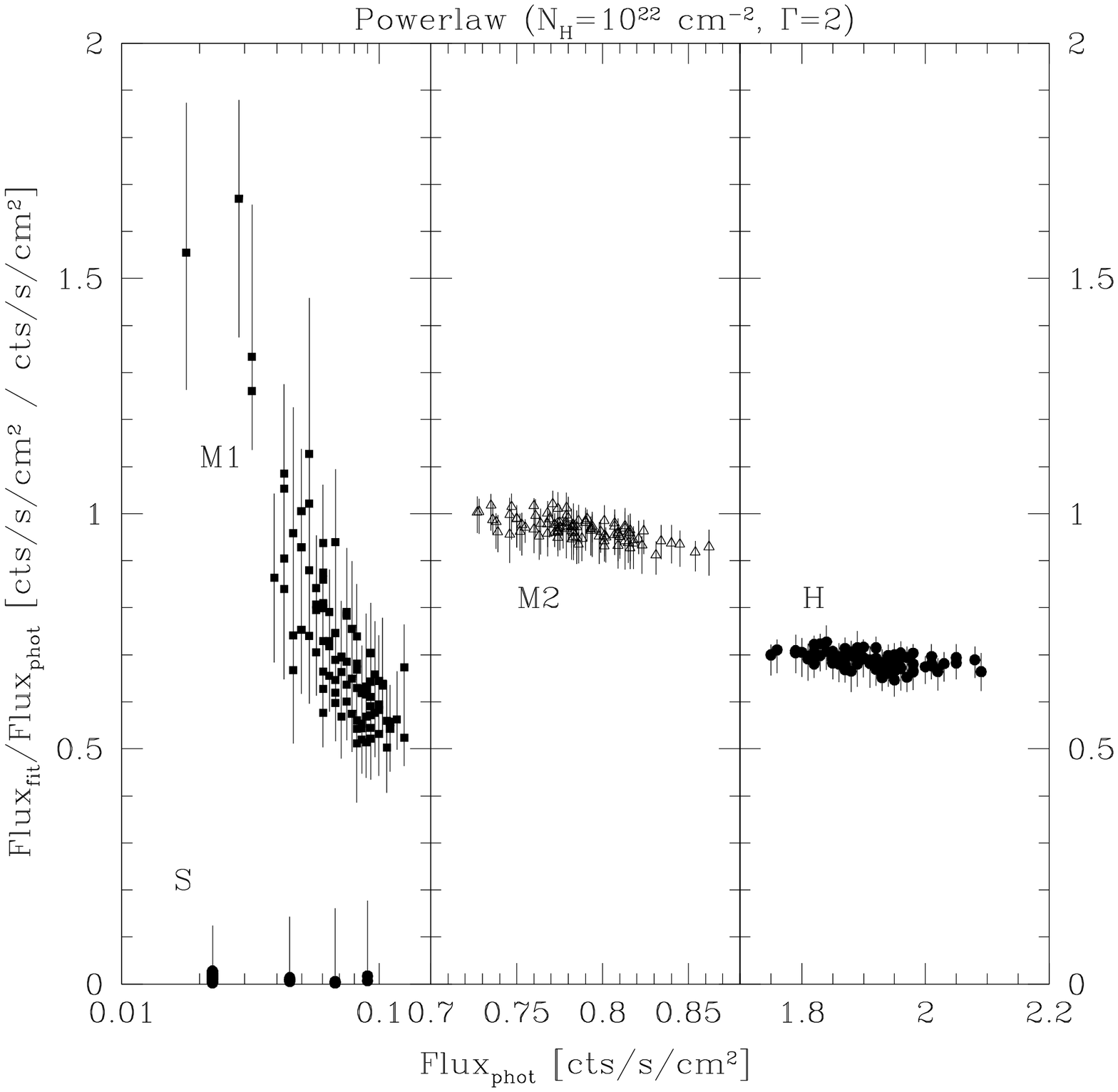}}
  \caption{Comparison between Xspec fluxes ($F_{fit}$) over photometric
fluxes ($F_{phot}$) computed with our method versus photometric
fluxes for two spectral shapes and four spectral parameter
combination. The different bands are noted in the figures, and the
right hand figures are cropped for clarity. Ideally all y-axis values
would be one. Error bars correspond to 68\% CL.\label{fig:fit_comp}}
\efige

It is clear from the figures that this photometry computes fluxes
within better than 30\% in most cases, even in the absence of band
ratio (color) corrections. That the medium-hard band (M2) has the best
correspondence between X-ray photometric fluxes and Xspec fluxes is
simply based on the fact that the medium-hard filter has the best
properties, very steep edges and a relatively flat top (variation of
$\sim$10\%) as shown in Fig. \ref{fig:filters}. Even in the worst
cases it is better than 50\%. It should be noted that even in the
worst cases the systematic error of 68\% CL covers the discrepancy
between the true and estimated fluxes. The large discrepancies are due
to several reasons: (1) a relatively strongly varying spectrum in the
energy band, (2) a varying ARF, and (3) differences between the
correction factor for that spectrum and the mode of the correction
factor distribution. The last factor is quite small, at least for the
spectral shapes chosen above. The main contribution to the discrepancy
comes from the combination of varying spectrum and ARF and a high
number of counts.

With a high number of counts ($>\sim$200) in a band, the changes in
the ARF over an energy band become important, particularly for the
broad hard band, and the resulting flux differences become significant
with respect to the statistical error.

\subsection{Real data}
The real data we use for our comparison are relatively bright ($>100$
source counts) sources from \chandra observations of M 33. We use two
approaches to estimate the flux independently from our method. First
we fit the source spectra with spectral models and use the {\it
flux} command in Xspec to get the photon flux; for best fit models see
\citet{grimm:07}.

Secondly we take each photon in an energy band, use the inverse of the
ARF at the photon energy, sum this inverse over all photons in the
energy band, and divide the result by the exposure time which gives
the photon flux. This is the simplest possible photometry
method. However, it requires knowledge of individual photon energies,
whereas our method works with integrated quantities. This means
that the accuracy of this method depends not only on the number of
counts, but for weak sources also on where in the energy band the few
photons fall. And moreover, it does not take into account effects of
the RMF which are important for spectra that produce very different
counts in different bands.

Results for sources with well fitted spectra are shown in
Fig. \ref{fig:fit_comp2}. The figure shows the flux estimates of our
methods as red points, Xspec results as green squares, and
summation of inverse effective area as blue triangles. The error bars
are 68\% CL statistical errors only. In general there is good
agreement between all three methods of flux estimation.
\bfig[h]
  \resizebox{\hsize}{!}{\includegraphics{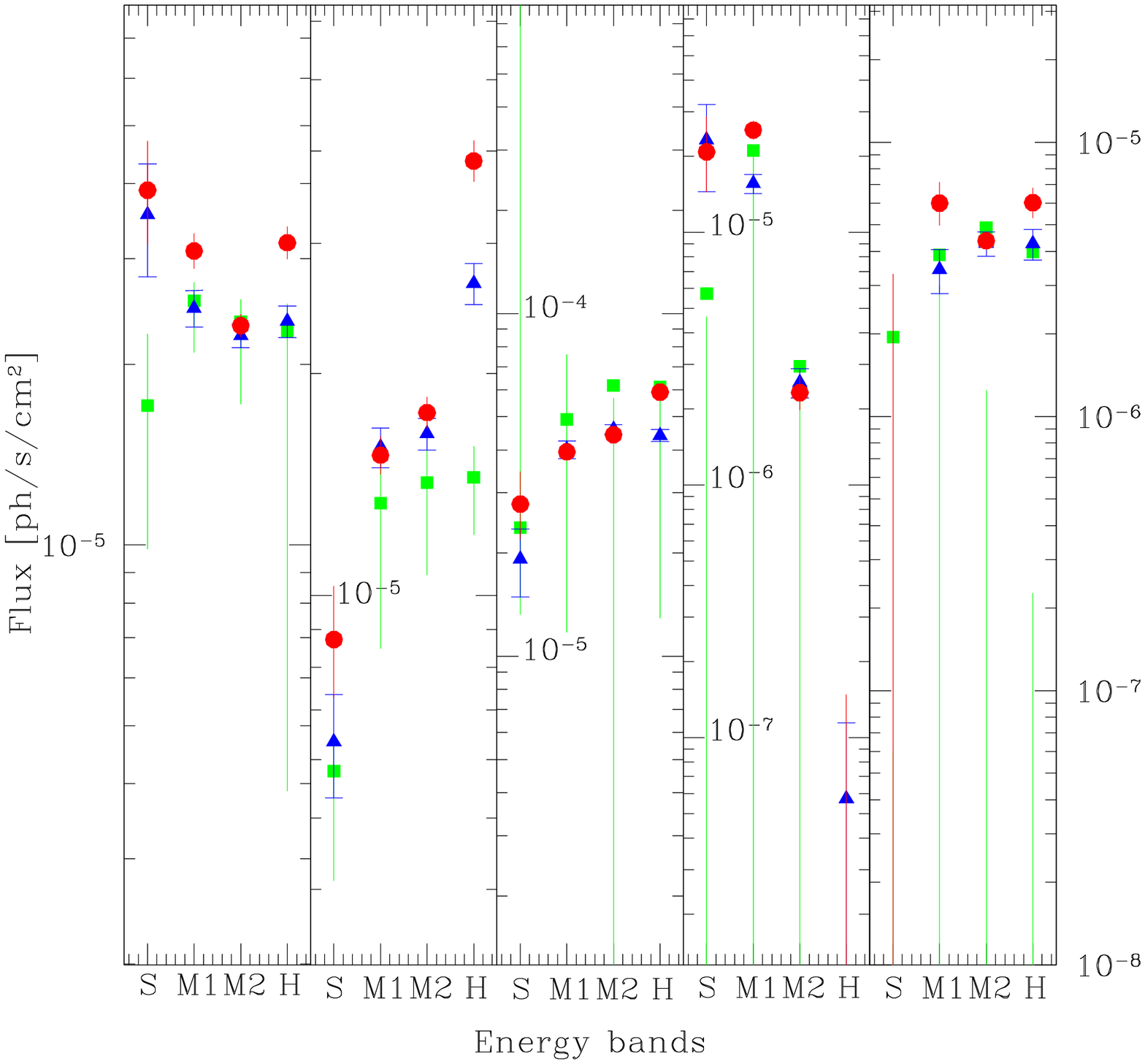}}
  \caption{Flux estimates of our X-ray photometry as (red) circles,
Xspec results are (green) squares, and summation of inverse effective
area as (blue) triangles versus energy band. Error bars are 68\% CL
statistical errors.\label{fig:fit_comp2}}
\efig

It is apparent from the figure that there are sometimes significant
discrepancies between the flux estimate from our photometry and Xspec
fluxes. However the summation of the inverse effective area and our
method agree quite well in general. So it is the spectral fitting that
results in relatively large deviations from the flux estimates obtained
by the other methods. This is probably due to several factors. First,
the spectrum is fitted in Xspec using the whole energy range. This can
result in an under- or overestimate of fluxes in smaller energy bands,
that do not strongly contribute to the overall fit. Furthermore, it
can be the case that the chosen spectrum is not a good representation
of the data but due to insufficient quality of the data this is not
apparent from bad fits. E.g. the spectrum used for the left panel in
Fig. \ref{fig:fit_comp2} has a reduced $\chi^2$ of 0.4 which clearly
indicates that the bremsstrahlung spectrum chosen for the fit is not
the true representation of the data.

\section{Conclusion}
We have presented a system of X-ray photometry for the \chandra
satellite. The system we have developed relies on a knowledge of
effective area and the energy-to-channel conversion to construct 
X-ray filters, but is unbiased by assumptions about the spectral shape
of a source. We have shown that the filters are comparable to filters
in the optical and infrared, and that our photometric system in X-rays
is able to estimate fluxes to within about 20\%. Even in the worst
cases it is better than 50\%. We have incorporated methods to
estimate systematic errors and consistently propagate statistical as
well as systematic errors.

Due to the construction method employed our filter system is very
flexible and can be adapted readily to other CCD X-ray detectors,
in particular to XMM-Newton EPIC. The code to compute fluxes is
available at a web page
\footnote{http://hea-www.cfa.harvard.edu/~jcm/xray/index.html}. A
table with a selection of correction factors for ACIS-S3 is given in
Appendix A in the electronic edition and available at the same
URL. The table contains only every fourth tile because of the
generally slowly varying correction factor with chip location. In the
future we will explore potential improvements to X-ray photometry by
means of:

\begin{itemize}
\item Making color corrections using band ratios.
A preliminary investigation suggests that extreme spectra (e.g. highly
absorbed or high photon indexes) will gain significantly in the
accuracy of flux estimates and even normal spectra will have a reduced
error range.

\item Optimizing the choice of bands.
The properties of our current method show that there is a
correlation between the accuracy of flux estimates and the filter
shape. The more ``boxy'' a filter is the better the flux estimate will
be. This suggests a limit for the width of a filter at which point
deviations from boxiness result in an accuracy of the flux estimate
below a certain value.

\item Comparing results for \chandra ACIS and XMM-Newton EPIC data.
\chandra and XMM have similar instrumental setups (ARF and RMF) and
overlapping science capabilities. And given the large numbers of
sources observed by these two X-ray missions it is very important to
be able to compare results for the two.
\end{itemize}

\section{Acknowledgments}
HJG thanks Ralph Kraft for providing the source code for computing
confidence limits, and Paul Plucinsky for discussions about the
RMF. This work has been supported by NASA grant GO2-3135X and CXC
contract NAS8-03060.

\bibliographystyle{apj}
\bibliography{ms}

\appendix
\bfig[h]
  \resizebox{\hsize}{!}{\includegraphics{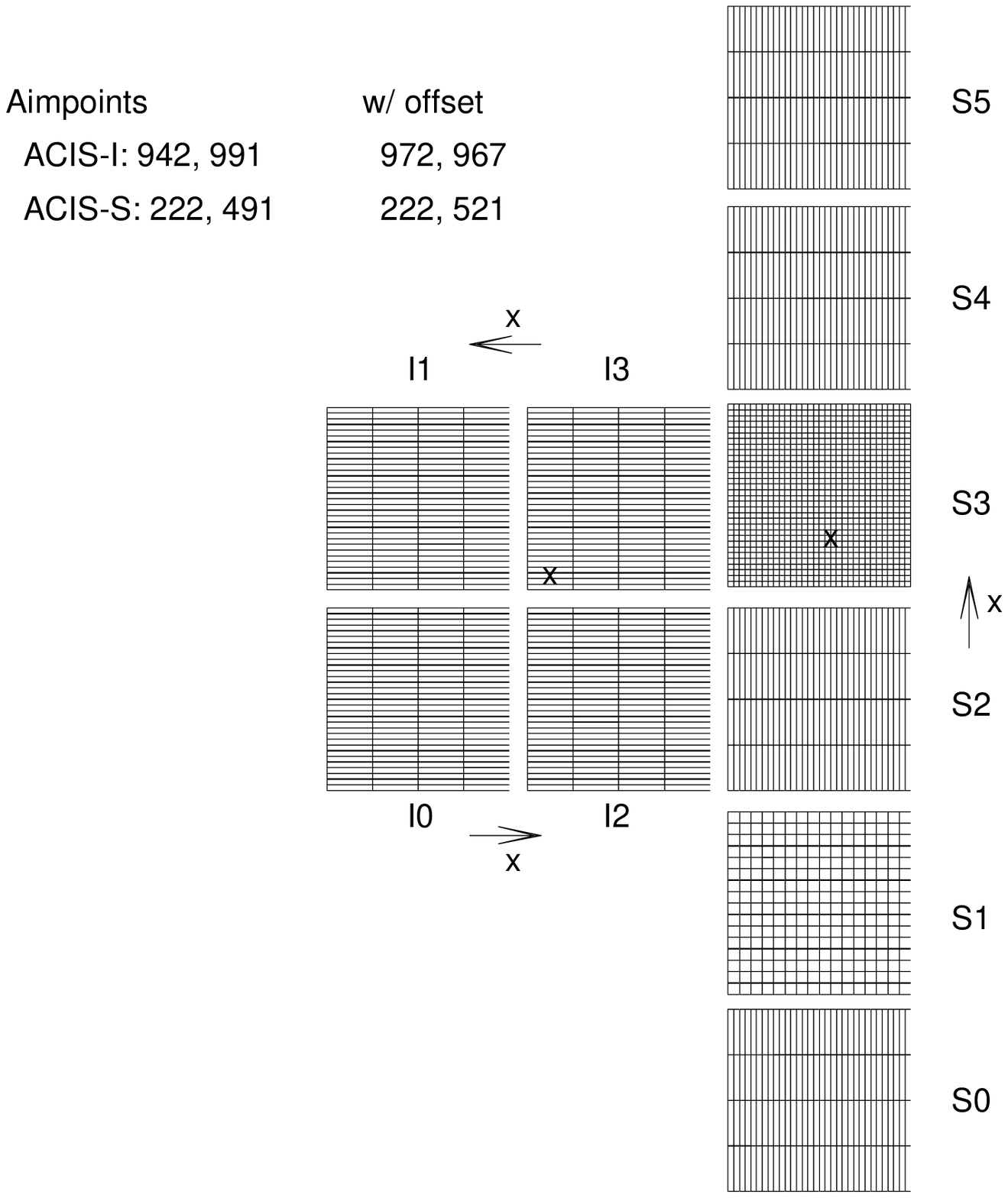}}
  \caption{Schematic of the tiling of the full ACIS
detector. Direction of the x-axis is indicated by the arrows. The
y-axis is counterclockwise to the x-axis. Aimpoints for the I3 and S3
chips are shown as crosses. The readout for chips is on the side of
the arrows.\label{fig:rmftiles}}
\efig
\clearpage
\input{tab1}
\newpage
\bfig[h]
  \resizebox{0.5\hsize}{!}{\includegraphics{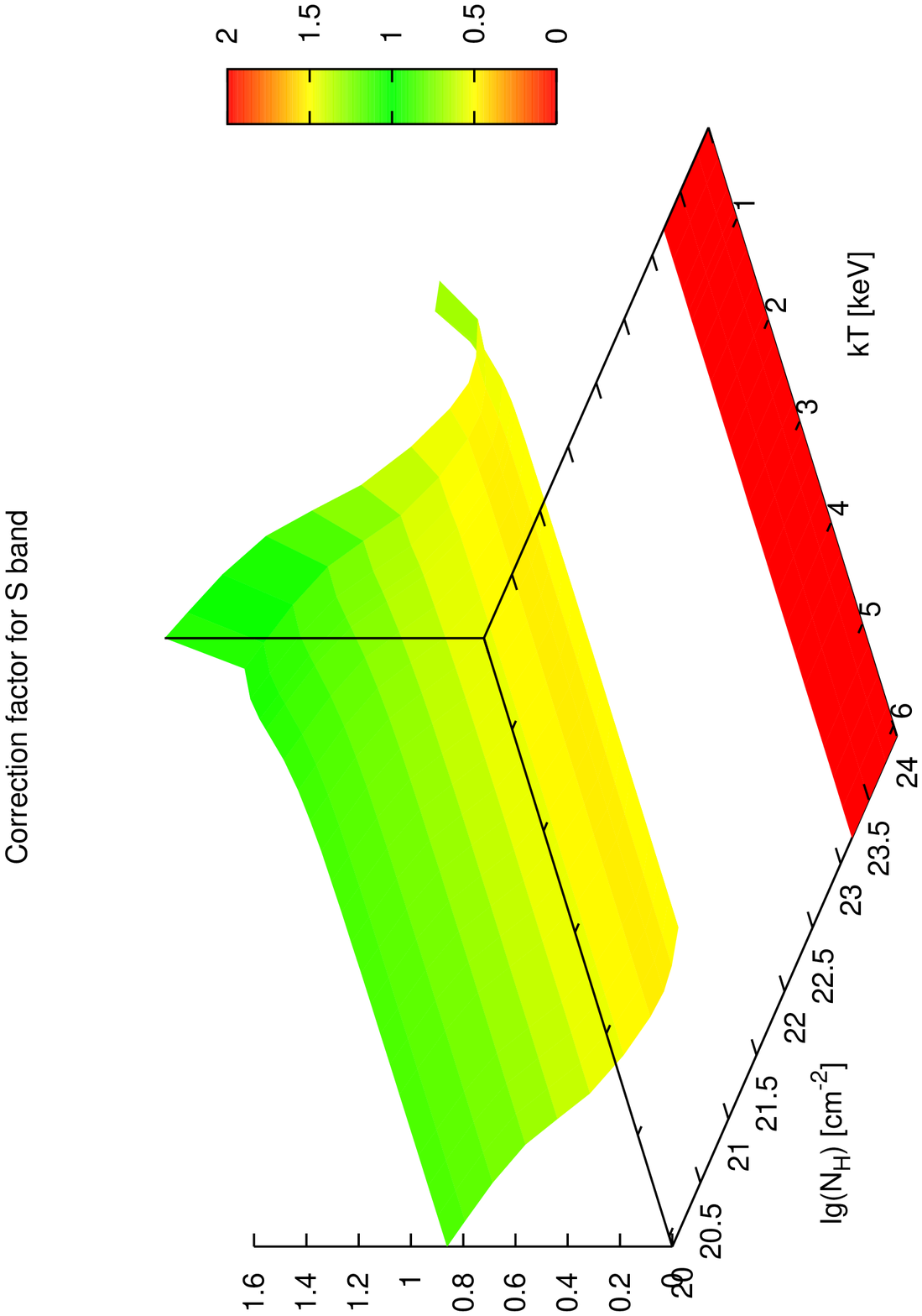}}
  \resizebox{0.5\hsize}{!}{\includegraphics{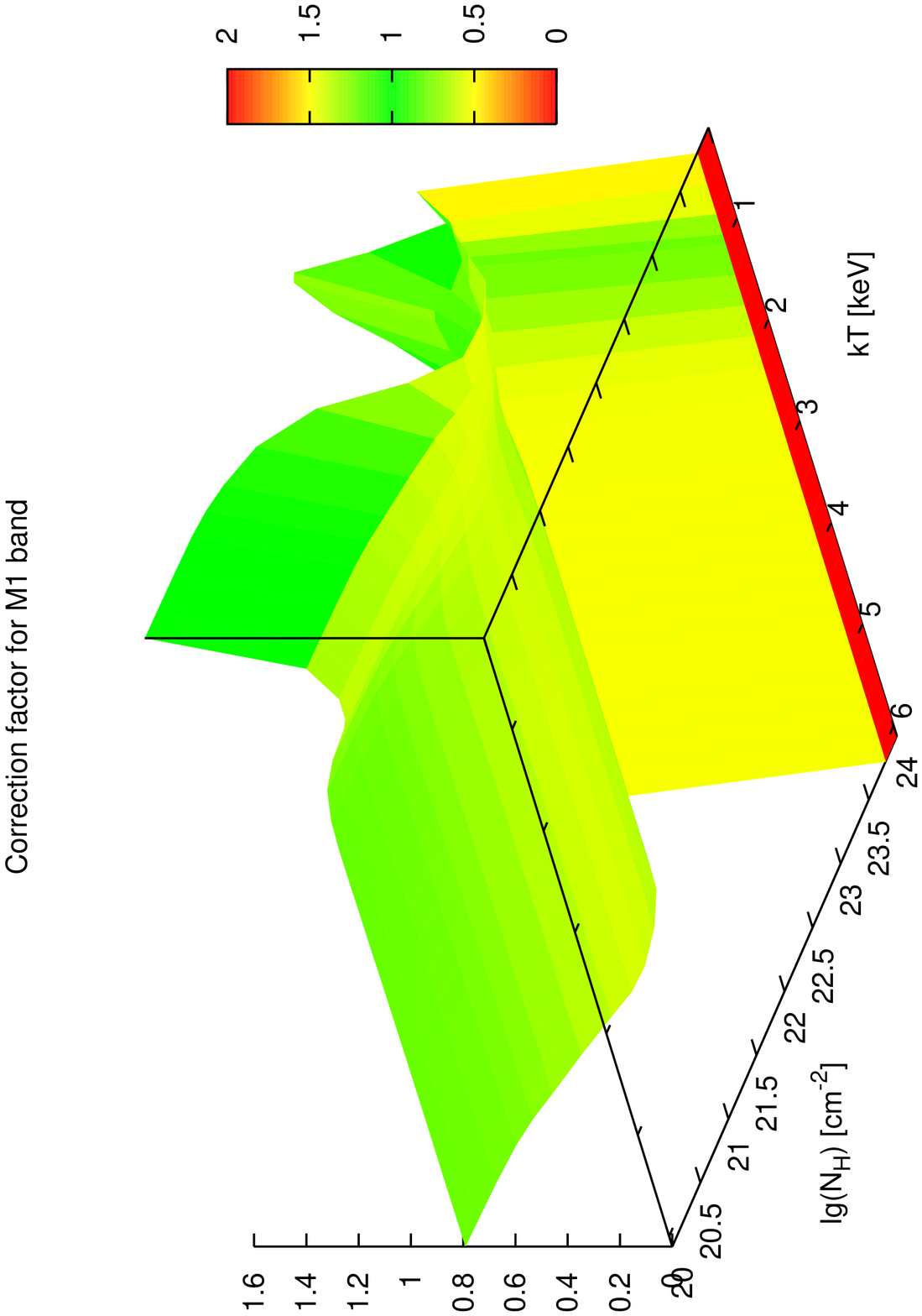}}
  \resizebox{0.5\hsize}{!}{\includegraphics{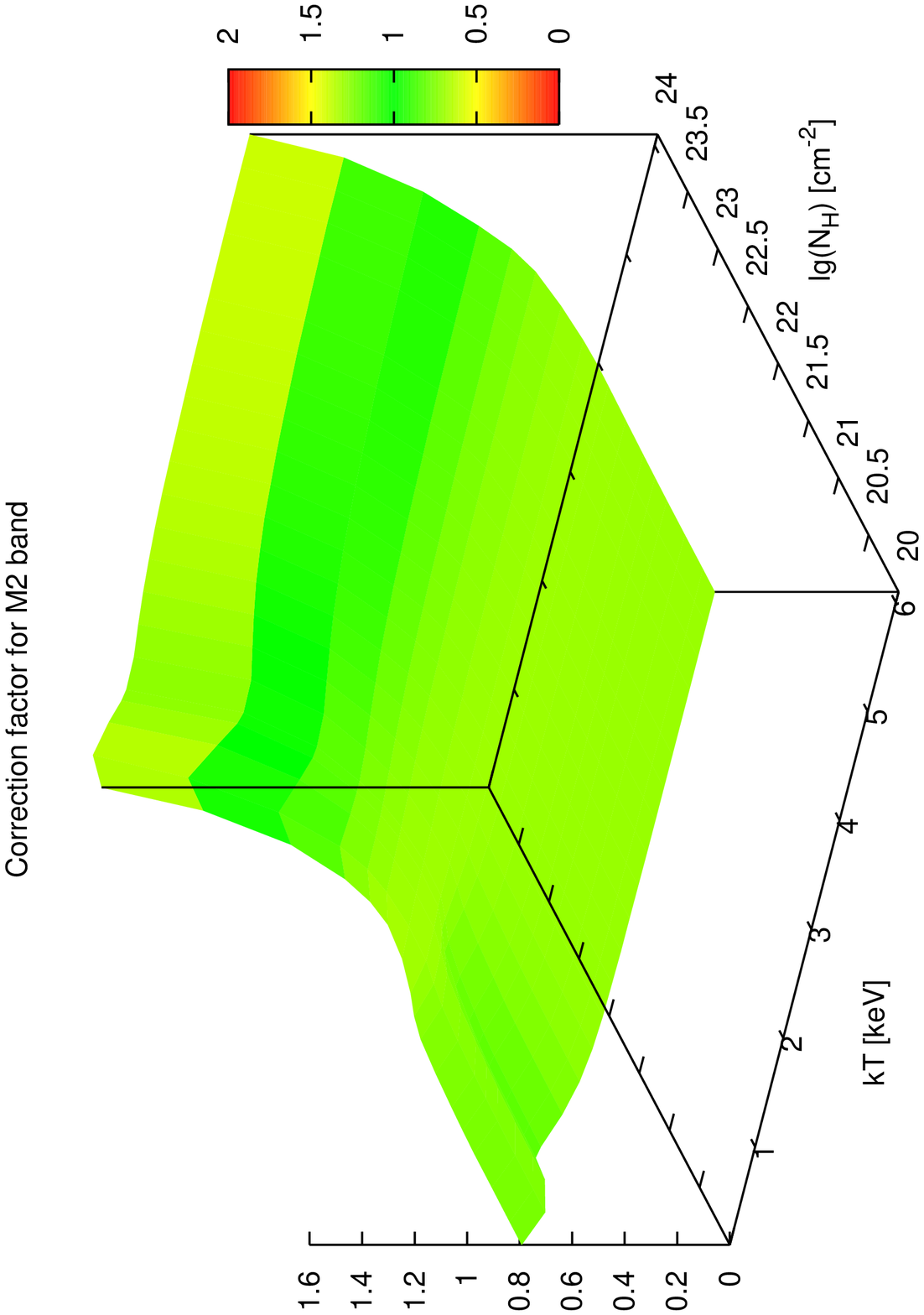}}
  \resizebox{0.5\hsize}{!}{\includegraphics{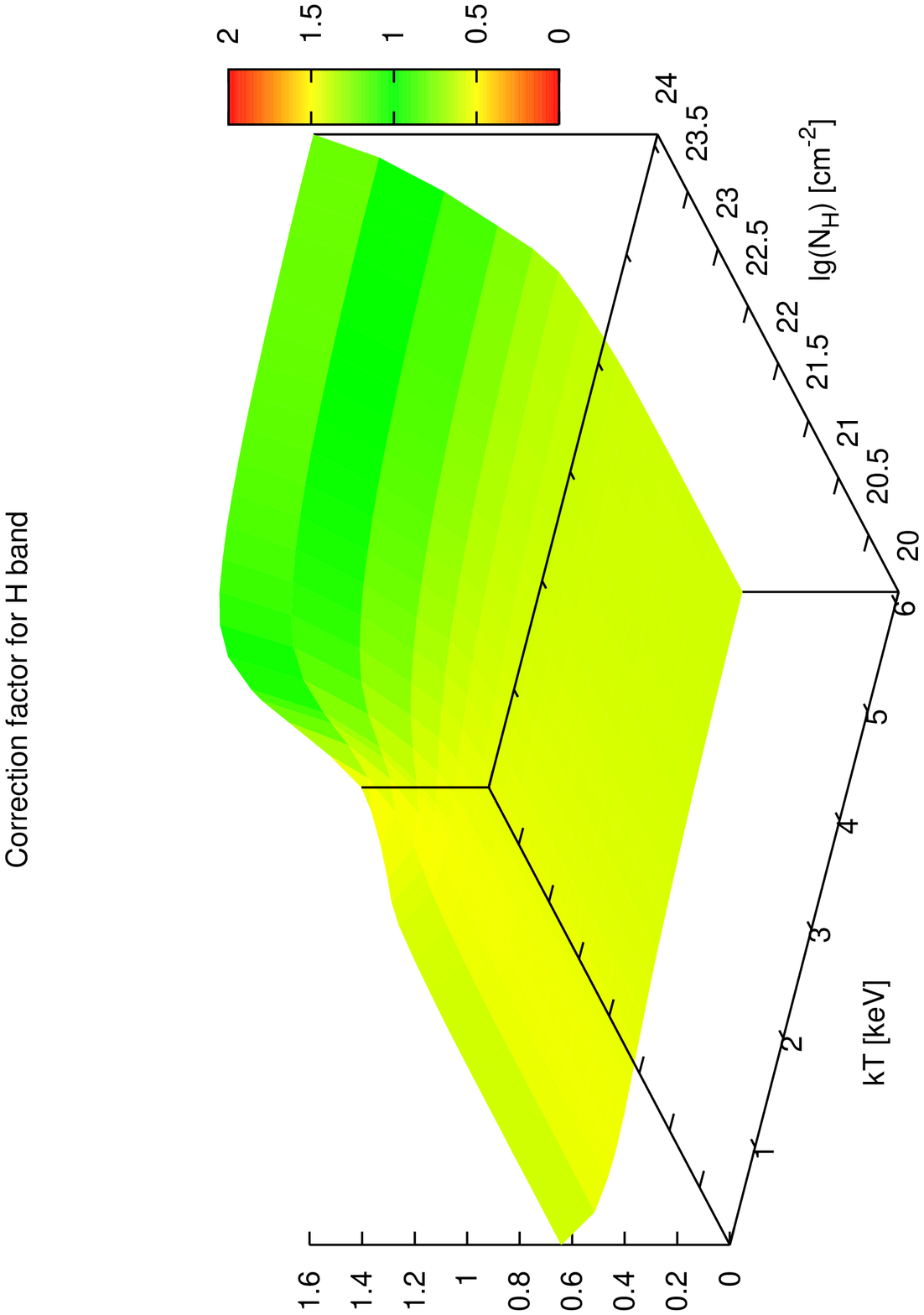}}
  \caption{Surface of correction factors versus spectral parameters
for an absorbed thermal plasma spectrum (APEC model) in the band
passes at the aimpoint of the FI ACIS-S3 chip in 2008.\label{fig:ap}}
\efig
\bfig[h]
  \resizebox{0.5\hsize}{!}{\includegraphics{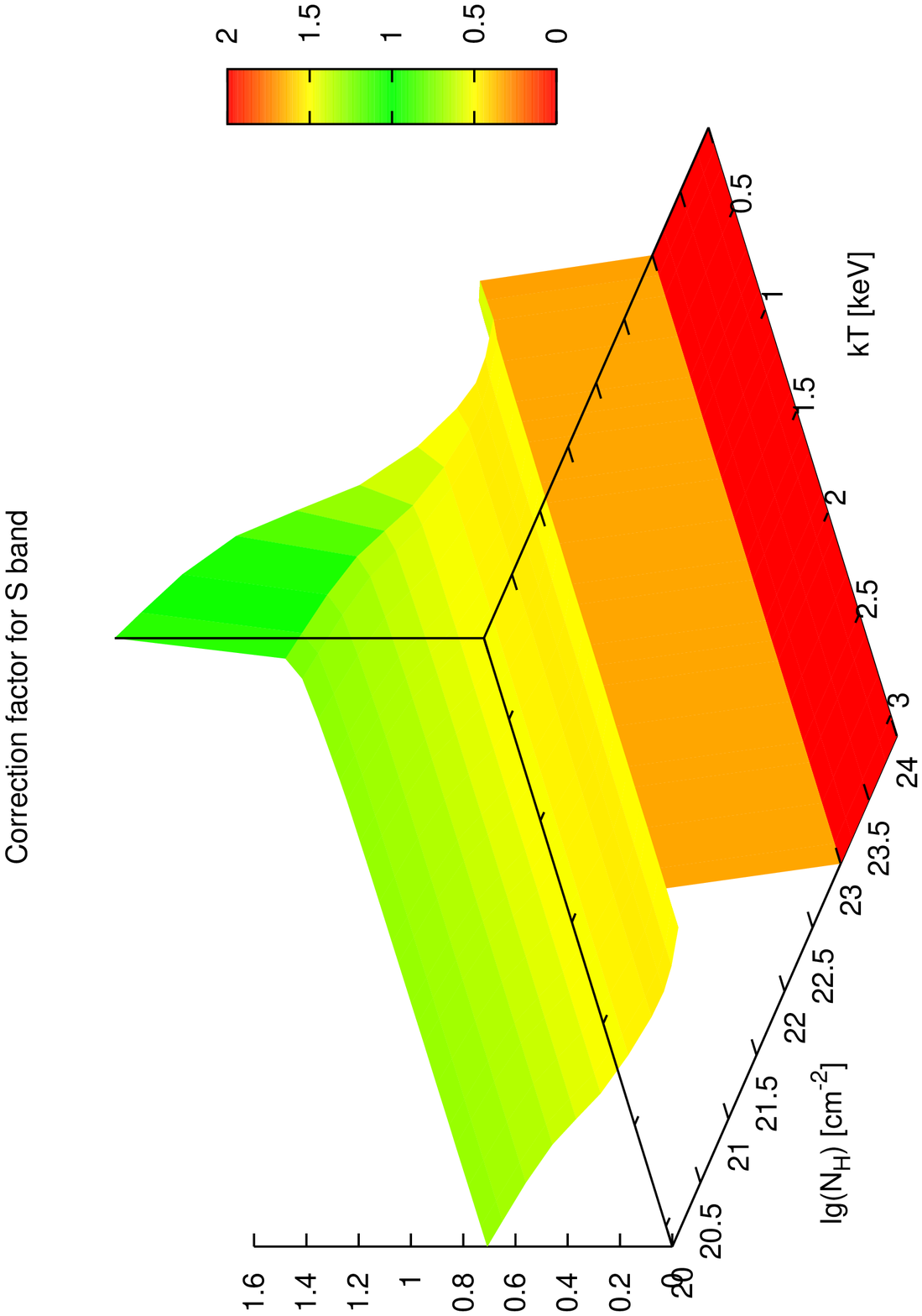}}
  \resizebox{0.5\hsize}{!}{\includegraphics{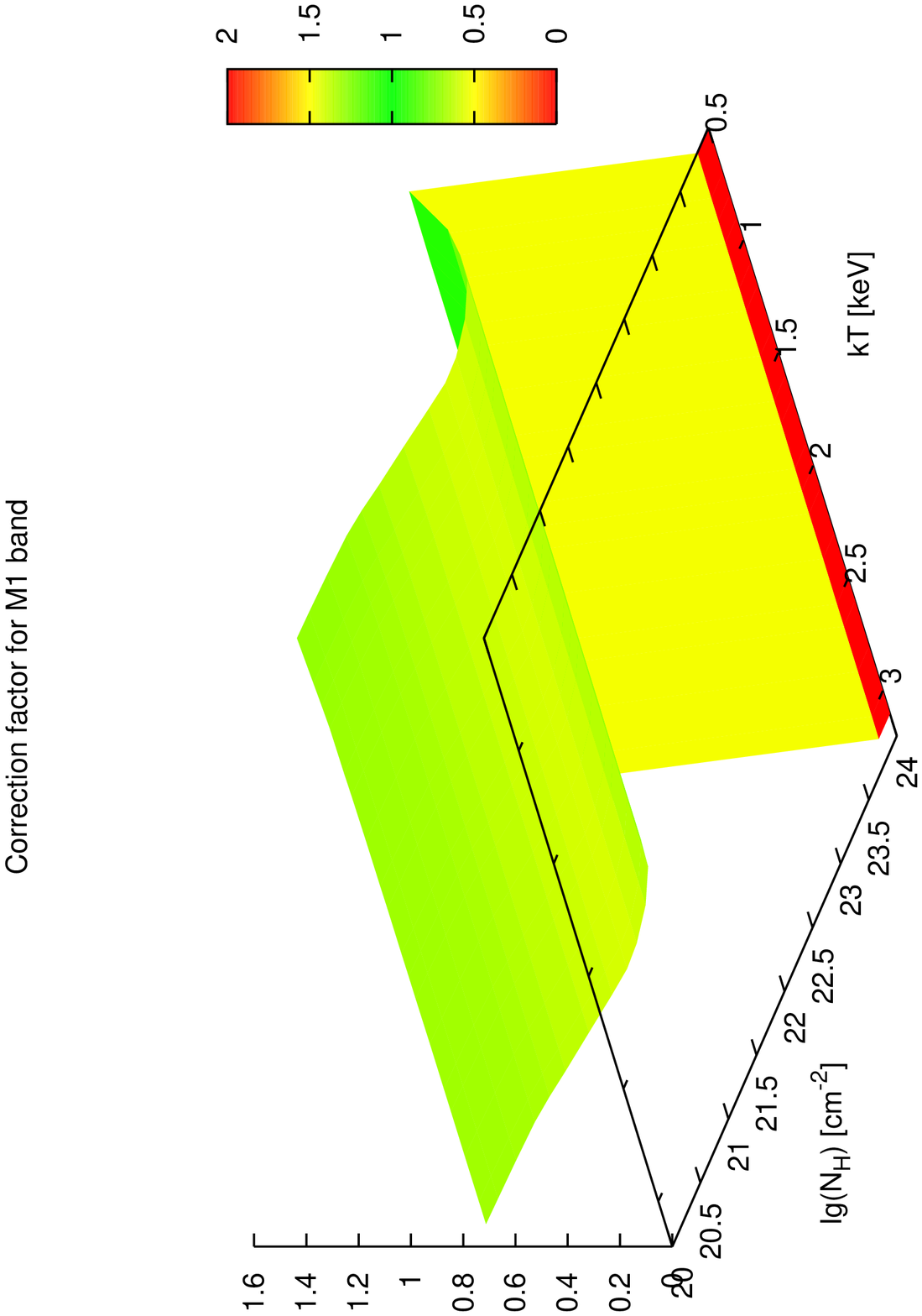}}
  \resizebox{0.5\hsize}{!}{\includegraphics{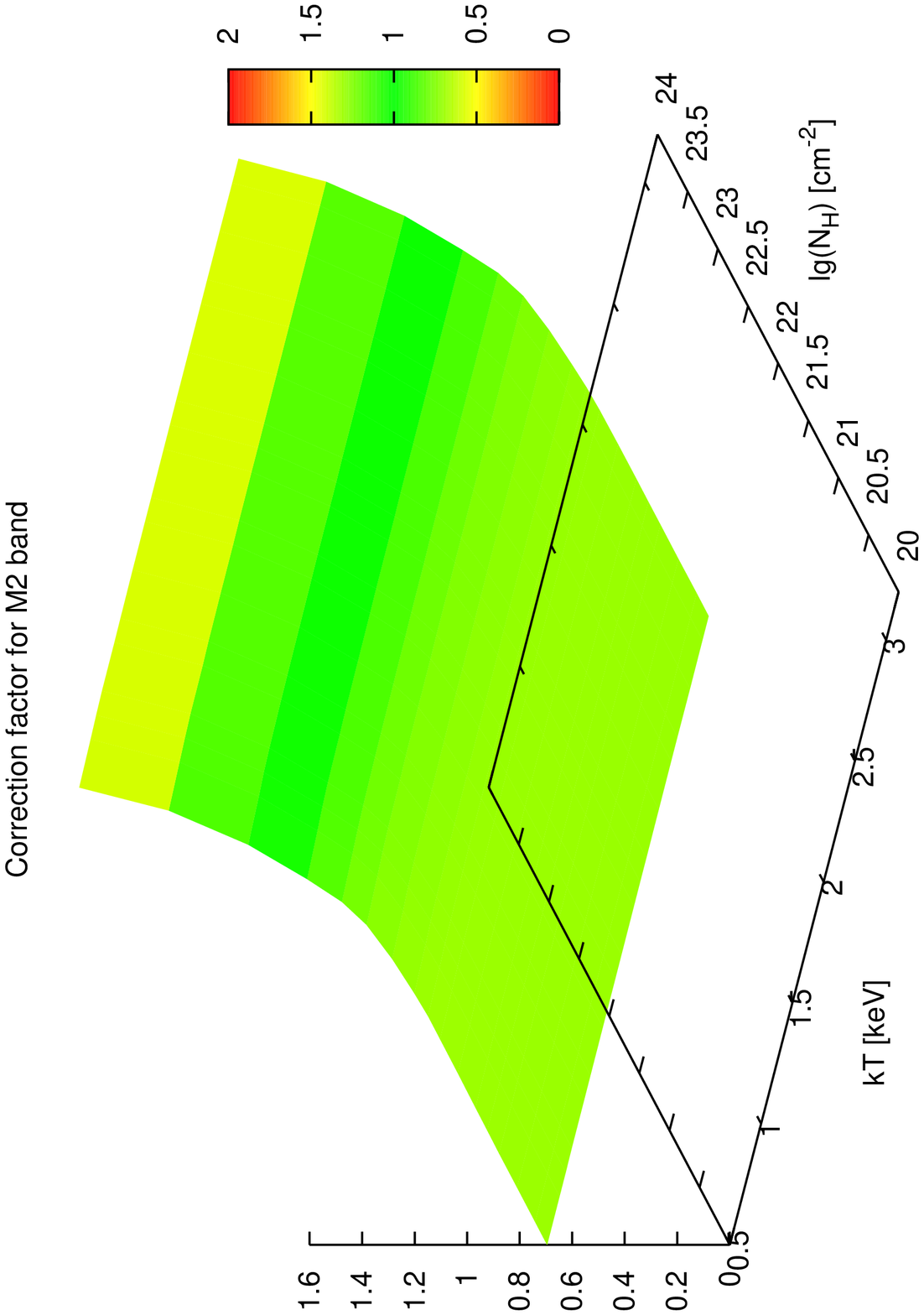}}
  \resizebox{0.5\hsize}{!}{\includegraphics{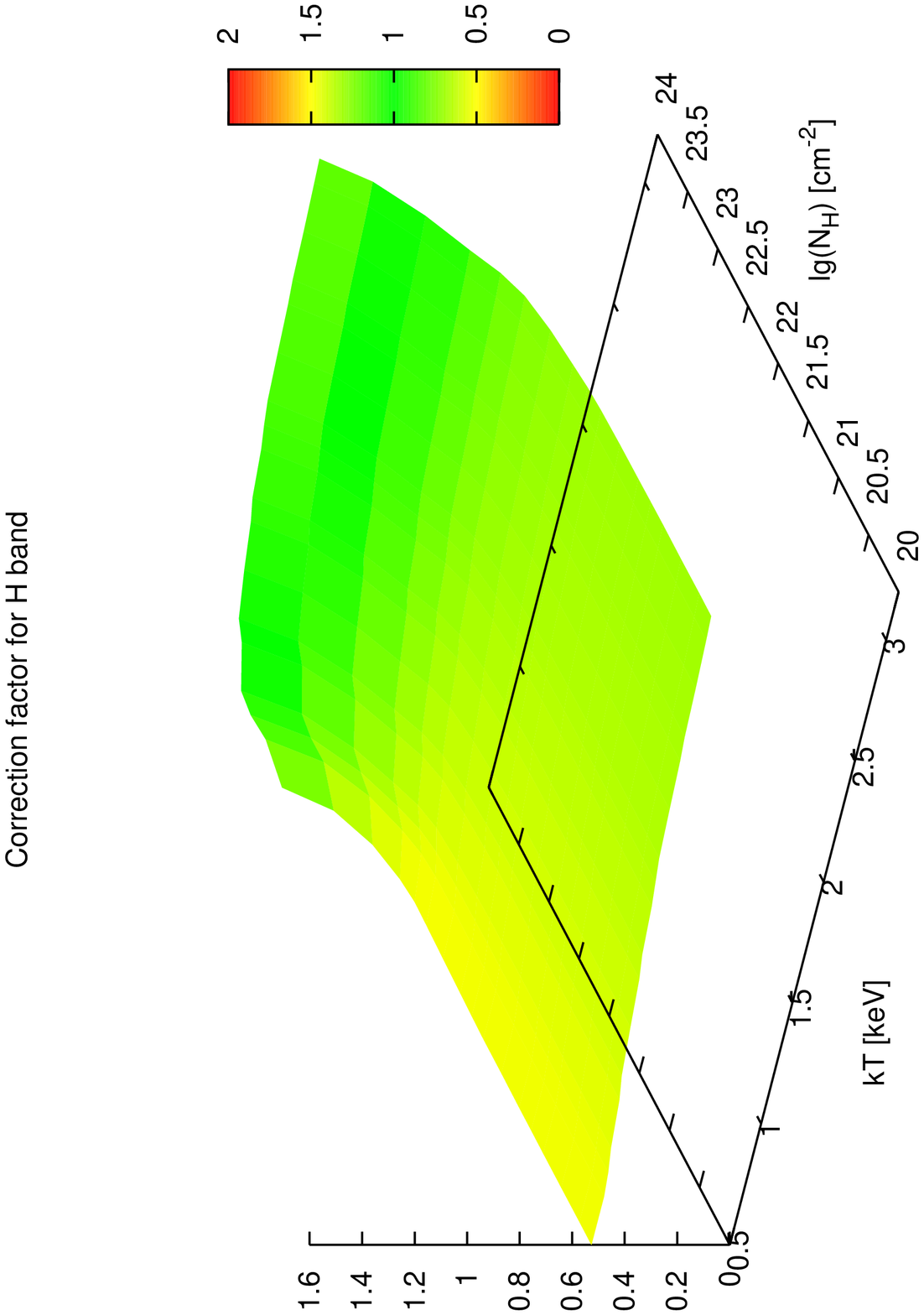}}
  \caption{Surface of correction factors versus spectral parameters
for an absorbed black body spectrum in the band passes at the aimpoint
of the FI ACIS-S3 chip in 2008.\label{fig:bb}}
\efig
\bfig[h]
  \resizebox{0.5\hsize}{!}{\includegraphics{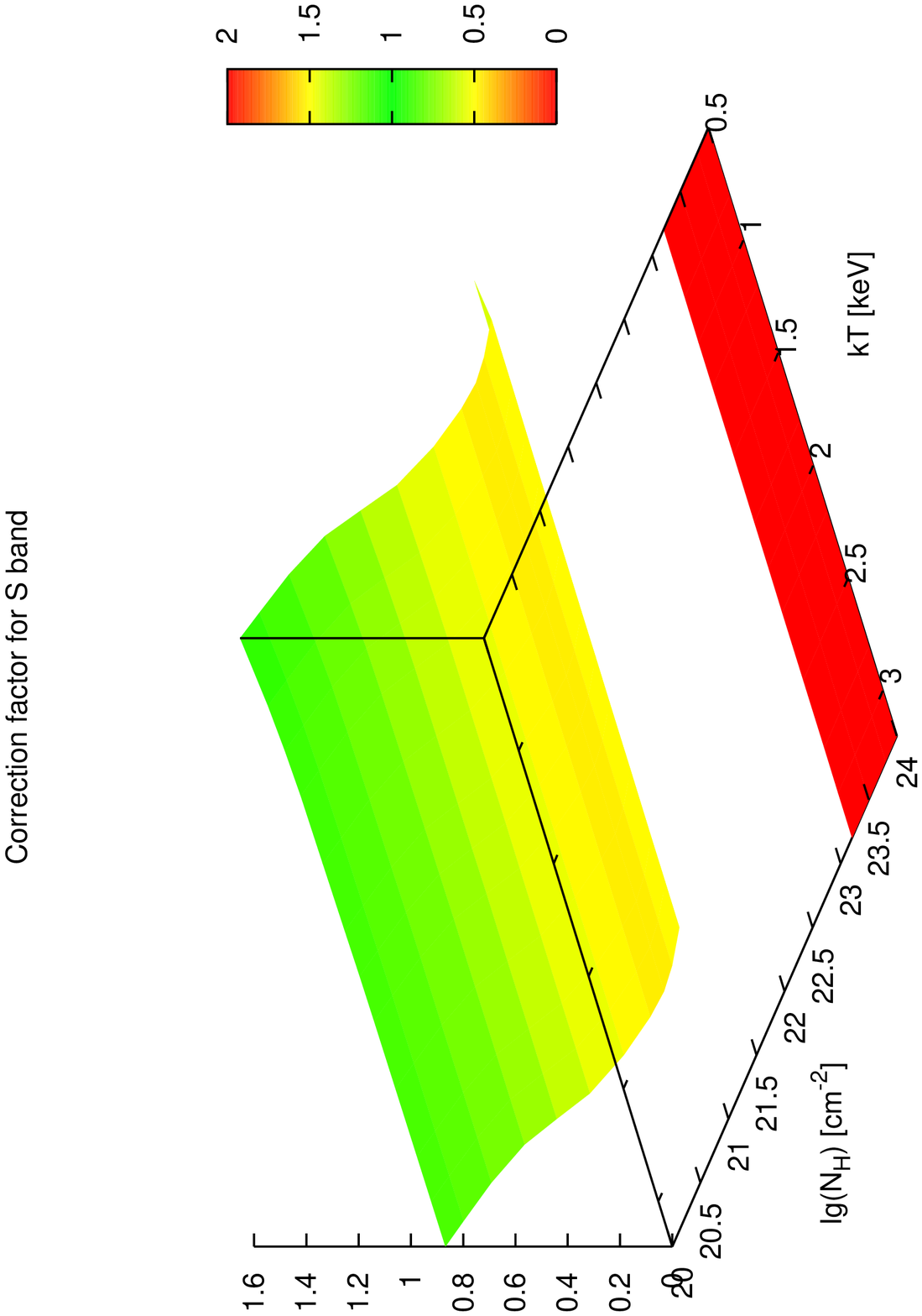}}
  \resizebox{0.5\hsize}{!}{\includegraphics{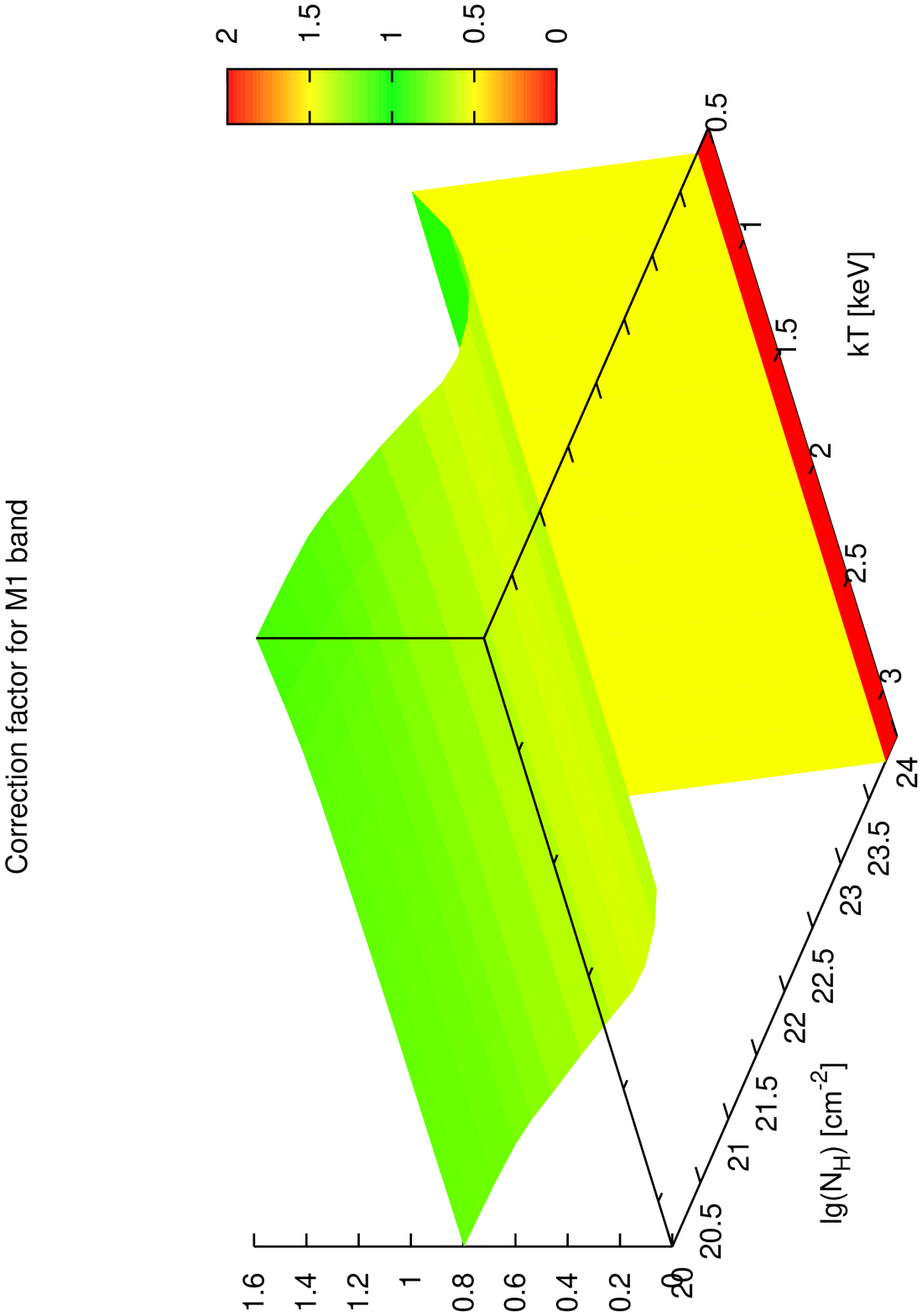}}
  \resizebox{0.5\hsize}{!}{\includegraphics{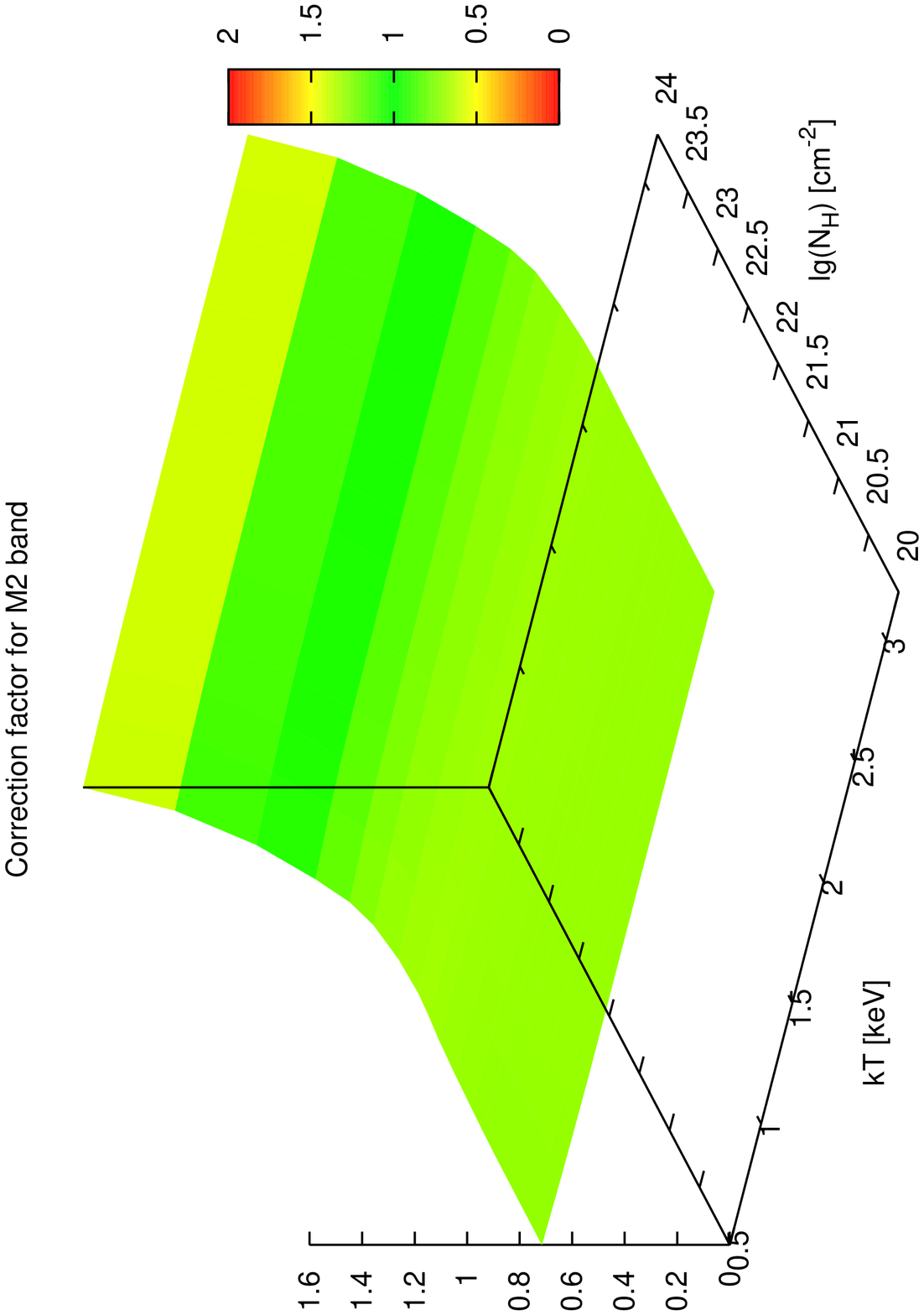}}
  \resizebox{0.5\hsize}{!}{\includegraphics{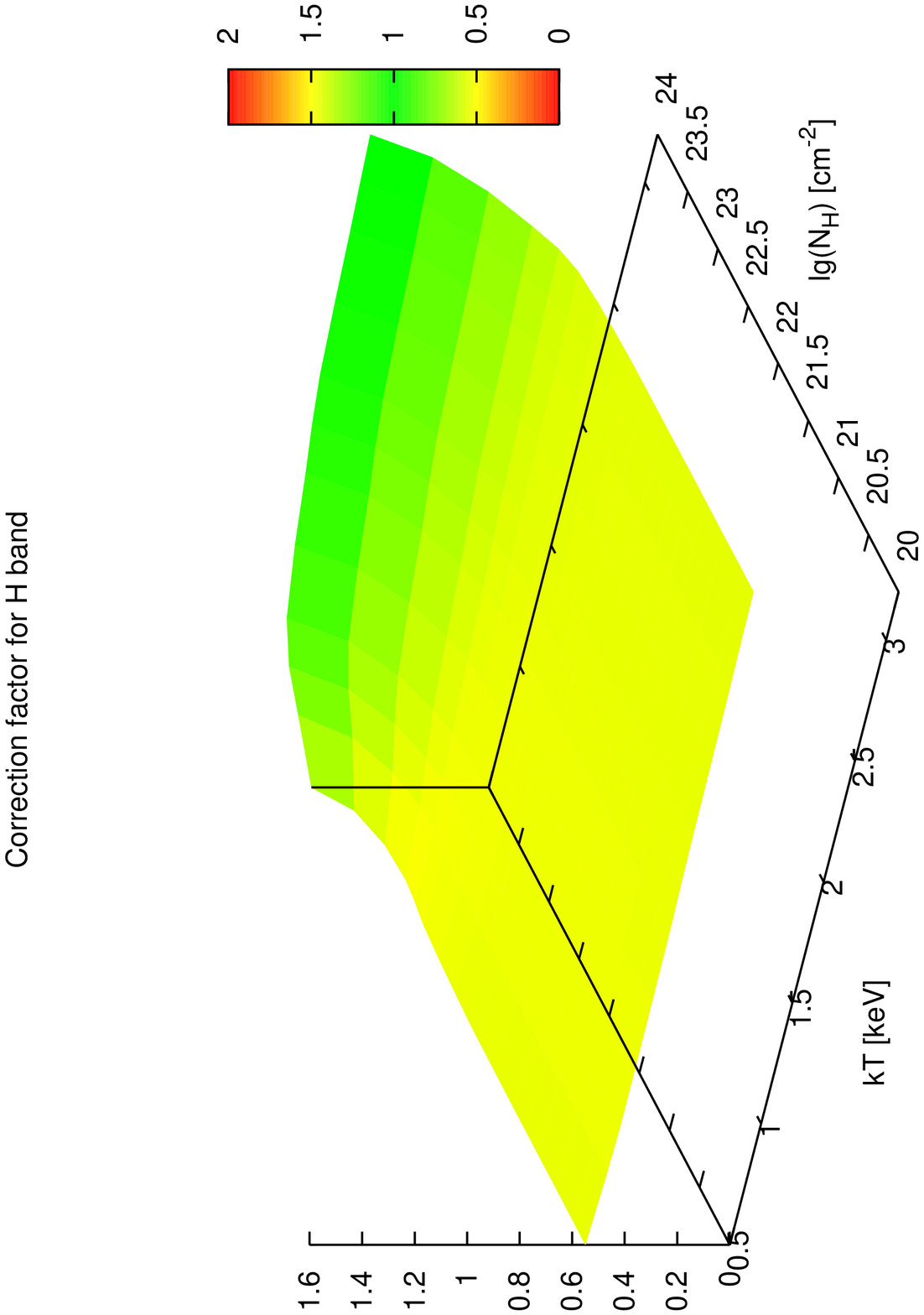}}
  \caption{Surface of correction factors versus spectral parameters
for an absorbed bremsstrahlung spectrum in the band passes at the
aimpoint of the FI ACIS-S3 chip in 2008.\label{fig:br}}
\efig

\end{document}

%% file: tab1.tex
\tiny
\begin{longtable}{rrcccc||rrcccc}
\caption{Correction factors for selected tiles on ACIS-S3.}\\

\hline
\multicolumn{2}{r}{Tile\tablenotemark{a}} &
\multicolumn{4}{c}{Correction factor\tablenotemark{b}[keV range]}&
\multicolumn{2}{c}{Tile} & \multicolumn{4}{c}{Correction factor [keV range]}\\
x & y & 0.3--0.5 & 0.5--1.0 & 1.0--2.1 & 2.1--7.5 &x & y & 0.3--0.5 & 0.5--1.0 & 1.0--2.1 & 2.1--7.5\\

\hline
\endfirsthead
\multicolumn{12}{c}{Table \ref{tab:corrections} (continued)}\\
\hline
\endhead
\hline
\endfoot
  16     &   16      & $0.39_{-0.00}^{+0.26}$ & $0.54_{-0.05}^{+0.18}$ & $0.69_{-0.07}^{+0.06}$ & $0.54_{-0.07}^{+0.14}$ &   592    &   16      & $0.40_{-0.00}^{+0.26}$ & $0.59_{-0.10}^{+0.13}$ & $0.69_{-0.06}^{+0.06}$ & $0.54_{-0.07}^{+0.15}$ \\
  16     &   112     & $0.45_{-0.01}^{+0.22}$ & $0.56_{-0.04}^{+0.19}$ & $0.70_{-0.07}^{+0.07}$ & $0.54_{-0.07}^{+0.14}$ &   592    &   112     & $0.45_{-0.01}^{+0.22}$ & $0.56_{-0.04}^{+0.19}$ & $0.70_{-0.07}^{+0.07}$ & $0.54_{-0.07}^{+0.15}$ \\
  16     &   208     & $0.47_{-0.01}^{+0.21}$ & $0.66_{-0.10}^{+0.11}$ & $0.70_{-0.07}^{+0.07}$ & $0.54_{-0.06}^{+0.15}$ &   592    &   208     & $0.47_{-0.01}^{+0.21}$ & $0.58_{-0.04}^{+0.17}$ & $0.70_{-0.07}^{+0.07}$ & $0.54_{-0.06}^{+0.15}$ \\
  16     &   304     & $0.47_{-0.01}^{+0.21}$ & $0.58_{-0.04}^{+0.17}$ & $0.70_{-0.07}^{+0.07}$ & $0.54_{-0.06}^{+0.15}$ &   592    &   304     & $0.47_{-0.00}^{+0.21}$ & $0.58_{-0.04}^{+0.17}$ & $0.70_{-0.07}^{+0.07}$ & $0.54_{-0.06}^{+0.15}$ \\
  16     &   400     & $0.47_{-0.01}^{+0.21}$ & $0.67_{-0.10}^{+0.11}$ & $0.70_{-0.07}^{+0.07}$ & $0.54_{-0.06}^{+0.15}$ &   592    &   400     & $0.48_{-0.01}^{+0.20}$ & $0.67_{-0.10}^{+0.11}$ & $0.70_{-0.07}^{+0.08}$ & $0.54_{-0.06}^{+0.15}$ \\
  16     &   496     & $0.47_{-0.01}^{+0.21}$ & $0.67_{-0.10}^{+0.11}$ & $0.70_{-0.07}^{+0.08}$ & $0.55_{-0.07}^{+0.14}$ &   592    &   496     & $0.48_{-0.01}^{+0.20}$ & $0.67_{-0.10}^{+0.11}$ & $0.70_{-0.07}^{+0.08}$ & $0.55_{-0.07}^{+0.14}$ \\
  16     &   592     & $0.48_{-0.01}^{+0.20}$ & $0.67_{-0.10}^{+0.10}$ & $0.70_{-0.07}^{+0.08}$ & $0.53_{-0.05}^{+0.16}$ &   592    &   592     & $0.48_{-0.01}^{+0.20}$ & $0.67_{-0.10}^{+0.10}$ & $0.70_{-0.07}^{+0.08}$ & $0.53_{-0.05}^{+0.16}$ \\
  16     &   688     & $0.48_{-0.01}^{+0.20}$ & $0.67_{-0.10}^{+0.10}$ & $0.70_{-0.07}^{+0.08}$ & $0.55_{-0.07}^{+0.14}$ &   592    &   688     & $0.48_{-0.01}^{+0.20}$ & $0.68_{-0.10}^{+0.11}$ & $0.70_{-0.07}^{+0.08}$ & $0.55_{-0.07}^{+0.14}$ \\
  16     &   784     & $0.48_{-0.01}^{+0.20}$ & $0.68_{-0.10}^{+0.11}$ & $0.70_{-0.08}^{+0.08}$ & $0.55_{-0.08}^{+0.13}$ &   592    &   784     & $0.48_{-0.01}^{+0.20}$ & $0.68_{-0.10}^{+0.11}$ & $0.70_{-0.07}^{+0.08}$ & $0.55_{-0.08}^{+0.14}$ \\
  16     &   880     & $0.47_{-0.01}^{+0.21}$ & $0.58_{-0.04}^{+0.17}$ & $0.70_{-0.08}^{+0.08}$ & $0.55_{-0.09}^{+0.14}$ &   592    &   880     & $0.47_{-0.01}^{+0.21}$ & $0.58_{-0.04}^{+0.17}$ & $0.70_{-0.07}^{+0.08}$ & $0.55_{-0.08}^{+0.13}$ \\
  16     &   976     & $0.44_{-0.01}^{+0.23}$ & $0.79_{-0.13}^{+1.29}$ & $0.70_{-0.08}^{+0.09}$ & $0.55_{-0.09}^{+0.14}$ &   592    &   976     & $0.44_{-0.01}^{+0.23}$ & $0.79_{-0.13}^{+1.30}$ & $0.70_{-0.08}^{+0.08}$ & $0.55_{-0.09}^{+0.14}$ \\
  16     &   1008    & $0.43_{-0.01}^{+0.23}$ & $0.56_{-0.04}^{+0.19}$ & $0.70_{-0.08}^{+0.09}$ & $0.55_{-0.09}^{+0.14}$ &   592    &   1008    & $0.43_{-0.01}^{+0.24}$ & $0.56_{-0.04}^{+0.19}$ & $0.70_{-0.08}^{+0.08}$ & $0.55_{-0.09}^{+0.14}$ \\
  112    &   16      & $0.40_{-0.01}^{+0.26}$ & $0.54_{-0.05}^{+0.18}$ & $0.69_{-0.07}^{+0.06}$ & $0.54_{-0.07}^{+0.14}$ &   688    &   16      & $0.40_{-0.01}^{+0.26}$ & $0.72_{-0.14}^{+1.77}$ & $0.69_{-0.06}^{+0.06}$ & $0.54_{-0.07}^{+0.15}$ \\
  112    &   112     & $0.45_{-0.01}^{+0.22}$ & $0.56_{-0.04}^{+0.19}$ & $0.70_{-0.07}^{+0.07}$ & $0.54_{-0.07}^{+0.14}$ &   688    &   112     & $0.45_{-0.01}^{+0.22}$ & $0.56_{-0.04}^{+0.19}$ & $0.70_{-0.07}^{+0.07}$ & $0.54_{-0.07}^{+0.15}$ \\
  112    &   208     & $0.47_{-0.01}^{+0.21}$ & $0.66_{-0.10}^{+0.11}$ & $0.70_{-0.07}^{+0.07}$ & $0.54_{-0.06}^{+0.14}$ &   688    &   208     & $0.47_{-0.01}^{+0.21}$ & $0.66_{-0.10}^{+0.11}$ & $0.70_{-0.07}^{+0.07}$ & $0.54_{-0.07}^{+0.15}$ \\
  112    &   304     & $0.47_{-0.00}^{+0.21}$ & $0.58_{-0.04}^{+0.17}$ & $0.70_{-0.07}^{+0.07}$ & $0.54_{-0.06}^{+0.15}$ &   688    &   304     & $0.48_{-0.01}^{+0.20}$ & $0.58_{-0.04}^{+0.17}$ & $0.70_{-0.07}^{+0.07}$ & $0.54_{-0.07}^{+0.15}$ \\
  112    &   400     & $0.48_{-0.01}^{+0.20}$ & $0.58_{-0.04}^{+0.17}$ & $0.70_{-0.07}^{+0.07}$ & $0.54_{-0.06}^{+0.15}$ &   688    &   400     & $0.48_{-0.01}^{+0.20}$ & $0.58_{-0.04}^{+0.17}$ & $0.70_{-0.07}^{+0.07}$ & $0.54_{-0.06}^{+0.15}$ \\
  112    &   496     & $0.48_{-0.01}^{+0.20}$ & $0.58_{-0.04}^{+0.17}$ & $0.70_{-0.07}^{+0.07}$ & $0.54_{-0.06}^{+0.15}$ &   688    &   496     & $0.48_{-0.01}^{+0.20}$ & $0.58_{-0.04}^{+0.17}$ & $0.70_{-0.07}^{+0.07}$ & $0.54_{-0.06}^{+0.15}$ \\
  112    &   592     & $0.48_{-0.01}^{+0.20}$ & $0.67_{-0.10}^{+0.11}$ & $0.70_{-0.07}^{+0.08}$ & $0.55_{-0.07}^{+0.14}$ &   688    &   592     & $0.48_{-0.01}^{+0.20}$ & $0.58_{-0.04}^{+0.17}$ & $0.70_{-0.07}^{+0.07}$ & $0.53_{-0.05}^{+0.16}$ \\
  112    &   688     & $0.48_{-0.01}^{+0.20}$ & $0.67_{-0.10}^{+0.11}$ & $0.70_{-0.07}^{+0.08}$ & $0.55_{-0.07}^{+0.14}$ &   688    &   688     & $0.48_{-0.01}^{+0.20}$ & $0.67_{-0.10}^{+0.11}$ & $0.70_{-0.07}^{+0.08}$ & $0.53_{-0.06}^{+0.16}$ \\
  112    &   784     & $0.47_{-0.01}^{+0.21}$ & $0.58_{-0.04}^{+0.17}$ & $0.70_{-0.07}^{+0.08}$ & $0.53_{-0.05}^{+0.16}$ &   688    &   784     & $0.47_{-0.00}^{+0.21}$ & $0.67_{-0.10}^{+0.11}$ & $0.70_{-0.07}^{+0.08}$ & $0.55_{-0.08}^{+0.14}$ \\
  112    &   880     & $0.47_{-0.02}^{+0.20}$ & $0.66_{-0.10}^{+0.11}$ & $0.70_{-0.08}^{+0.08}$ & $0.55_{-0.08}^{+0.14}$ &   688    &   880     & $0.46_{-0.00}^{+0.22}$ & $0.58_{-0.04}^{+0.17}$ & $0.70_{-0.07}^{+0.08}$ & $0.55_{-0.08}^{+0.14}$ \\
  112    &   976     & $0.44_{-0.01}^{+0.23}$ & $0.56_{-0.04}^{+0.19}$ & $0.70_{-0.08}^{+0.07}$ & $0.55_{-0.09}^{+0.14}$ &   688    &   976     & $0.44_{-0.01}^{+0.23}$ & $0.79_{-0.13}^{+1.31}$ & $0.70_{-0.07}^{+0.08}$ & $0.55_{-0.09}^{+0.14}$ \\
  112    &   1008    & $0.43_{-0.01}^{+0.23}$ & $0.55_{-0.04}^{+0.20}$ & $0.70_{-0.08}^{+0.08}$ & $0.55_{-0.09}^{+0.14}$ &   688    &   1008    & $0.43_{-0.01}^{+0.24}$ & $0.63_{-0.10}^{+0.13}$ & $0.70_{-0.08}^{+0.08}$ & $0.55_{-0.09}^{+0.14}$ \\
  208    &   16      & $0.40_{-0.01}^{+0.26}$ & $0.54_{-0.05}^{+0.18}$ & $0.69_{-0.07}^{+0.06}$ & $0.54_{-0.07}^{+0.14}$ &   784    &   16      & $0.41_{-0.00}^{+0.26}$ & $0.59_{-0.10}^{+0.13}$ & $0.69_{-0.07}^{+0.06}$ & $0.54_{-0.07}^{+0.15}$ \\
  208    &   112     & $0.44_{-0.01}^{+0.23}$ & $0.63_{-0.10}^{+0.12}$ & $0.70_{-0.07}^{+0.07}$ & $0.54_{-0.06}^{+0.14}$ &   784    &   112     & $0.46_{-0.01}^{+0.22}$ & $0.79_{-0.13}^{+1.28}$ & $0.70_{-0.07}^{+0.07}$ & $0.54_{-0.07}^{+0.15}$ \\
  208    &   208     & $0.46_{-0.01}^{+0.21}$ & $0.72_{-0.11}^{+0.18}$ & $0.70_{-0.07}^{+0.07}$ & $0.54_{-0.06}^{+0.14}$ &   784    &   208     & $0.48_{-0.01}^{+0.20}$ & $0.58_{-0.04}^{+0.17}$ & $0.70_{-0.07}^{+0.07}$ & $0.53_{-0.06}^{+0.16}$ \\
  208    &   304     & $0.47_{-0.01}^{+0.20}$ & $0.66_{-0.10}^{+0.11}$ & $0.70_{-0.07}^{+0.07}$ & $0.54_{-0.06}^{+0.14}$ &   784    &   304     & $0.48_{-0.01}^{+0.21}$ & $0.67_{-0.10}^{+0.10}$ & $0.70_{-0.07}^{+0.07}$ & $0.53_{-0.06}^{+0.16}$ \\
  208    &   400     & $0.47_{-0.01}^{+0.20}$ & $0.58_{-0.04}^{+0.17}$ & $0.70_{-0.07}^{+0.07}$ & $0.54_{-0.06}^{+0.14}$ &   784    &   400     & $0.48_{-0.01}^{+0.21}$ & $0.67_{-0.10}^{+0.10}$ & $0.70_{-0.07}^{+0.07}$ & $0.53_{-0.06}^{+0.16}$ \\
  208\tablenotemark{c}&  496    & $0.47_{-0.01}^{+0.21}$ & $0.58_{-0.04}^{+0.17}$ & $0.70_{-0.07}^{+0.07}$ & $0.56_{-0.09}^{+0.13}$ &   784    &   496     & $0.48_{-0.01}^{+0.21}$ & $0.62_{-0.07}^{+0.13}$ & $0.70_{-0.07}^{+0.07}$ & $0.55_{-0.08}^{+0.14}$ \\
  208    &   592     & $0.47_{-0.01}^{+0.21}$ & $0.58_{-0.04}^{+0.17}$ & $0.70_{-0.07}^{+0.07}$ & $0.55_{-0.07}^{+0.14}$ &   784    &   592     & $0.48_{-0.00}^{+0.21}$ & $0.68_{-0.10}^{+0.11}$ & $0.70_{-0.07}^{+0.07}$ & $0.55_{-0.08}^{+0.14}$ \\
  208    &   688     & $0.47_{-0.01}^{+0.21}$ & $0.58_{-0.04}^{+0.17}$ & $0.70_{-0.07}^{+0.07}$ & $0.55_{-0.07}^{+0.14}$ &   784    &   688     & $0.49_{-0.01}^{+0.20}$ & $0.68_{-0.10}^{+0.11}$ & $0.70_{-0.07}^{+0.08}$ & $0.55_{-0.08}^{+0.14}$ \\
  208    &   784     & $0.47_{-0.01}^{+0.21}$ & $0.58_{-0.04}^{+0.17}$ & $0.70_{-0.07}^{+0.08}$ & $0.55_{-0.07}^{+0.14}$ &   784    &   784     & $0.48_{-0.01}^{+0.21}$ & $0.59_{-0.04}^{+0.16}$ & $0.70_{-0.07}^{+0.08}$ & $0.55_{-0.09}^{+0.14}$ \\
  208    &   880     & $0.46_{-0.01}^{+0.21}$ & $0.81_{-0.13}^{+1.14}$ & $0.70_{-0.07}^{+0.07}$ & $0.53_{-0.06}^{+0.16}$ &   784    &   880     & $0.48_{-0.01}^{+0.20}$ & $0.58_{-0.04}^{+0.17}$ & $0.70_{-0.08}^{+0.08}$ & $0.55_{-0.09}^{+0.14}$ \\
  208    &   976     & $0.44_{-0.01}^{+0.23}$ & $0.56_{-0.04}^{+0.19}$ & $0.70_{-0.08}^{+0.07}$ & $0.55_{-0.08}^{+0.14}$ &   784    &   976     & $0.45_{-0.01}^{+0.23}$ & $0.79_{-0.13}^{+1.31}$ & $0.70_{-0.08}^{+0.07}$ & $0.54_{-0.07}^{+0.14}$ \\
  208    &   1008    & $0.43_{-0.01}^{+0.23}$ & $0.55_{-0.04}^{+0.20}$ & $0.70_{-0.08}^{+0.07}$ & $0.55_{-0.08}^{+0.14}$ &   784    &   1008    & $0.44_{-0.01}^{+0.23}$ & $0.56_{-0.04}^{+0.19}$ & $0.70_{-0.08}^{+0.07}$ & $0.52_{-0.06}^{+0.16}$ \\
  304    &   16      & $0.41_{-0.00}^{+0.26}$ & $0.53_{-0.04}^{+0.20}$ & $0.69_{-0.06}^{+0.06}$ & $0.54_{-0.06}^{+0.15}$ &   880    &   16      & $0.41_{-0.01}^{+0.25}$ & $0.72_{-0.14}^{+1.77}$ & $0.69_{-0.07}^{+0.06}$ & $0.55_{-0.09}^{+0.14}$ \\
  304    &   112     & $0.46_{-0.01}^{+0.22}$ & $0.79_{-0.13}^{+1.27}$ & $0.70_{-0.07}^{+0.07}$ & $0.54_{-0.06}^{+0.15}$ &   880    &   112     & $0.46_{-0.01}^{+0.21}$ & $0.79_{-0.13}^{+1.28}$ & $0.70_{-0.07}^{+0.07}$ & $0.55_{-0.09}^{+0.14}$ \\
  304    &   208     & $0.48_{-0.01}^{+0.21}$ & $0.58_{-0.04}^{+0.17}$ & $0.70_{-0.07}^{+0.07}$ & $0.54_{-0.06}^{+0.15}$ &   880    &   208     & $0.48_{-0.01}^{+0.20}$ & $0.58_{-0.04}^{+0.17}$ & $0.70_{-0.07}^{+0.07}$ & $0.55_{-0.09}^{+0.14}$ \\
  304    &   304     & $0.48_{-0.00}^{+0.21}$ & $0.67_{-0.10}^{+0.11}$ & $0.70_{-0.07}^{+0.07}$ & $0.54_{-0.06}^{+0.15}$ &   880    &   304     & $0.48_{-0.01}^{+0.21}$ & $0.67_{-0.10}^{+0.11}$ & $0.70_{-0.07}^{+0.07}$ & $0.55_{-0.09}^{+0.14}$ \\
  304    &   400     & $0.48_{-0.00}^{+0.22}$ & $0.67_{-0.10}^{+0.11}$ & $0.70_{-0.07}^{+0.07}$ & $0.54_{-0.06}^{+0.15}$ &   880    &   400     & $0.48_{-0.01}^{+0.21}$ & $0.67_{-0.10}^{+0.10}$ & $0.70_{-0.07}^{+0.07}$ & $0.55_{-0.09}^{+0.14}$ \\
  304    &   496     & $0.48_{-0.00}^{+0.22}$ & $0.67_{-0.10}^{+0.10}$ & $0.70_{-0.07}^{+0.07}$ & $0.56_{-0.08}^{+0.13}$ &   880    &   496     & $0.48_{-0.00}^{+0.21}$ & $0.67_{-0.10}^{+0.10}$ & $0.70_{-0.07}^{+0.07}$ & $0.55_{-0.09}^{+0.14}$ \\
  304    &   592     & $0.48_{-0.00}^{+0.22}$ & $0.68_{-0.10}^{+0.11}$ & $0.70_{-0.07}^{+0.07}$ & $0.56_{-0.08}^{+0.13}$ &   880    &   592     & $0.48_{-0.01}^{+0.21}$ & $0.67_{-0.10}^{+0.10}$ & $0.70_{-0.07}^{+0.07}$ & $0.55_{-0.09}^{+0.14}$ \\
  304    &   688     & $0.49_{-0.01}^{+0.21}$ & $0.68_{-0.10}^{+0.11}$ & $0.70_{-0.07}^{+0.08}$ & $0.56_{-0.08}^{+0.13}$ &   880    &   688     & $0.48_{-0.01}^{+0.21}$ & $0.67_{-0.10}^{+0.10}$ & $0.70_{-0.07}^{+0.08}$ & $0.55_{-0.09}^{+0.14}$ \\
  304    &   784     & $0.48_{-0.00}^{+0.22}$ & $0.68_{-0.10}^{+0.11}$ & $0.70_{-0.07}^{+0.08}$ & $0.56_{-0.09}^{+0.13}$ &   880    &   784     & $0.48_{-0.01}^{+0.21}$ & $0.67_{-0.10}^{+0.10}$ & $0.70_{-0.08}^{+0.08}$ & $0.54_{-0.07}^{+0.14}$ \\
  304    &   880     & $0.48_{-0.01}^{+0.21}$ & $0.58_{-0.04}^{+0.17}$ & $0.70_{-0.07}^{+0.07}$ & $0.55_{-0.07}^{+0.14}$ &   880    &   880     & $0.47_{-0.01}^{+0.21}$ & $0.58_{-0.04}^{+0.17}$ & $0.70_{-0.08}^{+0.08}$ & $0.55_{-0.09}^{+0.15}$ \\
  304    &   976     & $0.46_{-0.01}^{+0.22}$ & $0.79_{-0.13}^{+1.32}$ & $0.70_{-0.07}^{+0.07}$ & $0.53_{-0.06}^{+0.16}$ &   880    &   976     & $0.45_{-0.01}^{+0.23}$ & $0.79_{-0.13}^{+1.31}$ & $0.70_{-0.08}^{+0.07}$ & $0.55_{-0.08}^{+0.16}$ \\
  304    &   1008    & $0.45_{-0.01}^{+0.23}$ & $0.63_{-0.10}^{+0.13}$ & $0.70_{-0.07}^{+0.07}$ & $0.53_{-0.06}^{+0.16}$ &   880    &   1008    & $0.44_{-0.01}^{+0.23}$ & $0.63_{-0.10}^{+0.13}$ & $0.70_{-0.08}^{+0.07}$ & $0.55_{-0.08}^{+0.16}$ \\
  400    &   16      & $0.42_{-0.01}^{+0.25}$ & $0.59_{-0.10}^{+0.13}$ & $0.69_{-0.06}^{+0.06}$ & $0.54_{-0.07}^{+0.15}$ &   976    &   16      & $0.41_{-0.01}^{+0.25}$ & $0.59_{-0.10}^{+0.13}$ & $0.69_{-0.07}^{+0.06}$ & $0.55_{-0.09}^{+0.14}$ \\
  400    &   112     & $0.47_{-0.02}^{+0.21}$ & $0.79_{-0.13}^{+1.27}$ & $0.70_{-0.07}^{+0.07}$ & $0.54_{-0.06}^{+0.15}$ &   976    &   112     & $0.46_{-0.01}^{+0.22}$ & $0.79_{-0.13}^{+1.27}$ & $0.70_{-0.06}^{+0.07}$ & $0.55_{-0.09}^{+0.14}$ \\
  400    &   208     & $0.48_{-0.00}^{+0.21}$ & $0.58_{-0.04}^{+0.17}$ & $0.70_{-0.07}^{+0.07}$ & $0.54_{-0.06}^{+0.15}$ &   976    &   208     & $0.48_{-0.01}^{+0.21}$ & $0.58_{-0.04}^{+0.17}$ & $0.70_{-0.07}^{+0.07}$ & $0.55_{-0.09}^{+0.14}$ \\
  400    &   304     & $0.49_{-0.01}^{+0.21}$ & $0.67_{-0.10}^{+0.11}$ & $0.70_{-0.07}^{+0.07}$ & $0.54_{-0.06}^{+0.15}$ &   976    &   304     & $0.49_{-0.01}^{+0.21}$ & $0.67_{-0.10}^{+0.10}$ & $0.70_{-0.08}^{+0.07}$ & $0.54_{-0.08}^{+0.14}$ \\
  400    &   400     & $0.49_{-0.01}^{+0.21}$ & $0.67_{-0.10}^{+0.10}$ & $0.70_{-0.07}^{+0.07}$ & $0.56_{-0.08}^{+0.13}$ &   976    &   400     & $0.49_{-0.01}^{+0.21}$ & $0.67_{-0.10}^{+0.10}$ & $0.70_{-0.07}^{+0.08}$ & $0.55_{-0.09}^{+0.15}$ \\
  400    &   496     & $0.49_{-0.01}^{+0.21}$ & $0.67_{-0.10}^{+0.10}$ & $0.70_{-0.07}^{+0.07}$ & $0.56_{-0.08}^{+0.13}$ &   976    &   496     & $0.49_{-0.01}^{+0.21}$ & $0.68_{-0.10}^{+0.11}$ & $0.70_{-0.07}^{+0.08}$ & $0.55_{-0.09}^{+0.15}$ \\
  400    &   592     & $0.49_{-0.01}^{+0.21}$ & $0.68_{-0.10}^{+0.11}$ & $0.70_{-0.07}^{+0.07}$ & $0.56_{-0.09}^{+0.13}$ &   976    &   592     & $0.49_{-0.01}^{+0.21}$ & $0.68_{-0.10}^{+0.11}$ & $0.70_{-0.08}^{+0.08}$ & $0.55_{-0.08}^{+0.16}$ \\
  400    &   688     & $0.49_{-0.01}^{+0.21}$ & $0.68_{-0.10}^{+0.11}$ & $0.70_{-0.07}^{+0.08}$ & $0.55_{-0.07}^{+0.14}$ &   976    &   688     & $0.49_{-0.01}^{+0.21}$ & $0.68_{-0.10}^{+0.11}$ & $0.70_{-0.07}^{+0.09}$ & $0.55_{-0.08}^{+0.17}$ \\
  400    &   784     & $0.49_{-0.01}^{+0.21}$ & $0.68_{-0.10}^{+0.11}$ & $0.70_{-0.07}^{+0.08}$ & $0.56_{-0.09}^{+0.14}$ &   976    &   784     & $0.48_{-0.00}^{+0.22}$ & $0.68_{-0.10}^{+0.11}$ & $0.70_{-0.05}^{+0.09}$ & $0.50_{-0.04}^{+0.18}$ \\
  400    &   880     & $0.48_{-0.01}^{+0.21}$ & $0.58_{-0.04}^{+0.17}$ & $0.70_{-0.07}^{+0.08}$ & $0.53_{-0.05}^{+0.16}$ &   976    &   880     & $0.48_{-0.01}^{+0.21}$ & $0.67_{-0.10}^{+0.11}$ & $0.70_{-0.08}^{+0.09}$ & $0.50_{-0.04}^{+0.18}$ \\
  400    &   976     & $0.46_{-0.01}^{+0.22}$ & $0.57_{-0.04}^{+0.19}$ & $0.70_{-0.07}^{+0.08}$ & $0.55_{-0.08}^{+0.14}$ &   976    &   976     & $0.46_{-0.01}^{+0.22}$ & $0.57_{-0.04}^{+0.18}$ & $0.70_{-0.08}^{+0.08}$ & $0.50_{-0.04}^{+0.18}$ \\
  400    &   1008    & $0.45_{-0.01}^{+0.23}$ & $0.56_{-0.04}^{+0.19}$ & $0.70_{-0.08}^{+0.07}$ & $0.55_{-0.08}^{+0.14}$ &   976    &   1008    & $0.44_{-0.00}^{+0.23}$ & $0.56_{-0.04}^{+0.19}$ & $0.70_{-0.08}^{+0.08}$ & $0.50_{-0.04}^{+0.18}$ \\
  496    &   16      & $0.42_{-0.01}^{+0.25}$ & $0.53_{-0.04}^{+0.20}$ & $0.69_{-0.06}^{+0.06}$ & $0.54_{-0.07}^{+0.15}$ &   1008   &   16      & $0.40_{-0.00}^{+0.26}$ & $0.59_{-0.10}^{+0.13}$ & $0.69_{-0.07}^{+0.06}$ & $0.54_{-0.08}^{+0.15}$ \\
  496    &   112     & $0.47_{-0.01}^{+0.21}$ & $0.71_{-0.12}^{+0.13}$ & $0.70_{-0.07}^{+0.07}$ & $0.54_{-0.06}^{+0.15}$ &   1008   &   112     & $0.46_{-0.01}^{+0.22}$ & $0.71_{-0.11}^{+0.13}$ & $0.70_{-0.08}^{+0.07}$ & $0.52_{-0.06}^{+0.17}$ \\
  496    &   208     & $0.48_{-0.00}^{+0.22}$ & $0.67_{-0.10}^{+0.11}$ & $0.70_{-0.07}^{+0.07}$ & $0.54_{-0.06}^{+0.15}$ &   1008   &   208     & $0.48_{-0.00}^{+0.21}$ & $0.67_{-0.10}^{+0.11}$ & $0.70_{-0.07}^{+0.07}$ & $0.54_{-0.08}^{+0.14}$ \\
  496    &   304     & $0.49_{-0.01}^{+0.21}$ & $0.68_{-0.10}^{+0.11}$ & $0.70_{-0.07}^{+0.07}$ & $0.54_{-0.06}^{+0.15}$ &   1008   &   304     & $0.49_{-0.01}^{+0.21}$ & $0.68_{-0.10}^{+0.11}$ & $0.70_{-0.07}^{+0.08}$ & $0.54_{-0.08}^{+0.14}$ \\
  496    &   400     & $0.49_{-0.00}^{+0.21}$ & $0.59_{-0.04}^{+0.16}$ & $0.70_{-0.07}^{+0.08}$ & $0.55_{-0.07}^{+0.14}$ &   1008   &   400     & $0.49_{-0.01}^{+0.21}$ & $0.68_{-0.10}^{+0.11}$ & $0.70_{-0.08}^{+0.08}$ & $0.54_{-0.08}^{+0.14}$ \\
  496    &   496     & $0.49_{-0.01}^{+0.21}$ & $0.59_{-0.04}^{+0.17}$ & $0.70_{-0.07}^{+0.08}$ & $0.55_{-0.07}^{+0.14}$ &   1008   &   496     & $0.49_{-0.01}^{+0.21}$ & $0.59_{-0.04}^{+0.17}$ & $0.70_{-0.08}^{+0.08}$ & $0.55_{-0.08}^{+0.16}$ \\
  496    &   592     & $0.49_{-0.01}^{+0.21}$ & $0.59_{-0.04}^{+0.16}$ & $0.70_{-0.07}^{+0.08}$ & $0.55_{-0.07}^{+0.14}$ &   1008   &   592     & $0.49_{-0.01}^{+0.21}$ & $0.59_{-0.04}^{+0.17}$ & $0.70_{-0.08}^{+0.09}$ & $0.54_{-0.08}^{+0.14}$ \\
  496    &   688     & $0.49_{-0.01}^{+0.21}$ & $0.59_{-0.04}^{+0.17}$ & $0.70_{-0.07}^{+0.08}$ & $0.53_{-0.05}^{+0.16}$ &   1008   &   688     & $0.49_{-0.01}^{+0.21}$ & $0.59_{-0.04}^{+0.17}$ & $0.70_{-0.03}^{+0.10}$ & $0.50_{-0.04}^{+0.18}$ \\
  496    &   784     & $0.49_{-0.01}^{+0.21}$ & $0.68_{-0.10}^{+0.11}$ & $0.70_{-0.07}^{+0.08}$ & $0.55_{-0.07}^{+0.14}$ &   1008   &   784     & $0.49_{-0.01}^{+0.21}$ & $0.59_{-0.04}^{+0.17}$ & $0.70_{-0.02}^{+0.11}$ & $0.50_{-0.04}^{+0.18}$ \\
  496    &   880     & $0.48_{-0.01}^{+0.21}$ & $0.58_{-0.04}^{+0.17}$ & $0.70_{-0.07}^{+0.08}$ & $0.55_{-0.08}^{+0.14}$ &   1008   &   880     & $0.48_{-0.01}^{+0.21}$ & $0.68_{-0.10}^{+0.11}$ & $0.70_{-0.04}^{+0.10}$ & $0.50_{-0.04}^{+0.19}$ \\
  496    &   976     & $0.46_{-0.01}^{+0.22}$ & $0.57_{-0.04}^{+0.19}$ & $0.70_{-0.08}^{+0.08}$ & $0.55_{-0.09}^{+0.14}$ &   1008   &   976     & $0.46_{-0.01}^{+0.22}$ & $0.81_{-0.13}^{+1.26}$ & $0.70_{-0.08}^{+0.09}$ & $0.50_{-0.04}^{+0.18}$ \\
  496    &   1008    & $0.45_{-0.01}^{+0.23}$ & $0.56_{-0.04}^{+0.20}$ & $0.70_{-0.08}^{+0.08}$ & $0.55_{-0.09}^{+0.14}$ &   1008   &   1008    & $0.44_{-0.01}^{+0.23}$ & $0.79_{-0.13}^{+1.37}$ & $0.70_{-0.08}^{+0.09}$ & $0.54_{-0.09}^{+0.15}$
\label{tab:corrections}
\end{longtable}
\noindent$^{a}$ -- X and Y are the tile center in chip coordinates.\\
\noindent$^{b}$ -- Values are the mode of the correction factor
distribution (Eq.\ref{eq:corr_pdf}) with the 68\% CL error (see Sec.
\ref{sec:corrfac} \& \ref{sec:unc_sys}). The values should be used as
input in Eq. \ref{eq:flux}.\\
\noindent$^{c}$ -- Aimpoint on ACIS-S3\\

%% file: ms2.bbl
\begin{thebibliography}{23}
\expandafter\ifx\csname natexlab\endcsname\relax\def\natexlab#1{#1}\fi

\bibitem[{{Barlow}(2004)}]{barlow:04}
{Barlow}, R. 2004, ArXiv Physics e-prints

\bibitem[{{Bautz} {et~al.}(1999){Bautz}, {Prigozhin}, {Pivovaroff}, {Jones},
  {Kissel}, \& {Ricker}}]{bautz:99}
{Bautz}, M.~W., {Prigozhin}, G.~Y., {Pivovaroff}, M.~J., {Jones}, S.~E.,
  {Kissel}, S.~E., \& {Ricker}, G.~R. 1999, Nuclear Instruments and Methods in
  Physics Research A, 436, 40

\bibitem[{{Bessell}(2005)}]{bessel:05}
{Bessell}, M.~S. 2005, \araa, 43, 293

\bibitem[{{Burke} {et~al.}(1993){Burke}, {Mountain}, {Daniels}, {Cooper}, \&
  {Dolat}}]{burke:93}
{Burke}, B.~E., {Mountain}, R.~W., {Daniels}, P.~J., {Cooper}, M.~J., \&
  {Dolat}, V.~S. 1993, in Presented at the Society of Photo-Optical
  Instrumentation Engineers (SPIE) Conference, Vol. 2006, Proc. SPIE Vol. 2006,
  p. 272-285, EUV, X-Ray, and Gamma-Ray Instrumentation for Astronomy IV,
  Oswald H. Siegmund; Ed., ed. O.~H. {Siegmund}, 272--285

\bibitem[{CXC(2008)}]{pog:08}
CXC. 2008, The Chandra Proposers' Observatory Guide, 10th edn., Chandra X-ray
  Center, http://cxc.harvard.edu/proposer/POG/html/index.html

\bibitem[{{Drake} {et~al.}(2006){Drake}, {Ratzlaff}, {Kashyap}, {Edgar},
  {Izem}, {Jerius}, {Siemiginowska}, \& {Vikhlinin}}]{drake:06}
{Drake}, J.~J., {Ratzlaff}, P., {Kashyap}, V., {Edgar}, R., {Izem}, R.,
  {Jerius}, D., {Siemiginowska}, A., \& {Vikhlinin}, A. 2006, in Presented at
  the Society of Photo-Optical Instrumentation Engineers (SPIE) Conference,
  Vol. 6270, Observatory Operations: Strategies, Processes, and Systems. Edited
  by Silva, David R.; Doxsey, Rodger E.. Proceedings of the SPIE, Volume 6270,
  pp. 62701I (2006).

\bibitem[{{Fabbiano} {et~al.}(2007){Fabbiano}, {Evans}, {Evans}, {Glotfelty},
  {Hain}, {McCollough}, {Primini}, {Rots}, {Calderwood}, {Doe}, {Grier},
  {Harbo}, {Karovska}, {McDowell}, {Plummer}, {Paton}, {Tibbetts}, {Stone}, \&
  {Zografou}}]{fabbiano:07}
{Fabbiano}, G., {Evans}, I., {Evans}, J., {Glotfelty}, K., {Hain}, R.,
  {McCollough}, M., {Primini}, F., {Rots}, A., {Calderwood}, T., {Doe}, S.,
  {Grier}, J., {Harbo}, P., {Karovska}, M., {McDowell}, J., {Plummer}, D.,
  {Paton}, L., {Tibbetts}, M., {Stone}, D.~V., \& {Zografou}, P. 2007, in
  Astronomical Society of the Pacific Conference Series, Vol. 376, Astronomical
  Society of the Pacific Conference Series, ed. R.~A. {Shaw}, F.~{Hill}, \&
  D.~J. {Bell}, 172--+

\bibitem[{{Fan} {et~al.}(1999){Fan}, {Strauss}, {Schneider}, {Gunn}, {Lupton},
  {Yanny}, {Anderson}, {Anderson}, {Annis}, {Bahcall}, {Bakken}, {Bastian},
  {Berman}, {Boroski}, {Briegel}, {Briggs}, {Brinkmann}, {Carr}, {Colestock},
  {Connolly}, {Crocker}, {Csabai}, {Czarapata}, {Davis}, {Doi}, {Elms},
  {Evans}, {Federwitz}, {Frieman}, {Fukugita}, {Gurbani}, {Harris}, {Heckman},
  {Hennessy}, {Hindsley}, {Holmgren}, {Hull}, {Ichikawa}, {Ichikawa},
  {Ivezi{\'c} }, {Kent}, {Knapp}, {Kron}, {Lamb}, {Leger}, {Limmongkol},
  {Lindenmeyer}, {Long}, {Loveday}, {MacKinnon}, {Mannery}, {Mantsch},
  {Margon}, {McKay}, {Munn}, {Nash}, {Newberg}, {Nichol}, {Nicinski},
  {Okamura}, {Ostriker}, {Owen}, {Pauls}, {Peoples}, {Petravick}, {Pier},
  {Pordes}, {Prosapio}, {Rechenmacher}, {Richards}, {Richmond}, {Rivetta},
  {Rockosi}, {Sandford}, {Sergey}, {Sekiguchi}, {Shimasaku}, {Siegmund},
  {Smith}, {Stoughton}, {Szalay}, {Szokoly}, {Tucker}, {Vogeley}, {Waddell},
  {Wang}, {Weinberg}, {Yasuda}, \& {York}}]{fan:99}
{Fan}, X., {Strauss}, M.~A., {Schneider}, D.~P., {Gunn}, J.~E., {Lupton},
  R.~H., {Yanny}, B., {Anderson}, S.~F., {Anderson}, Jr., J.~E., {Annis}, J.,
  {Bahcall}, N.~A., {Bakken}, J.~A., {Bastian}, S., {Berman}, E., {Boroski},
  W.~N., {Briegel}, C., {Briggs}, J.~W., {Brinkmann}, J., {Carr}, M.~A.,
  {Colestock}, P.~L., {Connolly}, A.~J., {Crocker}, J.~H., {Csabai}, I.,
  {Czarapata}, P.~C., {Davis}, J.~E., {Doi}, M., {Elms}, B.~R., {Evans}, M.~L.,
  {Federwitz}, G.~R., {Frieman}, J.~A., {Fukugita}, M., {Gurbani}, V.~K.,
  {Harris}, F.~H., {Heckman}, T.~M., {Hennessy}, G.~S., {Hindsley}, R.~B.,
  {Holmgren}, D.~J., {Hull}, C., {Ichikawa}, S.-I., {Ichikawa}, T., {Ivezi{\'c}
  }, {\v Z}., {Kent}, S., {Knapp}, G.~R., {Kron}, R.~G., {Lamb}, D.~Q.,
  {Leger}, R.~F., {Limmongkol}, S., {Lindenmeyer}, C., {Long}, D.~C.,
  {Loveday}, J., {MacKinnon}, B., {Mannery}, E.~J., {Mantsch}, P.~M., {Margon},
  B., {McKay}, T.~A., {Munn}, J.~A., {Nash}, T., {Newberg}, H.~J., {Nichol},
  R.~C., {Nicinski}, T., {Okamura}, S., {Ostriker}, J.~P., {Owen}, R., {Pauls},
  A.~G., {Peoples}, J., {Petravick}, D., {Pier}, J.~R., {Pordes}, R.,
  {Prosapio}, A., {Rechenmacher}, R., {Richards}, G.~T., {Richmond}, M.~W.,
  {Rivetta}, C.~H., {Rockosi}, C.~M., {Sandford}, D., {Sergey}, G.,
  {Sekiguchi}, M., {Shimasaku}, K., {Siegmund}, W.~A., {Smith}, J.~A.,
  {Stoughton}, C., {Szalay}, A.~S., {Szokoly}, G.~P., {Tucker}, D.~L.,
  {Vogeley}, M.~S., {Waddell}, P., {Wang}, S.-I., {Weinberg}, D.~H., {Yasuda},
  N., \& {York}, D.~G. 1999, \aj, 118, 1

\bibitem[{{Gehrels}(1986)}]{gehrels:86}
{Gehrels}, N. 1986, \apj, 303, 336

\bibitem[{{Gierli{\'n}ski} \& {Newton}(2006)}]{gierlinski:06}
{Gierli{\'n}ski}, M. \& {Newton}, J. 2006, \mnras, 370, 837

\bibitem[{{Grimm} {et~al.}(2007){Grimm}, {McDowell}, {Zezas}, {Kim}, \&
  {Fabbiano}}]{grimm:07}
{Grimm}, H.-J., {McDowell}, J., {Zezas}, A., {Kim}, D.-W., \& {Fabbiano}, G.
  2007, \apjs, 173, 70

\bibitem[{{Hasinger} \& {van der Klis}(1989)}]{hasinger:89}
{Hasinger}, G. \& {van der Klis}, M. 1989, \aap, 225, 79

\bibitem[{{Johnson} \& {Morgan}(1953)}]{johnson:53}
{Johnson}, H.~L. \& {Morgan}, W.~W. 1953, \apj, 117, 313

\bibitem[{{Kraft} {et~al.}(1991){Kraft}, {Burrows}, \& {Nousek}}]{kraft:91}
{Kraft}, R.~P., {Burrows}, D.~N., \& {Nousek}, J.~A. 1991, \apj, 374, 344

\bibitem[{{Marshall} {et~al.}(2004){Marshall}, {Tennant}, {Grant}, {Hitchcock},
  {O'Dell}, \& {Plucinsky}}]{marshall:04}
{Marshall}, H.~L., {Tennant}, A., {Grant}, C.~E., {Hitchcock}, A.~P., {O'Dell},
  S.~L., \& {Plucinsky}, P.~P. 2004, in Presented at the Society of
  Photo-Optical Instrumentation Engineers (SPIE) Conference, Vol. 5165, X-Ray
  and Gamma-Ray Instrumentation for Astronomy XIII. Edited by Flanagan, Kathryn
  A.; Siegmund, Oswald H. W. Proceedings of the SPIE, Volume 5165, pp. 497-508
  (2004)., ed. K.~A. {Flanagan} \& O.~H.~W. {Siegmund}, 497--508

\bibitem[{{Plucinsky} {et~al.}(2003){Plucinsky}, {Schulz}, {Marshall}, {Grant},
  {Chartas}, {Sanwal}, {Teter}, {Vikhlinin}, {Edgar}, {Wise}, {Allen},
  {Virani}, {DePasquale}, \& {Raley}}]{plucinsky:03}
{Plucinsky}, P.~P., {Schulz}, N.~S., {Marshall}, H.~L., {Grant}, C.~E.,
  {Chartas}, G., {Sanwal}, D., {Teter}, M., {Vikhlinin}, A.~A., {Edgar}, R.~J.,
  {Wise}, M.~W., {Allen}, G.~E., {Virani}, S.~N., {DePasquale}, J.~M., \&
  {Raley}, M.~T. 2003, in Presented at the Society of Photo-Optical
  Instrumentation Engineers (SPIE) Conference, Vol. 4851, X-Ray and Gamma-Ray
  Telescopes and Instruments for Astronomy. Edited by Joachim E. Truemper,
  Harvey D. Tananbaum. Proceedings of the SPIE, Volume 4851, pp. 89-100
  (2003)., ed. J.~E. {Truemper} \& H.~D. {Tananbaum}, 89--100

\bibitem[{{Prestwich} {et~al.}(2003){Prestwich}, {Irwin}, {Kilgard}, {Krauss},
  {Zezas}, {Primini}, {Kaaret}, \& {Boroson}}]{prestwich:03}
{Prestwich}, A.~H., {Irwin}, J.~A., {Kilgard}, R.~E., {Krauss}, M.~I., {Zezas},
  A., {Primini}, F., {Kaaret}, P., \& {Boroson}, B. 2003, \apj, 595, 719

\bibitem[{{Prigozhin} {et~al.}(1998){Prigozhin}, {Rasmussen}, {Bautz}, \&
  {Ricker}}]{prigozhin:98}
{Prigozhin}, G.~Y., {Rasmussen}, A., {Bautz}, M.~W., \& {Ricker}, G.~R. 1998,
  in Presented at the Society of Photo-Optical Instrumentation Engineers (SPIE)
  Conference, Vol. 3444, Proc. SPIE Vol. 3444, p. 267-275, X-Ray Optics,
  Instruments, and Missions, Richard B. Hoover; Arthur B. Walker; Eds., ed.
  R.~B. {Hoover} \& A.~B. {Walker}, 267--275

\bibitem[{{Puschell} {et~al.}(1981){Puschell}, {Owen}, \&
  {Laing}}]{puschell:81}
{Puschell}, J.~J., {Owen}, F.~N., \& {Laing}, R.~A. 1981, in Bulletin of the
  American Astronomical Society, Vol.~13, Bulletin of the American Astronomical
  Society, 507--+

\bibitem[{{Schwartz} {et~al.}(2000){Schwartz}, {David}, {Donnelly}, {Edgar},
  {Gaetz}, {Graessle}, {Jerius}, {Juda}, {Kellogg}, {McNamara}, {Plucinsky},
  {Van Speybroeck}, {Wargelin}, {Wolk}, {Zhao}, {Dewey}, {Marshall}, {Schulz},
  {Elsner}, {Kolodziejczak}, {O'Dell}, {Swartz}, {Tennant}, \&
  {Weisskopf}}]{schwartz:00}
{Schwartz}, D.~A., {David}, L.~P., {Donnelly}, R.~H., {Edgar}, R.~J., {Gaetz},
  T.~J., {Graessle}, D.~E., {Jerius}, D., {Juda}, M., {Kellogg}, E.~M.,
  {McNamara}, B.~R., {Plucinsky}, P.~P., {Van Speybroeck}, L.~P., {Wargelin},
  B.~J., {Wolk}, S., {Zhao}, P., {Dewey}, D., {Marshall}, H.~L., {Schulz},
  N.~S., {Elsner}, R.~F., {Kolodziejczak}, J.~J., {O'Dell}, S.~L., {Swartz},
  D.~A., {Tennant}, A.~F., \& {Weisskopf}, M.~C. 2000, in Presented at the
  Society of Photo-Optical Instrumentation Engineers (SPIE) Conference, Vol.
  4012, Proc. SPIE Vol. 4012, p. 28-40, X-Ray Optics, Instruments, and Missions
  III, Joachim E. Truemper; Bernd Aschenbach; Eds., ed. J.~E. {Truemper} \&
  B.~{Aschenbach}, 28--40

\bibitem[{{Watson}(2007)}]{watson:07}
{Watson}, M. 2007, Astronomy and Geophysics, 48, 30

\bibitem[{{White} \& {Marshall}(1984)}]{white:84}
{White}, N.~E. \& {Marshall}, F.~E. 1984, \apj, 281, 354

\bibitem[{{Zhao} {et~al.}(2004){Zhao}, {Jerius}, {Edgar}, {Gaetz}, {Van
  Speybroeck}, {Biller}, {Beckerman}, \& {Marshall}}]{zhao:04}
{Zhao}, P., {Jerius}, D.~H., {Edgar}, R.~J., {Gaetz}, T.~J., {Van Speybroeck},
  L.~P., {Biller}, B., {Beckerman}, E., \& {Marshall}, H.~L. 2004, in Presented
  at the Society of Photo-Optical Instrumentation Engineers (SPIE) Conference,
  Vol. 5165, X-Ray and Gamma-Ray Instrumentation for Astronomy XIII. Edited by
  Flanagan, Kathryn A.; Siegmund, Oswald H. W. Proceedings of the SPIE, Volume
  5165, pp. 482-496 (2004)., ed. K.~A. {Flanagan} \& O.~H.~W. {Siegmund},
  482--496

\end{thebibliography}
